\documentclass[11pt]{article}

\usepackage{amsmath, amssymb, color, amsthm, epsfig, natbib, lscape, caption2, mathrsfs, amsfonts, fancyhdr, mathabx}
\usepackage{subfigure}
\usepackage{graphicx}
\usepackage{adjustbox}
\usepackage{pdflscape}
\usepackage{algpseudocode}
\makeatletter
\usepackage{multirow,xspace,setspace,booktabs,subfigure}
\usepackage{tabularx}
\usepackage{tabulary}

\usepackage{algorithm2e}

\allowdisplaybreaks

\DeclareMathOperator*{\argmin}{argmin}

\newcommand{\Date}[1]{\def\@Date{#1}}
\def\today{\number\day~\ifcase\month\or
	January\or February\or March\or April\or May\or June\or
	July\or August\or September\or October\or November\or December\fi~\number\year}

\def \b1{{\bf 1}}

\def \bq{{q}}

\def \bbeta{{\boldsymbol \beta}}

\def \btheta{{\boldsymbol \theta}}

\def \H{{\mathcal{H}}}
\def \P{{\mathbb{P}}}
\def \B{{\mathbb{B}}}
\def \bB{{\bar{B}}}
\def \C{{\mathbb{C}}}
\def \mE{{\mathbb{E}}}
\def \mI{{\mathbb{I}}}
\def \mR{{\mathbb{R}}}
\def \I{{\mathbb{I}}}
\def \tr{{\rm tr}}

\def \W{{\tilde W}}

\newtheorem{theorem}{Theorem}%[section]
\newtheorem{lemma}{Lemma}

\theoremstyle{definition}
\newtheorem{condition}{Condition}

\newtheorem{remark}{Remark}
\newtheorem{corollary}{Corollary}
\theoremstyle{remark}

\newtheorem{example}{Example}

\begin{document}
	
	\begin{center}
		{\bf\LARGE
			Statistical Inference for Ultrahigh Dimensional Location Parameter Based on Spatial Median % Signal-plus-noise Model\\
		}
	\end{center}
	
	%\bigskip
	
	\begin{center}
		Guanghui  Cheng$^a$, ~~~~~~~~Liuhua Peng$^b$, ~~~~~~~~Changliang Zou$^c$
	\end{center}

	\begin{center}
		{\sl $^a$Guangzhou Institute of International Finance, Guangzhou University, \\ 
			$^b$School of Mathematics and Statistics, The University of Melbourne, \\
			$^c$ School of Statistics and Data Science, Nankai University}%The University of Macau, The University of Melbourne}
	\end{center}

\begin{abstract}
	Motivated by the widely used geometric median-of-means estimator in machine learning, this paper studies statistical inference for ultrahigh dimensionality location parameter based on the sample spatial median under a general multivariate model, including simultaneous confidence intervals construction, global tests, and multiple testing with false discovery rate control. %, as well as asymptotic relative efficiency of the sample spatial median relative to the sample mean. 
	To achieve these goals, we derive a novel Bahadur representation of the sample spatial median with a maximum-norm bound on the remainder term, and establish Gaussian approximation for the sample spatial median over the class of hyperrectangles.
	In addition, a multiplier bootstrap algorithm is proposed to approximate the distribution of the sample spatial median.
	The approximations are valid when the dimension diverges at an exponentially rate of the sample size, which facilitates the application of the spatial median in the ultrahigh dimensional region.
	%Moreover, we extend the Gaussian and bootstrap approximations to the two-sample problem, when the difference between two location parameter is of interest. 
	The proposed approaches are further illustrated by simulations and analysis of a genomic dataset from a microarray study.
\end{abstract}

\noindent
{\sl keywords}:  Bootstrap approximation; Gaussian approximation;  high-dimensional; spatial median; FDR control

\section{Introduction}
\label{s:intro}
\textcolor{black}{Geometric median-of-means (GMOM) has been widely used for robust estimation of multivariate means, and it has been broadly adopted in machine learning \citep{Minsker2015,Hsu2016,Prasad2020}. The idea of GMOM is to first divide the data into disjoint subsamples and calculate the empirical means of each of the subsamples. Then the GMOM estimator is computed as the spatial median (also called geometric median) of the obtained empirical means. The previous studies on the GMOM focused on establishing its non-asymptotic error bounds under certain heavy-tailed assumptions. Its distributional properties, which are essential for statistical inference, remain unknown.}

High-dimensional data with the dimension increases to infinity as the number of observations goes to infinity have been encountered in many scientific disciplines.
There is a growing evidence of the multivariate normal distribution is problematic to model high-dimensional data due to the presents of heavy-tailedness and inadequate to accommodate tail dependence.
For example, the distributions of the microarray expression are observed to be non-normal and have heavy tails even after log transformation in many gene expression data \citep{Purdom2005,Wang2015}.
As another example, elliptical distributions, in particular the multivariate $t$-distribution and symmetric multivariate normal inverse Gaussian distribution, provided far superior models to the multivariate normal for daily and weekly US stock-return data \citep{McNeil2005}.
In such cases, the sample spatial median is favored against the sample mean for estimating the location parameter. 
\textcolor{black}{The above discussions strongly motivate studying the spatial median under high-dimensionality, especially its distributional properties and the implementation in statistical inference for high-dimensional location parameter.}

Let $X_1,\ldots,X_n$ be a sequence of independent and identically distributed (i.i.d.) $p$-dimensional random vectors from a population $X$ with cumulative distribution function $F_X$ in $\mathbb{R}^p$. 
In this paper, we work on a general multivariate model where $X$ admits the following stochastic representation:
\begin{eqnarray}\label{eq:Model_X}
	X = \btheta + \nu\Gamma U\,,
\end{eqnarray}
where $\btheta$ is the location parameter, $\nu$ is a nonnegative univariate random variable and $U$ is a $p$-dimensional random vector with independent components. Model \eqref{eq:Model_X} covers many commonly used multivariate models and distribution families, including the independent components model \citep{Yao2015} and the elliptical distribution family \citep{Fang1990}. We refer to Section \ref{sec:note} for more detailed discussions.

Spatial median, an extension of the univariate median to multivariate distributions, was proposed for robust inference of the location parameter \citep{Haldane1948,Weber1929}. 
The sample spatial median $\hat{\btheta}_{n}\in\mathbb{R}^{p}$ minimizes the empirical criteria function $L_n(\bbeta) =\sum_{i=1}^{n}(\|X_i-\bbeta\|-\|X_i\|)$, where $\|\cdot\|$ is the Euclidean norm. Equivalently,
\begin{eqnarray}\label{eq:btheta_criteria}
	\hat{\btheta}_{n} = \argmin_{\bbeta\in {\mathbb R}^p}L_{n}(\bbeta) = \argmin_{\bbeta\in {\mathbb R}^p} \sum_{i=1}^{n}(\|X_i-\bbeta\|-\|X_i\|)\,. 
\end{eqnarray}
The function $L_{n}(\bbeta)$ is convex, and $\hat{\btheta}_{n}$ is unique if the observations $\{X_i\}_{i=1}^{n}$ are not concentrated on a line in $\mathbb{R}^{p}$ when $p> 2$ \citep{Milasevic1987}.
When the dimension $p$ is fixed, the spatial median has been well studied in the literature. 
We refer to Chapter 6.2 of \citet{Oja2010} for a nice review. 

In the high-dimensional setting, where the dimension $p$ diverges to infinity as the number of observations $n\to\infty$, there are several existing works that study the asymptotic properties of the sample spatial median.
\citet{Zou2014} offered an expansion of $\hat{\btheta}_{n}$ under elliptical distributions with identical shape matrix, and \citet{Cheng2019} extended the result to a general shape matrix. 
As a recent work, \citet{LiXu2022} improved the expansion in \citet{Cheng2019} with a smaller order remainder term under stronger conditions, and established a central limit theorem for the squared Euclidean distance $\|\hat{\btheta}_{n}-\btheta\|^2$.
In \citet{Zou2014} and \citet{Cheng2019}, they both require that $p=O(n^2)$.
In addition, it is required in \citet{LiXu2022} that $p$ diverges at the same rate as $n$.
However, in modern areas such as genomics and proteomics, the dimension of the data may grow exponentially with the sample size, which lies in the ``ultrahigh dimensional'' region \citep{FanLv2008}.
The previous works with restrictions on the polynomial dimensionality limit the usage of the spatial median under ultrahigh-dimensionality.
Moreover, the previous results are all under elliptical distributions.
Thus, it is of great importance to establish asymptotic properties of the spatial median and investigate its applications under ultrahigh dimensionality and beyond elliptical distributions.

%Throughout the paper, we consider the high-dimensional scenario with $p,  n\geq 5$ and $p=p(n)\to\infty$ as $n\to\infty$.
%To solve this issue, we aim to establish valid statistical inference for $\btheta$ which allows a faster divergent $p$.
In this paper, we first establish Gaussian and bootstrap approximations hit hyperrectangles for the sample spatial median under the general model \eqref{eq:Model_X} beyond elliptical distributions, which are valid when the dimension diverges exponential with the sample size.
They serve as the theoretical foundations of statistical inference for the location parameter based on the sample spatial median under ultrahigh dimensionality. 
Consistent simultaneous confidence intervals (SCIs) and global tests for the location parameters are established.
We also study multiple testing for every component of $\btheta$ based on $\hat{\btheta}_n$. %, with false discovery rate (FDR) controlled combined with the Benjamini-Hochberg procedure. 
Motivated by simultaneous inference of $\btheta$, we define a high-dimensional asymptotic relative efficiency of the sample spatial median relative to the sample mean.
Most importantly, our theoretical results guarantee the validity of the proposed inferential methods for exponentially divergent $p$.
The advantages of our proposed approaches have been justified by simulations and a real data analysis.
%Finally, we extend the Gaussian and bootstrap approximations to the two-sample problem, when the difference between two location parameters is of interest. 

The main contributions of this paper are summarized as follow. 
Firstly, we establish SCIs for the location parameter $\btheta$ based on the sample spatial median $\hat{\btheta}_n$, which is new in the literature. 
The consistency of bootstrap approximation guarantees that the probability that the SCIs cover all components of the location parameter approaches the nominal confidence level under ultrahigh dimensionality.
We also propose a novel test for ultrahigh dimensional location parameter based on the maximum-norm of the sample spatial median. 
The proposed test not only maintains nominal significance level asymptotically for exponentially divergent $p$, but also is more powerful under sparse alternatives compared to those based on $L_2$-norms \citep{LiXu2022,Wang2015}.
As another major inference, we study multiple testing for every component of the location parameter, and the false discovery rate (FDR) can be well controlled combined with the Benjamini-Hochberg procedure based on the sample spatial median, which extends the existing methods based on the sample mean \citep{Liu2014}. 
In all inferential methods, the procedures based on the sample spatial median advances those based on the sample mean for heavy-tailed distributions.

Secondly, this paper serves as the first work that provides Gaussian and bootstrap approximations for the sample spatial median under ultrahigh dimensionality. 
Gaussian and bootstrap approximations for high-dimensional sample mean have received extensive attraction in the last decade. 
\citet{Cher2013} and \citet{Cher2017} established Gaussian and bootstrap approximations for the maxima of a sum of centered independent random vectors under Kolmogorov distance and on hyperrectangles, respectively. 
See also \citet{Chen2018}, \citet{Cher2019} and \citet{Cher2020} for related works.
Compared to the sample mean, which has a simple linear form, 
the theoretical difficulty for the sample spatial median lies in that it does not enjoy an explicit form.
This issue is addressed by deriving a novel Bahadur representation of the sample spatial median with a maximum-norm bound on the remainder term, which extends the results of \citet{Zou2014}, \citet{Cheng2019} and \citet{LiXu2022} under elliptical distributions and polynomial dimensionality.
%A potential relevant work is \citet{Imaizumi2021}, which studied Gaussian approximation for M-estimator. 
%However, their results cannot be applied in our framework, we refer to Remark \ref{remark:M-estimator} for a detailed discussion.
\textcolor{black}{Moreover, our results can be applied to the GMOM under reasonable conditions, and thus enhance the practice usage of GMOM.}

Thirdly, we propose a novel multiplier bootstrap method for the sample spatial median. % regarding the form of the criteria function $L_{n}(\bbeta)$. 
Instead of multiplying on the loss function, which is generally the case for M-estimator \citep{Imaizumi2021}, the multiplier is applied on the centralized $X_i$. Specifically, the bootstrap version of $\hat{\btheta}_n$ is defined as $\tilde{\btheta}_{n}= \argmin_{\bbeta\in {\mathbb R}^d}\sum_{i=1}^n\|Z_i(X_i-\hat{\btheta}_n)-\bbeta\|$, where $Z_1,\ldots,Z_n$ are the multipliers.
The multiplier bootstrap is consistent under ultrahigh dimensionality thanks to this novel formulation.
This is, however, different from the multiplier bootstrap method for the sample mean, which again has an explicit form \citep{Cher2013,Cher2017}.

The rest of the paper is organized as follows. 
Section \ref{sec:note} introduces model and assumptions. 
Section \ref{sec:main} establishes Gaussian and bootstrap approximations to the distribution of the sample spatial median. 
Statistical inference for the location parameter based on the sample spatial median is presented in Section \ref{sec:applications}.
%Section \ref{sec:two-sample} extends the results to the two-sample problem.
Section \ref{sec:numerical} reports numerical results including simulations and a real data analysis. 
Preliminary lemmas and proofs of main results are presented in Appendix A of the supplementary material.
%Concluding remarks are made in Section \ref{sec:conclusion}.
Proofs of preliminary lemmas and additional simulations are given in Appendices B and C of the supplementary material.

\textbf{Notation:}
%The following notation is used throughout.
Denote %$\|x\|_{\varrho}=\big(\sum_{j=1}^{d}|x_j|^{\varrho}\big)^{1/\varrho}$ as the vector $\varrho$-norm of $x=(x_1,\ldots,x_d)^{\top}$ for $\varrho\geq1$, and
$|x|_{\infty}=\max(|x_1|,\ldots,|x_d|)$ as the maximum-norm of $x=(x_1,\ldots,x_d)^{\top}$.
Denote $a_n\lesssim b_n$ if $a_n\leq Cb_n$ for a positive constant $C$, and $a_n\asymp b_n$ means $a_n\lesssim b_n$ and $b_n\lesssim a_n$. 
% In addition, $a \approx b$ means $a/b=1+o(1)$. 
For $\alpha>0$, let $\psi_{\alpha}(x)=\exp(x^{\alpha})-1$ be a function defined on $[0,\infty)$. Then the Orlicz norm $\|\cdot\|_{\psi_{\alpha}}$ of a random variable $X$ is defined as
$
\|X\|_{\psi_{\alpha}}=\inf\left\{t>0, \mE\{\psi_{\alpha}\left(|X|/t\right)\}\leq 1\right\}.
$
We use $\tr(\cdot)$ to denote the trace operator for square matrices.
Moreover, we denote $I_p$ as the $p\times p$ identity matrix.
%\textcolor{red}{define $\tr$ here}
For $a,b\in\mathbb{R}$, we write $a \wedge b=\min(a,b)$.

\section{Model and assumptions}\label{sec:note}

In this paper, we consider a general multivariate model for the distribution $F_X$ such that $X_i$ admits the following stochastic representation: %\citep{Fang1990}:
\begin{eqnarray}\label{eq:model}
	X_i = \btheta + %\epsilon_i, ~~ {\rm and}~~ \epsilon_i=
	\nu_i\Gamma U_i\,,
\end{eqnarray}
where $\btheta$ is the location parameter, $\Gamma$ is a nonrandom and invertible $p\times p$ matrix, 
$U_i$ is a $p$-dimensional random vector with independent~standardized components, and $\nu_i$ is a nonnegative univariate random variable independent with the spatial sign of $U_i$.
The distribution of $X_i$ depends on $\Gamma$ through the shape matrix $\Omega=\Gamma\Gamma^{\top}$. %, where $\Omega$ is the shape matrix of $F_{X}$.
%In addition, $\Omega$ relates to the covariance matrix $\Sigma=\mE(X_iX_i^{\top})$ by $\Sigma=\mE(\nu_i^2)\Omega$. 
%However, an additional parameter $\nu_i$ controls the tail behavior, which produces diverse characteristics.
%In particular, the multivariate normal distribution with mean $\btheta$ and covariance matrix $\Sigma$ is elliptically distributed with $\nu_i^2$ follows a chi-square distribution with $p$ degrees of freedom and $\Omega=\Sigma$.

\begin{remark}
	Model \eqref{eq:model} covers many commonly used multivariate models and distribution families. 
	First, the independent components model \citep{Yao2015} follows \eqref{eq:model} with $\nu_i$ being a nonnegative constant.
	Second, model \eqref{eq:model} also includes elliptical distributions by choosing $U_i\sim N(0, I_{p})$ and $\nu_i=\xi_i/\|U_{i}\|$ for some nonnegative random variable $\xi_i$ independent of $U_{i}$. In this case, $\nu_{i}$ is independent of the spatial sign of $U_{i}$, but not $U_{i}$.
	The independent components model has received great extension in high-dimensional data analysis as well as signal processing and machine learning \citep{Oja2001}.
	In addition, the elliptical distribution family covers many non-Gaussian distributions such as multivariate $t$-distribution, multivariate logistic distribution, and so on. It is commonly adopted in the literature on studying the sample spatial median
	\citep{Cheng2019,LiXu2022,Zou2014}. 
	\textcolor{black}{In terms of the GMOM, if the data are from the independent components model, the subsample means satisfy model \eqref{eq:model} clearly. In addition, some subfamilies of elliptical distributions are closed under convolution, and thus the subsample means also follow model \eqref{eq:model}. Our results can be applied to the GMOM estimator directly in those cases.}
	%More importantly, the family of elliptical distribution has been commonly used to model practical data sets, especially under non-normality and heavy-tailedness.  
	%In such cases, the sample spatial median is robust compared to the sample mean for estimating $\btheta$, and it is more efficient than the sample mean for heavy-tailed distributions.
\end{remark}

For $i=1,\ldots,n$, and  $k=1,2,3,4$, denote
\begin{eqnarray}\label{eq:nota}
	W_i = S(X_i-\btheta) \text{~~and~~} R_i=\|X_i-\btheta\|\,
\end{eqnarray}
as the spatial-sign and radius of $X_i-\btheta$, where $S(X)=\|X\|^{-1}X\mI(X\neq 0)$ is the multivariate sign function with $\mI(\cdot)$ being the indicator function. 
Thus, %the sample spatial median 
$\hat{\btheta}_n$ satisfies %the following equation 
%\begin{eqnarray}\label{eq:sm}
$\sum_{i=1}^{n}S(X_i-\hat{\btheta}_{n}) = 0\,. $
%\end{eqnarray}
%To obtain the asymptotic results for $\hat{\btheta}_{n}$, 

Denote $U_{i}=(U_{i,1},\ldots,U_{i,p})^{\top}$,
we impose the following three conditions.

\begin{condition}\label{c1}
	$U_{i,1},\ldots,U_{i,p}$ are i.i.d.~symmetric random variables with $\mE(U_{i,j})=0$, $\mE(U_{i,j}^2)=1$, and
	$\|U_{i,j}\|_{\psi_{\alpha}}\leq c_0$ with some constant $c_0>0$ and $1\leq \alpha\leq 2$.
\end{condition}

%{\color{red}
	\begin{condition}\label{c2}
		%For $k=1,\ldots,5$, there exist two positive constants $\underline{b}$ and $\bB$ such that ${\underline b}\leq  \limsup_{p}\mE(\nu_i^{-k})\leq \bB$.
		%In addition, for any fixed $p$-dimensional vector $a$ satisfies $a^{\top}a=1$, $
		%{\underline b}\leq  \limsup_{p}\mE(|a^{\top}U_{i}|^{-k})\leq \bB.
		%$
		The moments $\zeta_{k}=\mE(R_i^{-k})$ for $k=1,2,3,4$ exist for large enough $p$. In addition, there exist two positive constants $\underline{b}$ and $\bB$ such that
		$
		{\underline b}\leq  \limsup_{p}\mE(R_i/\sqrt{p})^{-k}\leq \bB
		$ for $k=1,2,3,4$.
	\end{condition}
	
	\begin{condition}\label{c3}
		The shape matrix $\Omega=(\omega_{j\ell})_{p\times p}$ satisfies $\tr(\Omega)=p$ and it belongs to the following class:
		\begin{eqnarray}
			\notag & & {\mathcal U}(a_0(p), \underline{m}, \bar{M})  = \left\{\Omega: \underline{m} \leq \omega_{jj}\leq \bar{M},~  \sum_{\ell=1}^{p}|\omega_{j\ell}|\leq a_0(p), \text{~~for all~} j=1,\ldots,p\right\}\,,
		\end{eqnarray}
		where $\underline{m}\leq \bar{M}$ are bounded positive constants.
	\end{condition}
	
	\begin{remark}
		In Condition \ref{c1}, the symmetric assumption is to ensure that $\btheta$ in model \eqref{eq:model} coincides with the population spatial median, which minimizes $L(\bbeta)=\mE(\|X-\bbeta\|-\|X\|)$.
		It is obvious that Condition \ref{c1} is satisfied by elliptical distributions with $U_i\sim N(0,I_p)$.
		The condition $\|U_{i,j}\|_{\psi_{\alpha}}\leq c_0$ implies that $U_{i,j}$ has a sub-exponential distribution.
		It is worth highlighting that with slight modification of the proofs of main theorems, the i.i.d.~condition on $U_{i,1},\ldots,U_{i,p}$ can be weaken by replacing Condition \ref{c1} with the following assumption:
		$U_{i,1},\ldots,U_{i,p}$ are independent symmetric random variables with $\mE(U_{i,j})=0$, $\mE(U_{i,j}^2)=1$ for all $j=1,\ldots,p$, and
		$\sup_{1\leq j\leq p}\|U_{i,j}\|_{\psi_{\alpha}}\leq c_0$ with some constant $c_0>0$ and $1\leq \alpha\leq 2$.
	\end{remark}
	
	\begin{remark}
		The condition ${\underline b}\leq  \limsup_{p}\mE(R_i/\sqrt{p})^{-k}\leq \bB$ indicates that $\zeta_{k}\asymp p^{-k/2}$ for $k=1,2,3,4$. It is introduced to avoid $X_i$ from concentrating too much near $\btheta$. For elliptical distributions, it is a generalization of Assumption 1 of \citet{Zou2014}, which is satisfied by many common distributions.
		For the independent components model, Condition \ref{c2} is equivalent to that
		$
		{\underline b}\leq  \limsup_{p}\mE(\|\Gamma U_{i}\|/\sqrt{p})^{-k}\leq \bB\,.
		$
		According to Lemma \ref{lemma:SM_moments} in Appendix A, $\mE(\|\Gamma U_{i}\|^{k}) = p^{k/2}\{1+o(1)\}$ for $k=1,2,3,4$. Then the Cauchy-Schwarz inequality implies that
		$\mE(\|\Gamma U_{i}\|^{-k}) \geq \{\mE(\|\Gamma U_{i}\|^{k})\}^{-1} = p^{-k/2}\{1+o(1)\}\,,$
		from which we know $\mE(\|\Gamma U_{i}\|^{-k})\gtrsim p^{-k/2}$. Furthermore, denote $\Gamma_{j}$ as the $j$th row of $\Gamma$, then by the inequality of harmonic and quadratic means,
		\begin{eqnarray}
			\notag \notag p^{2}\|\Gamma U_{i}\|^{-4} = \left\{\frac{p}{(\Gamma_{1}U_{i})^2+\cdots+(\Gamma_{p}U_{i})^2}\right\} \leq \frac{(\Gamma_{1}U_{i})^{-4}+\cdots+(\Gamma_{p}U_{i})^{-4}}{p}\,.
		\end{eqnarray}
		It follows that $\mE(\|\Gamma U_{i}\|^{-4})\lesssim p^{-2}$ if $\mE\{(\Gamma_{1}U_{i})^{-4}\},\ldots,\mE\{(\Gamma_{p}U_{i})^{-4}\}$ are uniformly bounded, and from which $\mE(\|\Gamma U_{i}\|^{-k})\lesssim p^{-k/2}$ by Jensen's inequality. Thus, Condition \ref{c2} is satisfied by the independent components models as long as $\Gamma_{1}U_{i},\ldots,\Gamma_{p}U_{i}$ are not concentrating too much near $0$.
		See also discussions in
		\citet{Cardot2013} on similar conditions.
	\end{remark}
	
	\begin{remark}
		It is noticed that the shape matrix $\Omega$ is only well defined up to a scalar multiple, the condition $\tr(\Omega)=p$ is used to regularize $\Omega$ to make model \eqref{eq:model} identifiable.
		The class ${\mathcal U}(a_0(p), \underline{m}, \bar{M})$
		covers a wide range of symmetric square matrices, and it is commonly adopted in the literature on high-dimensional analysis. 
		For example, a similar matrix class is introduced in \citet{Bic2008}. The condition $\underline{m} \leq \omega_{jj}\leq \bar{M}$ requires bounded diagonal elements. 
		The order of $a_0(p)$, which will be specified later, controls the orders of the off-diagonal elements of $\Omega$.
	\end{remark}
	
	\section{Gaussian and bootstrap approximations}\label{sec:main}
	
	\subsection{Bahadur representation and Gaussian approximation}\label{sec:02}
	
	%\subsection{Gaussian approximation}
	In this section, we establish Gaussian approximation for $\hat{\btheta}_n$, which is valid when $p$ diverges exponentially over $n$.
	The following lemma offers a Bahadur representation of $\hat{\btheta}_n$, and it severs as the foundation of the Gaussian approximation result in Theorem \ref{theo1}.
	
	\begin{lemma}\label{lem:Br}
		(Bahadur representation) Assume Conditions \ref{c1}, \ref{c2} and \ref{c3} with $a_0(p)\asymp p^{1-\delta}$ for some positive constant
		$\delta\leq 1/2$ hold. If
		$\log p=o(n^{1/3})$ and $\log n=o(p^{1/3 \wedge \delta})$, then
		\[
		n^{1/2}(\hat{\btheta}_{n}-\btheta)=n^{-1/2}\zeta_{1}^{-1}\sum_{i=1}^nW_i+C_n\,,
		\]
		where $|C_n|_{\infty}=O_p\{n^{-1/4}\log^{1/2}(np)+ p^{-(1/6 \wedge \delta/2)}\log^{1/2}(np)\}$.
	\end{lemma}
	
	\begin{remark}
		To the best of our knowledge, Lemma \ref{lem:Br} serves as the first result that offers the Bahadur representation of the sample spatial median with a maximum-norm bound on the remainder term. %, and it extends existing results beyond elliptical distributions.
		In \citet{Zou2014} and \citet{Cheng2019}, the same expansion with the remainder term $C_n$ satisfies $\|C_n\|=o_{p}(\zeta_{1}^{-1})$ was obtained, and their result was improved to $\|C_n\|=o_p(1)$ in \citet{LiXu2022}, by replacing $\zeta_{1}$ with $n^{-1}\sum_{i=1}^{n}R_i^{-1}$ in the linear term, but under a more restricted condition that $p$ and $n$ are of the same order.
		It is worth noticing that the previous results \citep{Cheng2019,LiXu2022,Zou2014} are all derived under elliptical distributions.
		%Note that $|C_n|_{\infty}\to 0$ as long as $\log (p)=o(n^{1/2})$ in Lemma \ref{lem:Br}.
	\end{remark}
	
	Let ${\mathcal A}^{\mathrm{re}}=\{\prod_{j=1}^{p}[a_j, b_j]: -\infty\leq a_j\leq b_j\leq \infty, j=1,\ldots,p\}$ be the class of rectangles in $\mathbb{R}^{p}$.
	With the Bahadur representation in Lemma \ref{lem:Br} on hand, we establish the following Gaussian approximation result for $\hat{\btheta}_n$ over hyperrectangles.
	%The following theorem derives a Gaussian approximation result.
	\begin{theorem}\label{theo1}
		\textit{(Gaussian approximation)}
		Assume Conditions \ref{c1}, \ref{c2} and \ref{c3} with $a_0(p)\asymp p^{1-\delta}$ for some positive constant
		$\delta\leq 1/2$ hold. If
		$\log p=o(n^{1/5})$ and $\log n=o(p^{1/3 \wedge \delta})$,
		%$
		%n^{-1}\log^{5}(np) \lor p^{-\delta} \log^{1/2}(np)\rightarrow 0,
		%$
		then
		\[
		\rho_n({\mathcal A}^{\mathrm{re}}) = \sup_{A \in {\mathcal A}^{{\rm re}}}\left|\P\{{n}^{1/2}(\hat{\btheta}_{n}-\btheta) \in A \}-\P\left(G \in A\right)\right|\rightarrow 0\,
		\]
		as $n\to\infty$, where $G\sim N(0, \zeta_{1}^{-2}\B)$ with $\B=\mE(W_1W_1^{\top})$.
		%}
\end{theorem}

The Gaussian approximation for $\hat{\btheta}_n$ %in Theorem \ref{theo1}
indicates that the probabilities $\P\{n^{1/2}(\hat{\btheta}_{n}-\btheta) \in A\}$ can be approximated by that of a centered Gaussian random vector with covariance matrix $\zeta_{1}^{-2}\B$ for hyperrectangles $A\in{\mathcal A}^{\mathrm{re}}$.
%The condition $n^{-1}\log^{5}(np) \lor p^{-\delta} \log^{1/2}(np)\rightarrow 0$ is satisfied when $\log p=o(n^{1/5})$ and $\log n=o(p^{2\delta})$. % with a bounded constant $\bB$.
%Thus, 
Theorem \ref{theo1} allows for an exponentially divergent $p$, which fits the ultrahigh dimensional setting. Compared to the asymptotic normality of %the squared Euclidean distance 
$\|\hat{\btheta}_{n}-\btheta\|^2$ in \citet{LiXu2022}, in which $p$ is assumed to have the same order as $n$, the Gaussian approximation result in Theorem \ref{theo1} requires much weaker conditions on the rates of $n$ and $p$.

\begin{remark}\label{remark:B}
	Let $\B_{j\ell}$ be the $(j,\ell)$th element of $\B$. According to Lemma \ref{lemma:05} (iii) in Appendix A, $\zeta_{1}^{-2}\B_{j\ell} = \zeta_{1}^{-2}p^{-1}\omega_{j,\ell} + O(p^{-\delta/2})$ for all $1\leq j,\ell\leq p$. Thus, the covariance matrix of $G$ in Theorem \ref{theo1} is asymptotically proportional to the shape matrix $\Omega$.
\end{remark}

\begin{remark}\label{remark:M-estimator}
	As the sample spatial median is a special M-estimator, 
	Gaussian approximation for M-estimator in \citet{Imaizumi2021} is potentially applicable to the spatial median under high-dimensionality.
	However, it is worth highlighting that the results in \citet{Imaizumi2021} cannot be applied to our framework. 
	To be precise, Assumption 1 (A3) in \citet{Imaizumi2021} assumes that there exist constants $C>0$ and $\alpha\in(0,2)$ such that $\log \mathcal{N}(\varepsilon, \Theta, \|\cdot\|)\leq C\varepsilon^{-\alpha}$ holds for all $\varepsilon\in(0,1)$, where $\Theta$ is the parameter space, and $\mathcal{N}(\varepsilon, \Theta, \|\cdot\|)$ is the $\varepsilon$-covering number of $\Theta$ under the Euclidean norm $\|\cdot\|$ \citep{VanderVaart1996}. 
	When $\Theta$ is a compact subset of $\mathbb{R}^{p}$, $\mathcal{N}(\varepsilon, \Theta, \|\cdot\|)$ is of order $O(\varepsilon^{-p})$. 
	In this case, $\log \mathcal{N}(\varepsilon, \Theta, \|\cdot\|)\leq C\varepsilon^{-\alpha}$ cannot be satisfied when $p\to\infty$. 
	Thus, our theoretical findings are independent of those in \citet{Imaizumi2021}.
\end{remark}

\iffalse
\begin{remark}\label{remark:04}
	When $p$ is fixed, the classical CLT for $\hat{\btheta}_n$ \citep{Oja2010} is
	$ n^{1/2}(\hat{\btheta}_{n}-\btheta) \to N(0, \C^{-1}\B \C^{-1}) $
	in distribution, where $\C=\mE\{R_i^{-1}(I_p-W_iW_i^{\top} )\}$. Compared to Theorem \ref{theo1}, the asymptotic covariance matrix changes from $\C^{-1}\B \C^{-1}$ to $\zeta_{1}^{-2}\B$ under high-dimensionality. 
	In fact, 
	%blessing of high dimensionality, as a intuition that is
	$
	\left\|\C-\zeta_{1}I_p \right\|_2= \left\|\mE(R_i^{-1}W_iW_i^{\top})\right\|_2 \lesssim p^{-1}\mE(R_i^{-1})=p^{-1}\zeta_{1}\,
	$
	in probability, where the last inequality can be shown similar to Lemmas A.5 and A.6 of \citet{Wang2015}.
	Note that $p^{-1}\zeta_{1}/ \|\zeta_{1}I_p\|_2\to0$ under Condition \ref{c2}, the distance between $\C$ and $\zeta_{1}I_p$, so as $\C^{-1}\B \C^{-1}$ and $\zeta_{1}^{-2}\B$, is negligible under the matrix 2-norm as $p\to\infty$.
\end{remark}
\fi

Theorem \ref{theo1} immediately implies the following corollary %, which shows that the distribution of $n^{1/2}|\hat{\btheta}_{n}-\btheta|_{\infty}$ can be well approximated by that of $|G|_{\infty}$,  
since the Kolmogorov distance of sup-norm is a subset of $\mathcal{A}^{\mathrm{re}}$ corresponding to max-hyperrectangles in $\mathbb{R}^{p}$. 
\begin{corollary}\label{co1}
	Under the conditions assumed in Theorem \ref{theo1}, as $n\to\infty$,
	\[
	\rho_n=\sup_{t\in\mR}\left|\P(n^{1/2}|\hat{\btheta}_{n}-\btheta|_{\infty}\leq t)- \P(|G|_{\infty}\leq t)\right|\rightarrow  0.
	\]
\end{corollary}
\subsection{Multiplier bootstrap approximation}\label{sec:boot}

Theorem \ref{theo1} allows us to approximate the distribution of $n^{1/2}(\hat{\btheta}_{n}-\btheta)$ by that of $G$ hit hyperrectangles, where $G\sim N(0, \zeta_{1}^{-2}\B)$. 
However, it cannot be used directly in statistical inference for $\btheta$ as
the quantity $\zeta_{1}$ and the matrix $\B$ depend on the underlying distribution $F_X$ and are thus unknown. 
To solve this issue, we propose an easy-to-implement bootstrap method to approximate the distribution of $n^{1/2}(\hat{\btheta}_{n}-\btheta)$. % over the class of hyperrectangles. 

%In this following, we will use a new
%bootstrap called the pivotal bootstrap to further estimate the distribution of $|\hat{\btheta}_{n}|_{\infty}$.
Let $Z_1,\ldots,Z_n$ be a sequence of i.i.d.~random variables with mean zero and unit variance. Define the bootstrap version of the sample spatial median as
\begin{align}\label{btheta}
	\tilde{\btheta}_{n}= \argmin_{\bbeta\in {\mathbb R}^d}\sum_{i=1}^n\|Z_i(X_i-\hat{\btheta}_n)-\bbeta\|\,.
\end{align}
Then, the distribution of $n^{1/2}\tilde{\btheta}_{n}$ conditional on $X_1,\ldots,X_n$ is used to approximate that of $n^{1/2}(\hat{\btheta}_{n}-\btheta)$.
This algorithm is called the multiplier bootstrap, and $Z_1,\ldots,Z_n$ are the multiplier weights.
%It is worth noting that $\tilde{\btheta}_{n}$ is the sample spatial median of $Z_1(X_1-\hat{\btheta}_{n}), \ldots, Z_{n}(X_i-\hat{\btheta}_{n})$.

Regarding the proof of Lemma \ref{lem:SM_boot_br} in Appendix B, it is preferred that the multiplier weights $Z_1,\ldots,Z_n$ are bounded and satisfy $\mE(Z_{i}^{-2})<\infty$. 
Thus, we choose the Rademacher variables as the multipliers \citep{Cher2019}, that is, $\P(Z_i=1)=\P(Z_i=-1)=1/2$.
%The next theorem shows the validity of the multiplier bootstrap. % method.
\begin{theorem}\label{theo:boot}
	\textit{(Bootstrap approximation)}
	Under the conditions assumed in Theorem \ref{theo1},
	\[
	\rho_n^{\mathrm{MB}}({\mathcal A}^{\mathrm{re}})=\sup_{A \in {\mathcal A}^{\mathrm{re}}}\left|\P\{n^{1/2}(\hat{\btheta}_{n}-\btheta)\in A\}- \P^{*}(n^{1/2}\tilde{\btheta}_{n} \in A)\right|\rightarrow 0
	\]
	in probability as $n\to\infty$,
	where $\P^{*}$ denotes the conditional probability given $X_1,\ldots,X_n$.
\end{theorem}

Under the same conditions on the divergence rates of $n$ and $p$ as in Theorem \ref{theo1}, Theorem \ref{theo:boot} validates that conditional on $X_1,\ldots,X_n$, the distribution of the bootstrap sample spatial median $\tilde{\btheta}_n$ approximates that of $\hat{\btheta}_n$ consistently over hyperrectangles.

\begin{remark}
	The proof of Theorem \ref{theo:boot} is nontrivial and does not follow directly from existing results since $\tilde{\btheta}_n$ has no explicit form, which is different from the multiplier bootstrap methods for high-dimensional sample mean that have been analysed in the literature.
	The key step in the proof is to obtain a Bahadur representation of $\tilde{\btheta}_n$ similar as $\hat{\btheta}_n$ in Lemma \ref{lem:Br}. Specifically, we show that $n^{1/2}\tilde{\btheta}_{n}=n^{-1/2}\zeta_{1}^{-1}\sum_{i=1}^n Z_i W_i+{\tilde C}_n$ with $|{\tilde C}_n|_{\infty}=O_p\{n^{-1/4}\log^{1/2} (np)+p^{-(1/6\wedge\delta/2)}\log^{1/2} (np)\}$ in Lemma \ref{lem:SM_boot_br} in Appendix A.
	%The proof of Theorem \ref{theo:boot} consists of obtaining a similar Bahadur representation for $\tilde{\btheta}_{n}$, and then applying the Rademacher bootstrap approximation result in \cite{Cher2019} on the linear term of the Bahadur representation.
\end{remark}

The next corollary is an immediate consequence of Theorem \ref{theo:boot}.
\begin{corollary}\label{co2}
	Under the conditions assumed in Theorem \ref{theo:boot}, as $n\to\infty$,
	\[
	\rho_n^{\mathrm{MB}}=\sup_{t\in\mR}\left|\P\{n^{1/2}|\hat{\btheta}_{n}-\btheta|_{\infty}\leq t\}- \P^{*}(n^{1/2}|\tilde{\btheta}_{n}|_{\infty}\leq t)\right|\rightarrow  0 \text{~~in probability.}
	\]
	
\end{corollary}

\section{Statistical inference}\label{sec:applications}

The Gaussian and multiplier bootstrap approximations for the sample spatial median %in Theorem \ref{theo:boot}
enable many statistical inferential methods for ultrahigh dimensional  population location parameter.
In this section, we present the following statistical inferences: simultaneous confidence intervals (SCIs) and global tests for the population location parameter, multiple testing for every component of $\btheta$, and high-dimensional asymptotic relatively efficient of the sample spatial median compared to the sample mean.

\subsection{Simultaneous confidence intervals}\label{sec:SCIs}

%Constructing simultaneous confidence intervals (SCIs) that preserve a nominal simultaneous confidence level is an important problem in multivariate data analysis.
We are interested in building SCIs for all components of $\btheta=(\theta_1,\ldots,\theta_p)^{\top}$.
Corollary \ref{co2} motivates the following way of constructing SCIs for $\btheta$.
Given a nominal confidence level $1-\tau$,
define the set ${\mathcal C}_{\tau}$ as
\[
{\mathcal C}_{\tau}=\left\{\btheta\in \mR^p, n^{1/2}|\hat{\btheta}_{n}-\btheta|_{\infty}<q^{B}_{1-\tau}\right\},
\]
where $q^{B}_{1-\tau}$ is the $(1-\tau)$th quantile of $n^{1/2}|\tilde{\btheta}_{n}|_{\infty}$ given $X_1,\ldots,X_n$.
%Then, $\mathcal{C}_{\tau}$ is an asymptotic $1-\tau$ SCIs for $\btheta$.
Denote $\hat{\btheta}_n=(\hat{\theta}_{n,1},\ldots,\hat{\theta}_{n,p})^{\top}$, the confidence intervals are $[\theta_{n,j}^{-},\theta_{n,j}^{+}]$ for $j=1,\ldots,p$, where
$$
\theta_{n,j}^{-}=\hat{\theta}_{n,j}-n^{-1/2}q^{B}_{1-\tau} \text{~~and~~} \theta_{n,j}^{+}=\hat{\theta}_{n,j}+n^{-1/2}q^{B}_{1-\tau}.
$$
The next theorem shows that $\mathcal{C}_{\tau}$ preserves the nominal simultaneous confidence level $1-\tau$ asymptotically under ultrahigh dimensionality.

\begin{theorem}\label{theo:SCIs}
	Suppose the conditions of Theorem \ref{theo:boot} hold, then $\P(\btheta\in {\mathcal C}_{\tau})\rightarrow 1-\tau$ as $n\to\infty$. Equivalently,
	$\P(\theta_j\in [\theta_{n,j}^{-},\theta_{n,j}^{+}] \text{~~for all~} 1\leq j\leq p)\rightarrow 1-\tau$ as $n\to\infty$.
\end{theorem}

\begin{remark}
	Unlike the fixed dimensional setting, $n^{1/2}|\tilde{\btheta}_{n}|_{\infty}$ and $n^{1/2}|\hat{\btheta}_{n}-\btheta|_{\infty}$ are maxima of divergent numbers of variables, and their quantiles are generally divergent as $p\to\infty$.
	Thus, Theorem \ref{theo:SCIs} is not a direct consequence of Corollary \ref{co2}.
	To ascertain the consistency of $\mathcal{C}_{\tau}$ theoretically, we show that, with probability approaching one, $q_{1-\tau}^{B}$ is bounded by two quantiles of $n^{1/2}|\hat{\btheta}_{n}-\btheta|_{\infty}$ with quantile levels close enough to $1-\tau$ using an anti-concentration inequality for divergent random sequences.
\end{remark}

\iffalse
We summarise the approach of building SCIs for $\btheta$ in Algorithm \ref{al1}.
\RestyleAlgo{ruled}
\begin{algorithm}\label{al1}
	\caption{Construction of SCIs for $\btheta$}
	\SetAlgoLined
	\KwData{Data $X_1, \ldots, X_n$; $\alpha$, confidence level; $B$, number of bootstrap iterations}
	\KwResult{Simultaneous confidence intervals: $[\theta_{n,j}^{-},\theta_{n,j}^{+}]$ for $j=1,\ldots,p$}
	\BlankLine
	
	Compute $\hat{\btheta}_n=(\hat{\theta}_{n,1},\ldots,\hat{\theta}_{n,p})^{\top}$ by minimizing $L_n(\bbeta)$ defined in \eqref{eq:btheta_criteria};

	\For{{$b\leftarrow 1$ \KwTo $B$}}{
		generate independent Rademacher random variables $Z_1^{(b)}, \ldots, Z_n^{(b)}$;
		
		compute $\tilde{\btheta}_{n}^{(b)}$ by minimizing \eqref{btheta} using $Z_1^{(b)}, \ldots, Z_n^{(b)}$ as the multiplier weights;
		
		compute $n^{1/2}|\tilde{\btheta}_{n}^{(b)}|_{\infty},$;
	}
	
	Compute $q^{B}_{1-\alpha}$ as the $(1-\alpha)$-th quantiles of $n^{1/2}|\tilde{\btheta}_{n}^{(1)}|_{\infty}, \ldots, |n^{1/2}\tilde{\btheta}_{n}^{(B)}|_{\infty}$;
	
	\For{{$j\leftarrow 1$ \KwTo $p$}}{
		compute $\theta_{n,j}^{-}=\hat{\theta}_{n,j}-n^{-1/2}q^{B}_{1-\alpha}$;
		
		compute $\theta_{n,j}^{+}=\hat{\theta}_{n,j}+n^{-1/2}q^{B}_{1-\alpha}$;
	}
\end{algorithm}
\fi

\begin{remark}\label{remark:SCIs_mean}
	The Gaussian approximation for  the sample mean $\bar{X}_n=n^{-1}\sum_{i=1}^{n}X_i$ \citep{Cher2013,Cher2017,Cher2019} indicate that if $\log p = o(n^{1/5})$,
	\begin{eqnarray}\label{eq:gaussian_approx_mean}
		%\notag
		\sup_{t\in\mR}\left|\P(n^{1/2}|\bar{X}_{n}-\btheta|_{\infty}\leq t)- \P(|G_0|_{\infty}\leq t)\right|\rightarrow  0
	\end{eqnarray}
	as $n\to\infty$ under some moderate conditions, where $G_0\sim N(0, \Sigma)$ with $\Sigma=\mE(XX^{\top})$.
	Define $X_{i}^{*} = Z_i(X_i-\bar{X}_n)$ for $i=1,\ldots,n$, where $Z_1,\ldots,Z_n$ are the Rademacher weights. Denote $\bar{X}_{n}^{*}=n^{-1}\sum_{i=1}^{n}X_{i}^{*}$,
	it has been shown in \citet{Cher2019} that %under some mild conditions,
	\begin{eqnarray}\label{eq:boot_mean}
		& \sup_{t\in\mR}\left|\P(n^{1/2}|\bar{X}_{n}-\btheta|_{\infty}\leq t)- \P^{*}(n^{1/2}|\bar{X}_{n}^{*}|_{\infty}\leq t)\right|\rightarrow  0 &
	\end{eqnarray}
	in probability as $n\to\infty$ when $\log p = o(n^{1/5})$. % under some mild conditions.
	Based on \eqref{eq:boot_mean}, define
	$$
	{\mathcal C}_{\tau}^{\prime}=\left\{\btheta\in \mR^p, n^{1/2}|\bar{X}_{n}-\btheta|_{\infty}<q^{B\prime}_{1-\tau}\right\},
	$$
	where $q^{B\prime}_{1-\tau}$ is the $(1-\tau)$th quantile of $n^{1/2}|\bar{X}_{n}^{*}|_{\infty}$ conditional on $X_1,\ldots,X_n$.
	Then $\mathcal{C}_{\tau}^{\prime}$ is also an asymptotic $1-\tau$ SCIs for $\btheta$.
	Based on the discussion in Section \ref{sec:ARE}, $\mathcal{C}_{\tau}$ has advantage (relative shorter intervals) over $\mathcal{C}_{\tau}^{\prime}$ under heavy-tailed distributions.
	We refer to Section \ref{sec:simulations} for finite-sample justifications on this.
\end{remark}

%We have the following theorem parallel to Theorem \ref{theo:SCIs}.
%\begin{theorem}\label{theo:SCIs_two-sample}
%	Suppose the conditions of Theorem \ref{theo:boottwo} hold, we have $\P(\btheta^{X}-\btheta^{Y}\in {\mathcal C}_{\tau}^{\mathrm{TS}})\rightarrow 1-\tau$ as $n\to\infty$.
%\end{theorem}
%\begin{remark}\label{remark:05}
%	\textbf{Comparison to the SCIs based on the sample mean.}
%\end{remark}

\subsection{Global tests for high-dimensional location parameters}\label{sec:global_tests}

In this section, we propose a novel approach for global tests on high-dimensional location parameters. %, and we focus on the one-sample problem first.
Let $\btheta_0$ be a known $p$-dimensional vector, we are interested in testing
\begin{eqnarray}\label{eq:one_sample}
	H_0: \btheta = \btheta_0 \text{~~versus~~} H_1:\btheta\neq\btheta_0.
\end{eqnarray}

Theorems \ref{theo1} and \ref{theo:boot} motivate us proposing a maximum-norm type test statistic.
Define
\begin{align}\label{test}
	T_{n} = n^{1/2}|\hat{\btheta}_{n}-\btheta_0|_{\infty}
\end{align}
as the test statistic, and $H_0$ is rejected when $T_{n}$ is larger than a critical value.
%Under $H_0$, %Corollary \ref{co1} implies that the distribution of $T_n$ can be approximated by that of $N(0, \zeta_{1}^{-1}\B)$.
%In addition,
We can use the multiplier bootstrap to approximate the distribution of $T_n$ under $H_0$.
Specifically, with a nominal significance level $\tau$, the null hypothesis is rejected if $T_n>q_{1-\tau}^{B}$. %, where $q_{1-\tau}^{B}$ is the $(1-\tau)$th quantile of $n^{1/2}|\tilde{\btheta}_n|_{\infty}$ conditional on $X_1,\ldots,X_n$. %, which can be approximated using Monte Carlo simulations.
%Corollary \ref{co2}  and
Theorem \ref{theo:SCIs} guarantees that the test based on $T_n$ maintains nominal significance level asymptotically under ultrahigh dimensionality, that is, $\P(T_n>q^B_{1-\tau}\mid H_0)\rightarrow \tau$ as $n\to\infty$ when $\log p = o(n^{1/5})$.

\begin{remark}\label{remark:06}
	%Regarding the bootstrap approximation result for the sample mean given in \eqref{eq:boot_mean}, a
	An alternative test for \eqref{eq:one_sample} can be constructed based on $\bar{X}_n$ by defining the test statistic as
	%\begin{eqnarray}
	$ T_{\mathrm{Mean}} = n^{1/2}|\bar{X}_{n}-\btheta_0|_{\infty}$.
	%\end{eqnarray}
	Then, the null hypothesis is rejected if $T_{\mathrm{Mean}}>q_{1-\tau}^{B\prime}$. %, where $q_{1-\tau}^{B\prime}$ is the $(1-\tau)$-th quantile of $n^{1/2}|\bar{X}_{n}^{*}|_{\infty}$ conditional on $X_1,\ldots,X_n$.
	The test based on $T_n$ can be deemed as a nonparametric extension of the test based on $T_{\mathrm{Mean}}$ .
	As $\hat{\btheta}_n$ is more efficient than $\bar{X}_n$ for simultaneous inference of $\btheta$ under heavy-tailed distributions as discussed in Section \ref{sec:ARE}, we expect that the proposed test based on $T_n$ is more powerful than that based on $T_{\mathrm{Mean}}$ in those cases.
	This has been reflected by the simulation results in Appendix C of the supplementary material.
\end{remark}

The next theorem summarises the asymptotic power of the proposed test based on $T_n$.
\begin{theorem}\label{theo:power}
	Suppose the conditions of Theorem \ref{theo:boot} hold.
	For any given $0<\tau<1$, if
	$|\btheta-\btheta_0|_{\infty}\geq C\log^{1/2}(\tau^{-1}) n^{-1/2}\log^{1/2}(np)$ for some large enough constant $C>0$,
	then $\P(T_n>q^B_{1-\tau}\mid H_1)\rightarrow 1$ as $n\rightarrow \infty$.
\end{theorem}

Theorem \ref{theo:power} indicates that the test based on $T_n$ achieves consistency when the maximum element of $n^{1/2}|\btheta-\btheta_0|$ has a magnitude much large than $\log^{1/2}(\tau^{-1})\log^{1/2}(np)$ for a fixed significant level $\tau$.

\begin{remark}\label{remark:07}
	\citet{Wang2015} proposed a $L_2$-norm type test (WPL test) for \eqref{eq:one_sample} with $\btheta_0=0$ based on
	%\begin{eqnarray}
	$ T_{\mathrm{WPL}} = \sum_{i=1}^{n}\sum_{j=1}^{i-1}W_{i}^{\top}W_i$.
	%\end{eqnarray}
	It has been argued in \citet{Wang2015} and \citet{LiXu2022} that the
	%. It has been shown that under the local alternatives, the asymptotic power of the WPL test is
	%$\beta_{{\rm WPL}} = \Phi\left(-z_{\tau}+\sqrt{\frac{n(n-1)}{2}}\frac{\btheta^{\top}\C^2\btheta(1+o(1))}{\sqrt{\tr(\B^2)}}\right)$,
	%where $\Phi(\cdot)$ and $z_{\tau}$ are the cumulative distribution function (cdf) and the $\tau$-th quantile of $N(0,1)$, respectively.
	%As the distance between $\C$ and $\zeta_{1}I_p$ is negligible under matrix $2$-norm when $p\to\infty$ as discussed in Remark \ref{remark:04}, the
	signal of the WPL test is determined by the magnitude of $\|\btheta\|$, which is the $L_2$-norm of $\btheta$.
	As a contrast, the power of the test based on $T_n$  depends on $|\btheta|_{\infty}$.
	Thus, the proposed test based on $T_n$ is expected to be more powerful under sparse alternatives, when $\btheta$ contains only a limited number of non-zero components and its maximum element has certain order of magnitude.
	In such cases, $\|\btheta\|$ is not big enough for the rejection of the WPL test.
	See Appendix C in the supplementary material and Section \ref{sec:real_data} for numerical justifications.
\end{remark}
\subsection{Multiple testing with FDR control in large-scale tests}\label{sec:FDR}

Multiple testing with false discovery rate (FDR) control has been applied to many real problems, such as detecting differentially expressed genes in genomic study.
In this section, we study multiple testing for every component of $\btheta$ based on the spatial median with the Benjamini and Hochberg (B-H) method for FDR control.
For $j=1,\ldots,p$, we are interested in testing
\begin{eqnarray}
	\notag H_{0j}:  \theta_j=\theta_{0,j}  \text{~~versus~~} H_{1j}:\theta_{j} \neq\theta_{0,j}
\end{eqnarray}
simultaneously, where $\theta_{0,1},\ldots,\theta_{0,p}$ are given values.

Define the test statistics as
\begin{eqnarray}
	\notag T_{n,j} = n^{1/2}(\hat{\theta}_{n,j}-\theta_{0,j})/s_{n,j}
\end{eqnarray}
for $j=1,\ldots,p$, where $s_{n,j}^2=\hat{\zeta}_1^{-2}\hat{\B}_{jj}$ with $\hat{\zeta}_1=n^{-1}\sum_{i=1}^n\|X_i-\hat{\btheta}_n\|^{-1}$, and $\hat{\B}_{jj}$ is the $j$th diagonal element of ${\hat \B}=n^{-1}\sum_{i=1}^n\|X_i-\hat{\btheta}_n\|^{-2}(X_i-\hat{\btheta}_n)(X_i-\hat{\btheta}_n)^{\top}$.

According to the proof of Theorem \ref{theo:FDR} in Appendix A, $T_{n,j}$ converges in distribution to a standard normal under $H_{0j}$ for $j=1,\ldots,p$.
Thus, we utilise the standard normal distribution to estimate the marginal $p$-values.
For $j=1,\ldots,p$, define the $p$-value for $H_{0j}$ as
$P_{j} = 2-2\Phi(|T_{n,j}|)$.
Denote $P_{(1)}\leq \cdots\leq P_{(p)}$ be the ordered $p$-values, and define
\begin{eqnarray}
	\notag \hat{k} = \max\left\{j=0,\ldots,p: P_{(j)}\leq \tau j/p\right\}
\end{eqnarray}
for a pre-specific significance level $\tau$.
Then, the B-H procedure rejects the null hypotheses for which $P_{j}\leq P_{(\hat{k})}$. Denote $\mathcal{H}_{R}=\{j:P_{j}\leq P_{(\hat{k})}\}$ as the set of indices $j$ such that $H_{0j}$ is rejected by the B-H method, and let $|\mathcal{H}_{R}|$ be the cardinality of $\mathcal{H}_{R}$ that equals the total number of rejected null hypotheses. % by the B-H method.

Let $\mathcal{H}_0\subset\{1,\ldots,p\}$ be the set of indices $j$ corresponding to the true null hypotheses $H_{0j}$.
The false discovery proportion (FDP) and false discovery rate (FDR) of the B-H method are defined as
\begin{eqnarray}
	\notag \mathrm{FDP}_{M}=\frac{|\mathcal{H}_{0}\cap\mathcal{H}_{R}|}{|\mathcal{H}_{R}|\vee 1} \text{~~and~~} \mathrm{FDR}_{M} = \mathbb{E}(\mathrm{FDP}_{M}).
\end{eqnarray}

Regarding that $T_{n,1},\ldots,T_{n,p}$ are dependent, we impose the following condition on the weak dependence between any two components of $W_i$. Define $(r_{j\ell})_{p\times p}=\{\mathrm{diag}(\B)\}^{-1/2}\B\{\mathrm{diag}(\B)\}^{-1/2}$ as the correlation matrix, where $\mathrm{diag}(\B)$ is the diagonal matrix of $\B$.

\begin{condition}\label{c5}
	Suppose $\max_{1\leq j , \ell \leq p} |r_{j\ell}|\leq r$ with some constant $0< r< 1$. In addition, $\sum_{j=1}^p\I(r_{j\ell}=0)=O(p^{\eta})$ for some constant $0<\eta<(1-r)/(1+r)$.
\end{condition}

Similar conditions are assumed in \citet{Liu2014} and \citet{Bell2018}.
Let $p_0=|\mathcal{H}_0|$ be the number of true null hypotheses and $\B_{jj}$ be the $j$th diagonal element of $\B$.

\begin{theorem}\label{theo:FDR}
	Suppose Condition \ref{c5} and the conditions of Theorem \ref{theo1} hold. In addition, there exists $\mathcal{H}\subset\{1,\ldots,p\}$ such that
	${\mathcal H}=\big\{j: \zeta_{1}\B_{jj}^{-1/2}n^{1/2}|\theta_{j}-\theta_{0,j}|\geq 2\log^{1/2} (p)\big\}$
	and $|\mathcal H|\geq \log \log p \rightarrow \infty$ as $p\rightarrow \infty$.  Assume that the number of false null hypotheses $p_1\leq p^{\varpi}$ for some $0<\varpi<1$. Then, $\mathrm{FDR}_{M}/(\tau p_0/p)\to1$ as $n \rightarrow \infty$.
\end{theorem}

Theorem \ref{theo:FDR}  shows the B-H procedure based on $P_1, \ldots, P_p$ controls the FDR asymptotically, and it extends Theorem 4.1 in \citet{Liu2014} to spatial median-based test statistic.

\subsection{High-dimensional asymptotic relative efficiency}\label{sec:ARE}

%For spherical multivariate normal distribution, \citet{Brown1983} showed that the asymptotic efficiency of the sample spatial median $\hat{\btheta}_n$ relative to the sample mean $\bar{X}_n$, denoted as $\mathrm{ARE}(\hat{\btheta}_{n},\bar{X}_n)$, exceeds the usual univariate case $2/\pi$. 
%In addition, $\mathrm{ARE}(\hat{\btheta}_{n},\bar{X}_n)$ increases as the dimension increases, and it approaches to $1$ as $p$ tends to be sufficient large \citep{Magyar2011}. 
%For a heavy-tailed distribution, such as multivariate $t$-distribution, $\hat{\btheta}_n$ is more efficient than $\bar{X}_n$ through the classical central limit theorem (CLT). 
%$\mathrm{ARE}(\hat{\btheta}_{n},\bar{X}_n)$ is not easy to quantify as there is no obvious ``final'' limit distributions for $\hat{\btheta}_{n}$ and $\bar{X}_n$.
%Motivated by simultaneous inference of the location parameter, we define 
%a high-dimensional $\mathrm{ARE}(\hat{\btheta}_{n},\bar{X}_n)$ via the maximum-norms of $\hat{\btheta}_n$ and $\bar{X}_n$ based on the Gaussian approximation results for them.
%See more detailed discussion in Section \ref{sec:ARE}.

As two candidate estimators of the location parameter $\btheta$, it is of interest to study the asymptotic relative efficiency (ARE) of the sample spatial median $\hat{\btheta}_n$  relative to the sample mean $\bar{X}_n$. % under high-dimensionality.
When $p$ is fixed, for spherical multivariate normal distribution, \citet{Brown1983} showed that the asymptotic efficiency of $\hat{\btheta}_n$ relative $\bar{X}_n$, denoted as $\mathrm{ARE}(\hat{\btheta}_{n},\bar{X}_n)$, exceeds the usual univariate case $2/\pi$. 
In addition, $\mathrm{ARE}(\hat{\btheta}_{n},\bar{X}_n)$ increases as the dimension increases, and it approaches to $1$ as $p$ tends to be sufficient large \citep{Magyar2011}. 
However, when $p\to\infty$, the ARE is not straightforward to quantify as there are no obvious ``final'' limit distributions for $\hat{\btheta}_{n}$ and $\bar{X}_n$.
Motivated by the discussions in Sections \ref{sec:SCIs} and \ref{sec:global_tests}, we compare $\hat{\btheta}_n$ and $\bar{X}_n$ in terms of their efficiencies in simultaneous inference for $\btheta$, which are determined by the variations of $|\hat{\btheta}-\btheta|_{\infty}$ and $|\bar{X}_n-\btheta|_{\infty}$.
According to Corollary \ref{co1} and \eqref{eq:gaussian_approx_mean}, we define the high-dimensional ARE of $\hat{\btheta}_n$ compared to $\bar{X}_n$ in simultaneous inference for $\btheta$ as
\begin{align}\label{AE}
	\mathrm{ARE}(\hat{\btheta}_{n}, \bar{X}_n)=\mathrm{Var}(|G_0|_{\infty})/\mathrm{Var}(|G|_{\infty})\,,
\end{align}
which approximates $\mathrm{Var}(|\bar{X}_n-\btheta|_{\infty})/\mathrm{Var}(|\hat{\btheta}_{n}-\btheta|_{\infty})$. If $\lim_{p\to\infty}\mathrm{ARE}(\hat{\btheta}_{n}, \bar{X}_n)>1$, we say that $\hat{\btheta}_n$ is more efficient than $\bar{X}_n$ in simultaneous inference for $\btheta$ under high-dimensionality.

As discussed in Remark \ref{remark:B}, $G\sim N(0, \zeta_{1}^{-2}\B)$ with $\zeta_{1}^{-2}\B_{j\ell}=\zeta_{1}^{-2}p^{-1}\omega_{i\ell}$ for all $1\leq j,\ell\leq p$.
Moreover, we can show that $\Sigma_{j\ell} = \mE(\nu_{i}^{2})\omega_{j\ell} + O(p^{-1/2})$ similar to the proof of Lemma \ref{lemma:SM_Q} in Appendix B of the supplementary material, where $\Sigma_{j\ell}$ is the $(j,\ell)$th element of $\Sigma$.
%For a spherical distribution with $\Omega=I_{p}$, $G\sim N(0, p^{-1}\{\mE(\nu_i^{-1})\}^{-2}I_{p})$ and $G_0\sim N(0, p^{-1}\mE(\nu_i^2)I_p)$. Then 
%$\mathrm{ARE}(\hat{\btheta}_{n}, \bar{X}_n) = \mE(\nu_i^2)\{\mE(\nu_i^{-1})\}^{2}.$
%For a general elliptical distribution, $\zeta_{1}^{-2}\mE(W_{i,j}W_{i,\ell})\asymp \zeta_{1}^{-2}p^{-1}\omega_{j\ell}$ similar to the proof of Lemma \ref{lemma:SM_Q} in the supplementary material, where $W_{i,j}$ is the $j$th component of $W_i$ for $j=1,\ldots,p$. 
Thus, both the covariance matrix $\Sigma$ and $\zeta_{1}^{-2}\B$ are proportional to $\Omega$ asymptotically, and $\mathrm{ARE}(\hat{\btheta}_{n}, \bar{X}_n)$ is approximately $\mE(\nu_{i}^{2})\zeta_{1}^{2}p$.

%When $\Sigma$ and $\zeta_{1}^{-2}\B$ are known, the quantity $\mathrm{Var}(|G_0|_{\infty})/\mathrm{Var}(|G|_{\infty})$ may be computed or approximated via Monte Carlo simulations. However, this is 
As $\Sigma$ and $\zeta_{1}^{-2}\B$ are rarely known in practice, we use bootstrap approximation to estimate the value of $\mathrm{Var}(|G_0|_{\infty})/\mathrm{Var}(|G|_{\infty})$.
Combining Corollary \ref{co2} and \eqref{eq:boot_mean}, we propose using
$$
\mathrm{Var}^{*}(|\bar{X}_{n}^{*}|_{\infty})/ \mathrm{Var}^{*}(|\tilde{\btheta}_{n}|_{\infty}),
$$
%which can be approximated via Monte Carlo simulations, 
to estimate $\mathrm{ARE}(\hat{\btheta}_{n}, \bar{X}_n)$.

\begin{example}
	Suppose $X_1, \ldots X_n$ are i.i.d.~from $N(\btheta, I_p)$, then $\nu_i^2$ follows a chi-square distribution with $p$ degrees of freedom. 
	It follows that $\mE(\nu_i^2)=p$ and $\mE(\nu_i^{-1})=\Gamma(p/2-1/2)/\{2^{1/2}\Gamma(p/2)\}$, where $\Gamma(\cdot)$ is the gamma function.. So the ARE is
	$
	\mathrm{ARE}(\hat{\btheta}_{n}, \bar{X}_n)=p\{\Gamma(p/2-1/2)\}^2/\{2^{1/2}\Gamma(p/2)\}^2.
	$
	Using Stirling's formula,
	$ \lim_{p\to\infty} \mathrm{ARE}(\hat{\btheta}_{n}, \bar{X}_n) = 1. $
	Thus, for high-dimensional Gaussian data, the sample spatial median has the same asymptotically efficiency as the sample mean in simultaneous inference for $\btheta$.
\end{example}

\begin{example}
	When the data are from the multivariate $t$-distribution with degrees of freedom $v>2$ and shape matrix $\Omega=I_{p}$, $\nu_i^2/p\sim F_{p, v}$, where $F_{p,v}$ is the $F$ distribution with parameters $p$ and $v$.
	Then, $\mE(\nu_i^2)=pv/(v-2)$ and $\mE(\nu_i^{-1})=\Gamma(v/2+1/2)\Gamma(p/2-1/2)/\{v^{1/2}\Gamma(v/2)\Gamma(p/2)\}$. Thus, the ARE is
	$\mathrm{ARE}(\hat{\btheta}_{n}, \bar{X}_n)=(v-2)^{-1}p\{\Gamma(v/2+1/2)\Gamma(p/2-1/2)\}^2/\{\Gamma(v/2)\Gamma(p/2)\}^2. $
	It is clear that $\mathrm{ARE}(\hat{\btheta}_{n},~ \bar{X})>1$ for large enough $p$.~ In addition, 	
	$
	\lim_{p\to\infty} \mathrm{ARE}(\hat{\btheta}_{n}, \bar{X}_n) = 2(v-2)^{-1}\{\Gamma(v/2+1/2)\}^2/\{\Gamma(v/2)\}^2>1\,.
	$
	Thus, for high-dimensional $t$-distribution, the sample spatial median is asymptotically more efficient than the sample mean in simultaneous inference for $\btheta$.
\end{example}

\begin{figure}[hbt]
	\centerline{
		\includegraphics[width=14cm]{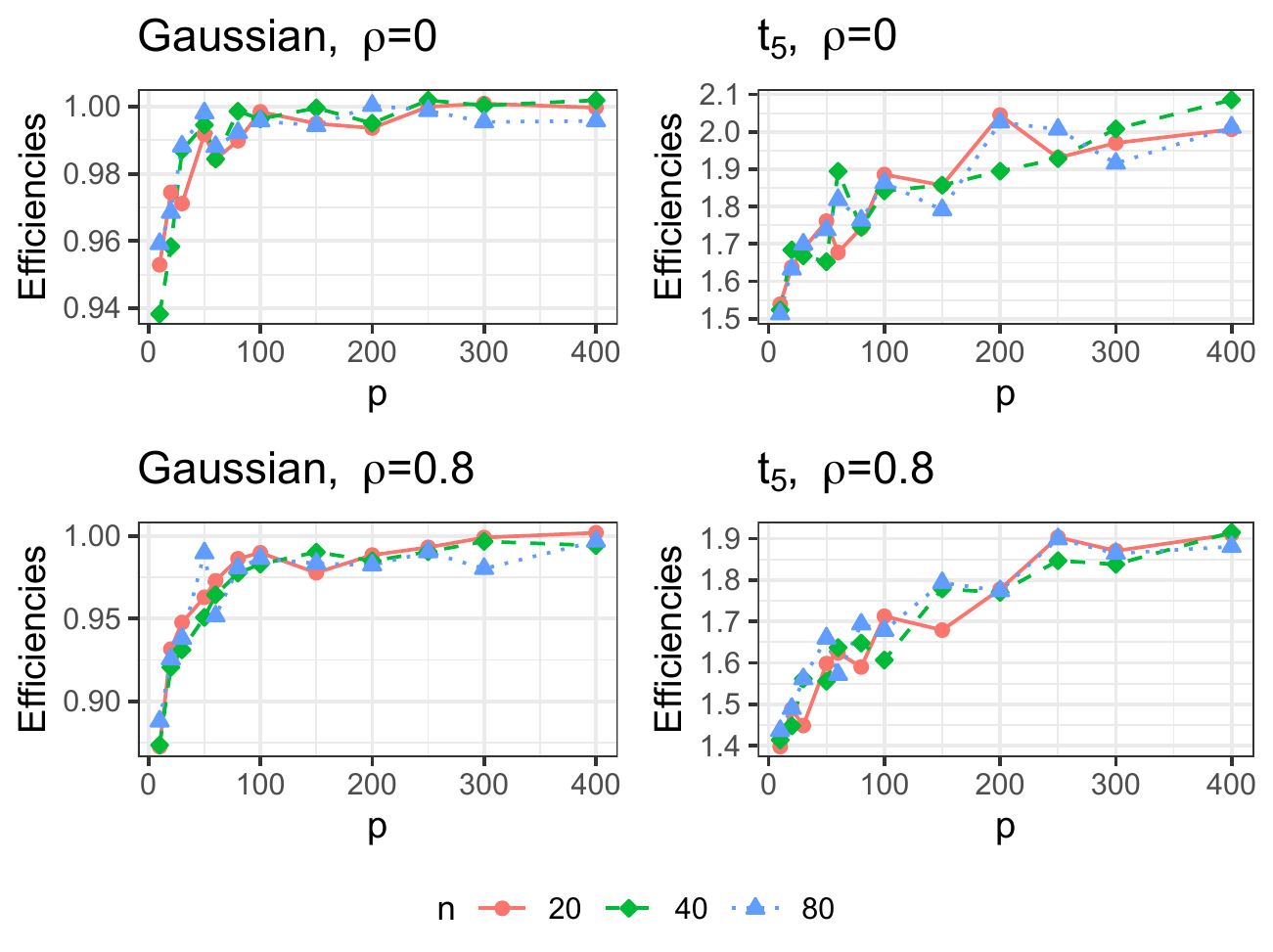}
	}
	\caption{Finite sample relative efficiency of $|\hat{\btheta}_{n}|_{\infty}$ compared to $|
		\bar{X}_n|_{\infty}$ based on $5000$ replications,  the data are generated from multivariate normal distribution (Gaussian) and $t$-distribution with $5$ degrees of freedom ($t_{5}$). The shape matrix $\Omega=(\rho^{|j-\ell|})_{p\times p}$ with $\rho=0$ and $0.8$.}
	\label{Fig1}
\end{figure}

Figure \ref{Fig1} plots the simulated values of $\mathrm{Var}(|\bar{X}_n|_{\infty})/\mathrm{Var}(|\hat{\btheta}_n|_{\infty})$ with a range of dimensions and sample sizes under different models.
For Gaussian data, the relative efficiency kept increasing in $p$, and it approached $1$ as $p$ getting larger.
For the data simulated from multivariate $t$-distribution, the relative efficiency was greater than $1$ for all combinations of $n$ and $p$. This indicates that the sample spatial median is more efficiency than the sample mean for $t$-distribution.
The results were consistent under different covariance structure considered in the simulation.

\section{Numerical studies}\label{sec:numerical}

In this section, we report Monte Carlo simulations on simultaneous confidence intervals and multiple testing with FDR control, along with a real data analysis, to demonstrate the performance of the proposed approaches. Additional simulations on global tests can be found in Appendix C of the supplementary material.
In the simulations, all results were based on $2500$ replications. In the bootstrap implementation, the number of bootstrap iterations was set to $B=400$. 

\subsection{Simulations on simultaneous confidence intervals}\label{sec:simulations}

We first examine the performance of the SCIs based on $\hat{\btheta}_n$, and compare it with the SCIs based on $\bar{X}_n$.
The sample size $n$ is taken to be $100$ or $200$, and the dimensions $p= 100$ and $1000$ are considered for each sample size. 
%as discussed in Remark \ref{remark:05}. 
%To save space, we only present the results for $n=100$ and $p\in\{1000, 2000\}$ here. The results for $n=200$ and $p\in\{50, 100\}$ can be found in the supplement material.
Two types of commonly used elliptical distributions are considered: (I) the multivariate normal distribution $N(\btheta,\Sigma)$; (II) the multivariate $t$-distribution with $3$ degrees of freedom, mean vector $\btheta$, and covariance matrix $\Sigma$.
In addition, we include the following independent components model: (III) $X_i=\btheta + \Sigma^{1/2}Z_i$, where each component of $Z_i$ are i.i.d.~from the standard Laplace distribution. %The random vectors generated from model III are not sub-gaussian.
We set $\Sigma=(\rho^{|j-\ell|})$ with $\rho=0, 0.2, 0.5$ and $0.8$.
To save space, we present the results for $\rho=0$ and $0.8$ here. 
The results for $\rho\in\{0.2,0.5\}$ are similar and are reported in the supplementary material. 
We consider both sparse and dense case scenarios for $\btheta$: (i) $\btheta_{1}=(2,-2,3,0,\ldots, 0)$; (ii) $\btheta_{2}=(0.2,\ldots,0.2_{\lfloor p/4\rfloor},0,\ldots,0)$. Here $\lfloor\cdot\rfloor$ is the floor function.

Table \ref{tab1} reports the coverage probability and median length of the SCIs based on $\hat{\btheta}_n$, the results of the SCIs based on $\bar{X}_n$ are presented in parentheses. 
For Models I and II from elliptical distributions, we observe that the SCIs based on $\hat{\btheta}_n$ and $\bar{X}_n$ both achieve satisfying coverage probability for different choices for $\rho$, $\btheta$, $n$ and $p$. 
For the data simulated from the multivariate normal distribution, the median length of the SCIs based on $\hat{\btheta}_n$ is very close to that of the the SCIs based on $\bar{X}_n$. 
These results indicate that $\hat{\btheta}_n$ has similar asymptotic efficiency as $\bar{X}_n$ in simultaneous inference for $\btheta$ under high-dimensional Gaussian model as discussed in Section \ref{sec:ARE}.
For the multivariate $t$-distribution, the SCIs based on $\hat{\btheta}_n$ is much narrower than the SCIs based on $\bar{X}_n$.
These results suggest that the SCIs based on $\hat{\btheta}_n$ is more efficient than the SCIs based on $\bar{X}_n$ for multivariate $t$-distribution, which is heavy-tailed.
This is consistent with the asymptotic analysis in Section \ref{sec:ARE}.
Moreover, the results for Model III, which does not belong to the elliptical distribution family, shows the robustness of the SCIs based on the spatial median, and it performs similar to the SCIs based on the sample mean.
We also note that the median length of the SCIs decreases when $n$ increases or $p$ decreases for each model.

\begin{table}[ht]
	\footnotesize
	\centering
	{
		\caption{Coverage probability (in $\%$) and median length of the SCIs based on $\hat{\btheta}_n$, the results of the SCIs based on $\bar{X}_n$ are in parentheses.}
		\label{tab1}
		\setlength\tabcolsep{3pt}
		\resizebox{\textwidth}{!}{
			\begin{tabular}{@{}cccccccccccccc@{}}
				
				\hline
				&&&&\multicolumn{4}{c}{$\btheta=\btheta_1$}& &\multicolumn{4}{c}{$\btheta=\btheta_2$}\cr
				&&&&\multicolumn{2}{c}{Coverage probability}&\multicolumn{2}{c}{Median length}&&\multicolumn{2}{c}{Coverage probability}&\multicolumn{2}{c}{Median length}\cr
				
				Model & $\rho$ & $n$&$p$& 90\% & 95\% & 90\% & 95\%& &90\% & 95\% & 90\% & 95\%\cr
				
				\hline
				I & 0 &100&~$100$ & 89.6 (89.9) & 94.4 (94.4) & 0.65 (0.65) & 0.69 (0.69)
				&& 88.9 (88.8) & 94.1 (93.9) & 0.65 (0.65) & 0.69 (0.69) \cr
				
				&&&$1000$& 89.5 (89.6) & 94.7 (94.4) & 0.77 (0.77) & 0.80 (0.80)
				&& 89.5 (89.5) & 94.0 (94.0) & 0.77 (0.77) & 0.81 (0.80) \cr
				
				&&200&~$100$& 89.8 (89.8) & 95.1 (95.1) & 0.46 (0.46) & 0.49 (0.49)
				&& 88.6 (88.8) & 94.4 (94.7) & 0.46 (0.46) & 0.49 (0.49) \cr
				
				&&&$1000$& 89.7 (89.7) & 94.4 (94.6) & 0.55 (0.55) & 0.57 (0.57)
				&& 89.1 (89.2) & 94.7 (94.6) & 0.55 (0.55) & 0.57 (0.57) \cr
				
				%\cline{2-13}
				
				\cline{2-13}
				
				& 0.8 &100&~$100$ & 89.1 (88.7) & 94.6 (94.6) & 0.64 (0.63) & 0.68 (0.67)
				&& 88.4 (88.6) & 93.7 (94.1) & 0.64 (0.63) & 0.68 (0.67) \cr
				
				&&&$1000$& 88.4 (88.4) & 93.8 (93.7) & 0.76 (0.76) & 0.80 (0.79)
				&& 89.0 (89.2) & 94.6 (94.6) & 0.76 (0.76) & 0.80 (0.79) \cr
				
				&&200&~$100$& 90.5 (90.1) & 95.2 (94.9) & 0.45 (0.45) & 0.48 (0.48)
				&& 89.6 (89.6) & 94.0 (94.1) & 0.45 (0.45) & 0.48 (0.48) \cr
				
				&&&$1000$& 90.4 (90.4) & 94.5 (94.4) & 0.54 (0.54) & 0.56 (0.56)
				&& 88.4 (88.5) & 93.6 (93.8) & 0.54 (0.54) & 0.56 (0.56) \cr
				
				\hline
				
				II & 0 &100&~$100$ & 89.7 (88.6) & 94.7 (93.7) & 0.71 (1.05) & 0.75 (1.11)
				&& 88.8 (88.8) & 94.5 (94.2) & 0.71 (1.05) & 0.75 (1.11) \cr
				
				&&&$1000$& 89.4 (91.0) & 95.8 (95.0) & 0.84 (1.25) & 0.88 (1.30) 
				&& 89.1 (89.0) & 94.4 (94.5) & 0.84 (1.25) & 0.88 (1.31) \cr
				
				&&200&~$100$& 88.6 (89.1) & 94.2 (95.1) & 0.50 (0.76) & 0.53 (0.81)
				&& 89.5 (89.7) & 94.4 (94.8) & 0.50 (0.76) & 0.53 (0.80) \cr
				
				&&&$1000$& 89.6 (88.7) & 94.8 (94.6) & 0.59 (0.90) & 0.62 (0.94)
				&& 90.1 (89.5) & 94.8 (93.9) & 0.59 (0.90) & 0.62 (0.94) \cr
				
				\cline{2-13}

				& 0.8 &100&~$100$ & 89.1 (90.7) & 94.4 (94.9) & 0.69 (1.02) & 0.74 (1.09)
				&& 89.4 (89.7) & 94.2 (94.4) & 0.69 (1.02) & 0.74 (1.09) \cr
				
				&&&$1000$& 89.3 (89.1) & 94.6 (94.4) & 0.83 (1.23) & 0.87 (1.29)
				&& 89.8 (88.8) & 94.7 (94.4) & 0.83 (1.23) & 0.87 (1.29) \cr
				
				&&200&~$100$& 87.6 (87.7) & 93.4 (93.6) & 0.49 (0.73) & 0.52 (0.78)
				&& 90.3 (90.1) & 94.9 (95.2) & 0.49 (0.73) & 0.52 (0.78) \cr
				
				&&&$1000$& 88.7 (89.7) & 94.7 (94.6) & 0.59 (0.88) & 0.61 (0.92)
				&& 90.2 (90.8) & 94.7 (95.7) & 0.59 (0.89) & 0.61 (0.93) \cr

				\hline
				
				III&0&100&$100$& 89.8 (89.4) & 94.6 (94.5) & 0.65 (0.66) & 0.69 (0.70)
				&& 89.1 (89.0) & 94.4 (94.4) & 0.65 (0.66) & 0.69 (0.70) \cr
				
				&&&$1000$ & 88.3 (88.2) & 93.6 (93.7) & 0.78 (0.78) & 0.82 (0.82)
				&& 89.1 (89.0) & 94.2 (93.8) & 0.78 (0.78) & 0.82 (0.82) \cr
				
				%\cline{2-12}
				
				&&200&~$100$ & 90.6 (91.1) & 95.0 (95.0) & 0.46 (0.46) & 0.49 (0.49)
				&& 90.6 (90.1) & 95.2 (95.2) & 0.46 (0.46) & 0.49 (0.49) \cr
				
				&&&$1000$ & 90.1 (90.4) & 95.0 (94.6) & 0.55 (0.55) & 0.57 (0.58)
				&& 88.7 (89) & 93.6 (93.8) & 0.55 (0.55) & 0.57 (0.58)\cr
				
				\cline{2-13}
				
				&0.8&100&$100$& 90.4 (89.7) & 95.0 (94.8) & 0.63 (0.63) & 0.68 (0.68)
				&& 89.0 (88.9) & 95.0 (94.9) & 0.63 (0.63) & 0.67 (0.68) \cr
				
				&&&$1000$ & 88.7 (88.9) & 93.8 (94.0) & 0.77 (0.77) & 0.80 (0.80)
				&& 89.0 (89.0) & 94.6 (94.3) & 0.76 (0.76) & 0.80 (0.80) \cr
				
				%\cline{2-12}
				
				&&200&~$100$ & 88.8 (89.1) & 94.2 (94.0) & 0.45 (0.45) & 0.48 (0.48)
				&& 90.2 (89.7) & 94.8 (95.0) & 0.45 (0.45) & 0.48 (0.48) \cr
				
				&&&$1000$ & 90.0 (90.3) & 95.0 (95.0) & 0.54 (0.54) & 0.57 (0.57)
				&& 88.8 (89.1) & 94.2 (94.1) & 0.54 (0.54) & 0.57 (0.57) \cr

				\hline
			\end{tabular}
		}
	}
\end{table}

\subsection{Simulations on multiple testing with FDR control}

In this section, we examine the performance of the sample spatial median-based B-H method introduced in Section \ref{sec:FDR}, and compare it to the B-H procedure based on the sample mean with p-values calculated from $N(0,1)$ in \citet{Liu2014}. We set $\theta_{0,j}=0$ for all $j=1,\ldots,p$. The data are generated from Models I and II with $p=1000$. For $\btheta=(\theta_1,\ldots,\theta_{p})^{\top}$, let $\theta_j=2(\log p/n)^{1/2}$ for $1\leq j\leq p_1$ and $\theta_j=0$ for $(p_1+1)\leq j\leq p$, where %$\kappa$ is chosen from $1$ to $5$ and
$p_1=0.1p$.

Table \ref{tab2} reports the empirical FDR and power for the sample spatial median-based ($\mathrm{FDR}_{M}$ and $\mathrm{power}_{M}$) and the sample mean-based ($\mathrm{FDR}_{A}$ and $\mathrm{power}_{A}$) B-H procedures \citep{Liu2014} with nominal level $\alpha=0.1$ and $0.2$. The results indicate that the FDR are well controlled by both methods. For the multivariate normal distribution, the B-H procedures based on the spatial median and the sample mean have similar performance. However, the sample spatial median-based B-H method outperforms the sample mean-based B-H procedure in terms of empirical power under multivariate $t$-distribution, which is heavy-tailed.

\begin{table}[ht]
	\centering
	{
		\caption{Empirical FDR and power for the spatial median-based ($\mathrm{FDR}_{M}$ and $\mathrm{power}_{M}$) and the sample mean-based ($\mathrm{FDR}_{A}$ and $\mathrm{power}_{A}$) in \citet{Liu2014} via B-H procedures.}
		\label{tab2}
		\setlength\tabcolsep{3pt}
		%\resizebox{\textwidth}{!}{
			\begin{tabular}{@{}ccccccccccccc@{}}
				\hline
				
				%\hline
				&&&\multicolumn{4}{c}{$\alpha=0.1$}& &\multicolumn{4}{c}{$\alpha=0.2$}\cr
				%&&&&\multicolumn{2}{c}{$c_0=0.05$}&\multicolumn{2}{c}{$c_0=0.1$}&&\multicolumn{2}{c}{$c_0=0.05$}&\multicolumn{2}{c}{$c_0=0.1$}\cr
				
				Model & $\rho$ & $n$& $\mathrm{FDR}_{M}$ & $\mathrm{FDR}_{A}$ & $\mathrm{power}_{M}$ & $\mathrm{power}_{A}$ & & $\mathrm{FDR}_{M}$ & $\mathrm{FDR}_{A}$ & $\mathrm{power}_{M}$ & $\mathrm{power}_{A}$ \cr
				\hline
				
				%\hline
				I & 0 &50& 0.124 & 0.124 & 0.996 & 0.996
				&& 0.224 & 0.222 & 0.999 & 0.999 \cr
				
				&&100& 0.107 & 0.106 & 0.997 & 0.997
				&& 0.202 & 0.201 & 0.999 & 0.999 \cr
				
				%\cline{2-12}
				
				& 0.2 &50& 0.125 & 0.124 & 0.996 & 0.996
				&& 0.224 & 0.223 & 0.999 & 0.999 \cr
				
				&&100& 0.107 & 0.106 & 0.997 & 0.997
				&& 0.202 & 0.201 & 0.999 & 0.999 \cr
				
				%\cline{2-12}
				
				& 0.5 &50& 0.125 & 0.124 & 0.996 & 0.996
				&& 0.225 & 0.223 & 0.999 & 0.999 \cr
				
				&&100& 0.107 & 0.105 & 0.997 & 0.997
				&& 0.202 & 0.201 & 0.999 & 0.999 \cr
				
				%\cline{2-12}
				
				& 0.8 &50& 0.127 & 0.124 & 0.996 & 0.996
				&& 0.227 & 0.223 & 0.999 & 0.999 \cr
				
				&&100& 0.108 & 0.105 & 0.997 & 0.997
				&& 0.204 & 0.199 & 0.999 & 0.999 \cr
				
				%\cline{2-12}
				
				\hline
				
				II & 0 &50& 0.117 & 0.099 & 0.984 & 0.728
				&& 0.215 & 0.193 & 0.992 & 0.805 \cr
				
				&&100& 0.103 & 0.088 & 0.987 & 0.710
				&& 0.197 & 0.179 & 0.994 & 0.795 \cr
				
				%\cline{2-12}
				
				& 0.2 &50& 0.117 & 0.098 & 0.984 & 0.727
				&& 0.215 & 0.194 & 0.992 & 0.805 \cr
				
				&&100& 0.103 & 0.087 & 0.987 & 0.709
				&& 0.198 & 0.179 & 0.994 & 0.795 \cr
				
				%\cline{2-12}
				
				& 0.5 &50& 0.118 & 0.099 & 0.984 & 0.727
				&& 0.216 & 0.194 & 0.992 & 0.803 \cr
				
				&&100& 0.103 & 0.087 & 0.987 & 0.708
				&& 0.198 & 0.178 & 0.994 & 0.794 \cr
				
				%\cline{2-12}
				
				& 0.8 &50& 0.120 & 0.098 & 0.984 & 0.724 
				&& 0.218 & 0.192 & 0.992 & 0.800 \cr
				
				&&100& 0.104 & 0.087 & 0.987 & 0.705
				&& 0.199 & 0.177 & 0.994 & 0.791 \cr
				
				\hline 
				
			\end{tabular}
			%}
	}
\end{table}

\subsection{Real data analysis}\label{sec:real_data}

Type 2 diabetes%, the most common type of diabetes, 
is a disease in which the body becomes resistant to normal effects of insulin and gradually loses the capacity to produce enough insulin. %, which results in blood glucose levels rise too high. 
Because skeletal muscle is the main tissue for insulin-stimulated glucose disposal, skeletal muscle insulin resistance is commonly viewed as the critical component of whole-body insulin resistance, and thus is critical to the pathogenesis of Type 2 diabetes.
%Type 2 diabetes is the result of interaction between environmental factors and a strong hereditary component, and the genetics of Type 2 diabetes are complicated.
%A better understanding of how insulin alters expression of genes can provide a deeper insight on Type 2 diabetes and improve the way the disease is managed and treated clinically. 
To investigate the effects of insulin on gene expression in skeletal muscle, a microarray study was performed in 15 diabetic patients using the Affymetrix Hu95A chip of muscle biopsies both before and after insulin treatment \citep{Wu2007}. 
In this paper, we are interested in the gene expression alteration, that is, the change of the gene expression level, due to the treatment. 
The data are available at \texttt{https://www.ncbi.nlm.nih.gov/geo/query/acc.cgi?acc=GSE22309}.
The data were normalized by quantile normalization by the \texttt{normalizeQuantiles} function in the \texttt{limma} R package.
Follow \citet{Wang2015}, we focused on $2547$ curated gene sets with at least 15 genes, which are from the C2 collection of the GSEA online pathway databases. 
The gene expression values are consolidated by taking the average when multiple probes are associated with the same gene.

We implemented the Median test based on $T_n$ on the $2519$ gene sets. % to see whether there is a significant changes in gene expression levels for each gene set. 
This is equivalent to testing whether the median change vector of gene expression levels is equal to 0.
The number of bootstrap iterations is $B=10^5$.
With the Bonferroni correction, there are $1242$ gene sets identified as significant at $5\%$ level.
For comparison, we applied the WPL test \citep{Wang2015} and the CQ test \citep{Chen2010} on the same gene sets. 
For the WPL test, $1060$ gene sets are selected as significant; and for the CQ test, $630$ gene sets are identified as significant. 
Out of the $630$ gene sets selected by the CQ test, $605$ of them are also identified by our proposed method, and $629$ of them are identified by the WPL test.
It has been argued in \citet{Wang2015} that some gene expression levels have heavy tails as their kurtosises are much larger than the kurtosis of a normal distribution, 3. 
Thus, the methods based on the spatial median (Median test and the WPL test) are expected to be more robust and efficient than those based on moments (CQ test).
In addition, out of the $1060$ gene sets identified by the WPL test, $958$ of them are significant based on our proposed approach.

\begin{figure}[htp]
	\centerline{
		\includegraphics[width=14cm,height=16cm]{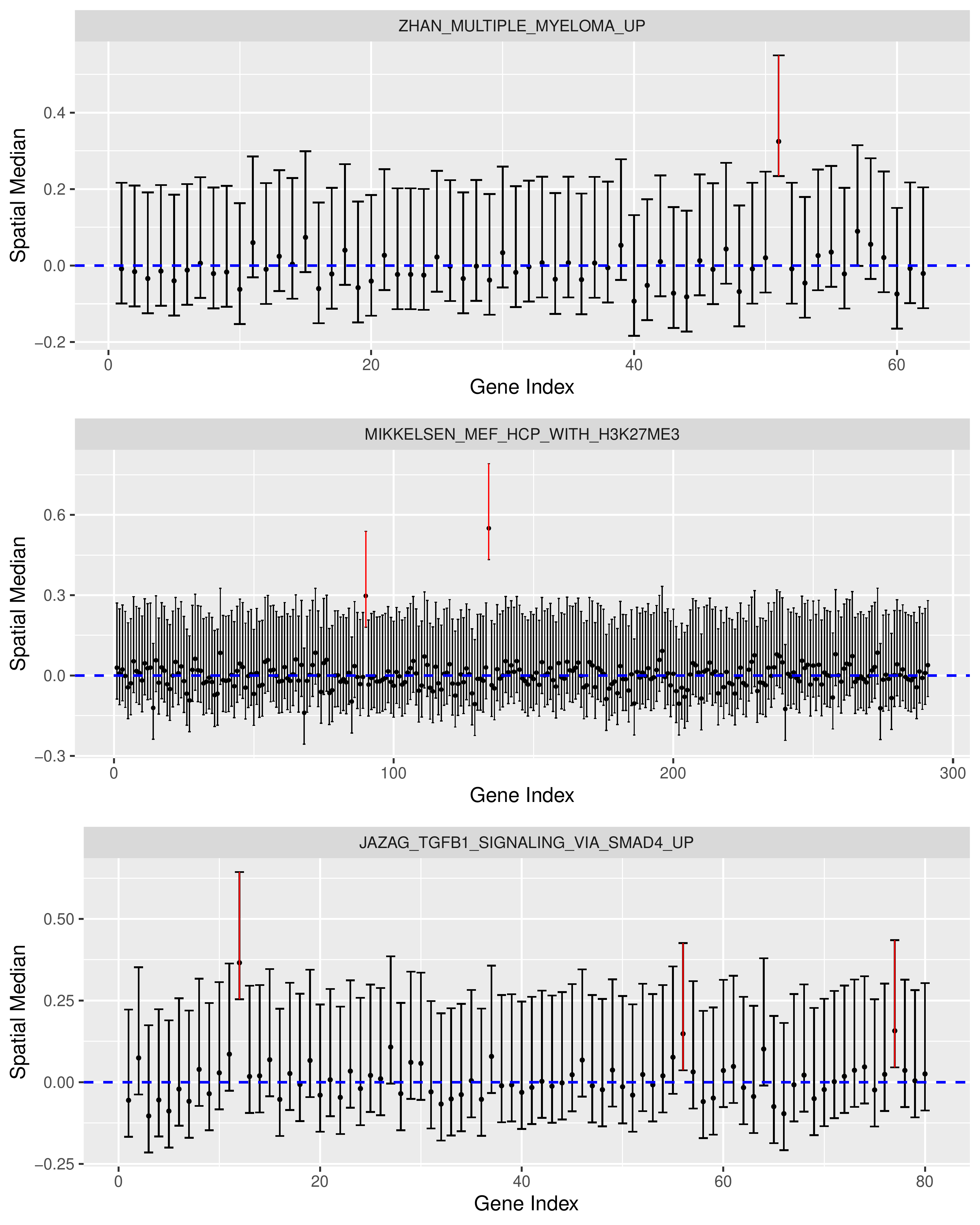}
	}
	\caption{Simultaneous Confidence intervals (SCIs) for spatial medians of three gene sets. 
		%: ZHAN\_MULTIPLE\_MYELOMA\_UP, MIKKELSEN\_MEF\_HCP\_WITH\_H3K27ME3, JAZAG\_TGFB1\_SIGNALING\_VIA\_SMAD4\_UP\\  and MIKKELSEN\_NPC\_HCP\_WITH\_H3K4ME3\_AND\_H3K27ME3.
	}
	\label{Fig_real_data}
\end{figure}

As argued in Remark \ref{remark:07}, the Median test based on $T_n$ is more powerful in detecting strong sparse signal compared to the WPL test.
To see this, we look into the following three gene sets: \\
(1) ZHAN\_MULTIPLE\_MYELOMA\_UP; \\
(2) MIKKELSEN\_MEF\_HCP\_WITH\_H3K27ME3; \\
(3) JAZAG\_TGFB1\_SIGNALING\_VIA\_SMAD4\_UP. \\
The p-values of the WPL test for these three gene sets are $0.41$, $0.31$, $0.27$, respectively.
However, the p-values of the Median test are all less than $1.0\times 10^{-5}$ with $B=10^5$ bootstrap iterations for these three gene sets.
Figure \ref{Fig_real_data} 
plots the SCIs for the spatial median vectors of the change of gene expression levels for these three gene sets.
The confidence intervals that do not cover $0$ are colored in red.
It is very clear that the only one or two big values in the spatial median results in a rejection of the Median test, while the signals from other dimensions are not strong enough to land a rejection by the the WPL test.

Finally, we use the spatial median-based B-H procedure to perform multiple testing with FDR control on the three gene sets to detect differentially expressed genes (DEG), which is one of the most important targets in genomic analysis. %We choose the . 
Table \ref{tab3} reports the detected differentially expressed genes (DEG) in each gene set with nominal level $\alpha=0.1$, along with the corresponding marginal p-value $P_{j}=2-2\Phi(|T_{n,j}|)$ and the confidence interval in the SCIs for the selected genes.
It can be seen that for all the selected genes, the marginal p-values are very small, and the corresponding confidence intervals do not cover $0$.

\begin{table}[htp]
	\centering
	{
		\caption{Detected differentially expressed genes (DEG) by the spatial median-based B-H procedure for three gene sets with $\alpha=0.1$; ``$p$-value'' refers to the marginal $p$-value $P_{j}=2-2\Phi(|T_{n,j}|)$, and ``CI'' refers to the confidence interval in the SCIs for the selected genes.}
		\label{tab3}
		\setlength\tabcolsep{3pt}
		\resizebox{\textwidth}{!}{
			\begin{tabular}{@{}lccc@{}}
				\hline
				
				Gene set & DEG & p-value & CI \\
				
				\hline 
				
				ZHAN\_MULTIPLE\_MYELOMA\_UP & CDKN1A & 0.00082 & (0.234, 0.550) \\
				
				MIKKELSEN\_MEF\_HCP\_WITH\_H3K27ME3 & MYOD1 & $<0.00001$ & (0.433, 0.791) \\
				
				JAZAG\_TGFB1\_SIGNALING\_VIA\_SMAD4\_UP & HDAC4 & 0.00058 & (0.254, 0.644) \\
				
				\hline 
				
			\end{tabular}
		}
	}
\end{table}

\section{Discussion}\label{sec:conclusion}

In this paper, we established one-sample and two-sample Gaussian and bootstrap approximations for ultrahigh dimensional sample spatial median under a general model beyond elliptical distributions.
It is of interest to study whether our results are potentially extendable to some other distribution families. 
We leave this to a future work.
In addition, the proposed test based on the maxima of the sample spatial median is more powerful under sparse alternatives compared to those based on $L_2$-norms.
It is well known that the $L_2$-norm type tests are more powerful under dense alternatives. Thus, it is of interest to consider combining the test based on the maximum-norm and $L_2$-norm, which could be potentially powerful under both sparse and dense alternatives. We also leave this to a future study.

\section*{Supplementary material}
The supplementary material %available at Biometrika online 
includes all the technical proofs and  some additional numerical results.

\bibliographystyle{agsm}
\bibliography{paper_ref.bib}

\newpage

\begin{center}
{\bf\LARGE
Supplement to ``Statistical Inference for Ultrahigh Dimensional Location Parameter Based on Spatial Median'' % 
}
\end{center}

\bigskip

\iffalse
\begin{center}
Guanghui  Cheng$^a$, ~~~~~~~~Liuhua Peng$^b$
\end{center}

\begin{center}
{\sl $^a$Guangzhou Institute of International Finance, Guangzhou University, \\ 
$^b$School of Mathematics and Statistics; School of Mathematics and Statistics, The University of Melbourne}%The University of Macau, The University of Melbourne}
\end{center}

\begin{abstract}
		The supplementary material contains two parts. Technical proofs, including preliminary lemmas and proof of main theorems, are presented in Section \ref{sec:SM_tech}. Section \ref{sec:SM_simulation} reports additional simulations.
	\end{abstract}
\fi

\appendix

\renewcommand{\theequation}{S.\arabic{equation}}
\renewcommand\thetable{A\arabic{table}}
\renewcommand\thefigure{A\arabic{figure}}
%  To get the journal style of heading for an appendix, mimic the following.

\section*{Appendix A: Technical Proofs}

\renewcommand\thesection{A}

We first introduce and recall some notation. 
%We write $\lambda_{\max}(M)=\lambda_{1}(M)\geq\cdots\geq \lambda_{d}(M)=\lambda_{\min}(M)$ for the eigenvalues of a $d\times d$ matrix $M$.
For a $d_1\times d_2$ matrix $M=(m_{j\ell})_{d_1\times d_2}$, its matrix $\varrho$-norm is $\|M\|_{\varrho} = \sup\{\|Mx\|_{\varrho}:\|x\|_{\varrho}=1\}$.
%For $\varrho=1,2$, and $\infty$, the corresponding 
Specifically, the $1$-, $2$-, and $\infty$-norms of $M$ are $\|M\|_1=\max_{1\leq \ell\leq d_2}\sum_{j=1}^{d_1}|m_{j\ell}|$, $\|M\|_2=\{\lambda_{\max}(M^{\top}M)\}^{1/2}$, and $\|M\|_{\infty}=\max_{1\leq j\leq d_1}\sum_{\ell=1}^{d_2}|m_{j\ell}|$. %, respectively.
The Frobenius norm of $M$ is $\|M\|_{F}=\{\sum_{j=1}^{d_1}\sum_{\ell=1}^{d_2}m_{j\ell}^2\}^{1/2}$. 

Define a random $p\times p$ matrix $Q=n^{-1}\sum_{i=1}^nR_i^{-1}W_iW_i^{\top}$ such that $\mE(Q)=\mE(R_i^{-1}W_iW_i^{\top})$, and denote $Q_{j\ell}$ as the $(j,\ell)$th element of $Q$. 
Denote $\mE^{*}(\cdot)$ and $\mathrm{Var}^{*}(\cdot)$ be the expectation and variance conditional on $X_1,\ldots,X_n$, respectively.
Recall that $W_{i,j}$ is the $j$th element of $W_i$ for $i=1,\ldots,n$ and $j=1,\ldots,p$; $\omega_{j\ell}$ is the $(j,\ell)$th element of $\Omega$; and $\Gamma_j$ is the $j$th row of $\Gamma$.
Finally, we will denote various positive absolute constants by $ C_1, C_2,C_3, \ldots$ without mentioning this explicitly.

\subsection{Preliminary lemmas}\label{sec:appendix_lem}

\renewcommand\thelemma{A\arabic{lemma}}
\setcounter{lemma}{0}

In this section. we present several preliminary lemmas, whose proof can be found in online Appendix B.
\begin{lemma}\label{lem:norm}
	(Concentration of norms) Suppose that Conditions \ref{c1} and \ref{c3} hold with $a_0(p)\asymp p^{1-\delta}$ for some positive constant $\delta\leq 1/2$.
	Then, for sufficient large $p$, there exist positive constants $c_1$ and $c_2$ such that
	\begin{eqnarray}
		\notag \P\left\{p-\epsilon p^{(1+\delta)/2}\leq \|U_1\|^2\leq p+\epsilon p^{(1+\delta)/2}\right\} & \geq & 1-c_1\exp\{-c_2p^{\delta\alpha/(4\alpha+4)}\} 
	\end{eqnarray}
	and
	\begin{eqnarray}
		\notag \P\left\{(1-\epsilon)\tr(\Omega)\leq \|\Gamma U_1\|^2\leq (1+\epsilon)\tr(\Omega)\right\} & \geq & 1-c_1\exp\{-c_2p^{\delta\alpha/(4\alpha+4)}\}
	\end{eqnarray}
	\iffalse
	and 
	\begin{eqnarray}
		\notag \P\left\{1-\epsilon \leq \|\Gamma S(U_1)\|^2\leq 1+\epsilon\right\} \geq 1-c_1\exp\{-c_2p^{\delta\alpha/(4\alpha+4)}\} 
	\end{eqnarray}
	\fi
	for any fixed $0<\epsilon<1$.
\end{lemma}

\begin{lemma}\label{lemma:SM_moments}
	Suppose that Conditions \ref{c1}, \ref{c2} and \ref{c3} hold with $a_0(p)\asymp p^{1-\delta}$ for some positive constant $\delta\leq 1/2$. Then, for any $i=1,\ldots,n$, \\ % and $1\leq i_1\neq i_2\leq n$, \\
	(i) $\mE(\|U_i\|^4) = p\mE(U_{i,j}^4) + p(p-1)$,
	\begin{eqnarray}
		\notag \mE(\|U_i\|^6) & = & p\mE(U_{i,j}^6) + 3p(p-1)\mE(U_{i,j}^4) + p(p-1)(p-2), \\
		\notag \mE(\|U_i\|^8) & = & p\mE(U_{i,j}^8) + 4p(p-1)\mE(U_{i,j_1}^6) + 3p(p-1)\{\mE(U_{i,j_1}^4)\}^2 \\
		\notag & & + 3p(p-1)\mE(U_{i,j}^4) + p(p-1)(p-2)(p-3)\,.
	\end{eqnarray}
	In addition, $\mE(\|U_i\|^{2k}) = p^k +  O(p^{k-1})$ and $\mE(\|U\|^{k})=p^{k/2}+O(p^{k/2-1})$ for any positive integer $k$.
	
	(ii) $\mE(\|\Gamma U_{i}\|^4) = p^2+O(p^{2-\delta})$, $\mE(\|\Gamma U_i\|^{6}) =p^3 + O(p^{3-\delta})$.
	%\begin{eqnarray}
	%	\notag \mE(\|\Gamma U_i\|^{6}) & \asymp & \{\tr(\Omega)\}^3, \\
	%	\notag \mE(\|\Gamma U_i\|^8) & \asymp & \{\tr(\Omega)\}^4 \,.
	%\end{eqnarray}
	In addition, $\mE(\|\Gamma U_{i}\|) = p^{1/2} + O(p^{1/2-\delta})$ and $\mE(\|\Gamma U_i\|^3)=p^{3/2} + O(p^{3/2-\delta})$. \\
	(iii) $\mE\{\|\Gamma S(U_i)\|^2\} = 1 + O(p^{-1/2})$
	and $\mE\{\|\Gamma S(U_i)\|^4\} = 1 + O(p^{-1/3})$. \\
	(iv) $\mE(\nu_{i}^{-k}) \lesssim \zeta_{k}p^{k/2}$ for $k=1,2,3$.
	\iffalse
	(iii) $\mE\{(\Gamma U_{i_1})^{\top}(\Gamma U_{i_2})\}=0$, $\mE[\{(\Gamma U_{i_1})^{\top}(\Gamma U_{i_2})\}^2]=\tr(\Omega^2)$ and $\mE[\{(\Gamma U_{i_1})^{\top}(\Gamma U_{i_2})\}^4]\lesssim \{\tr(\Omega^2)\}^2 + \tr(\Omega^4)$; \\
	(iv) $\mE\{(W_{i_1}^{\top}W_{i_2})^2\} \asymp \tr(\Omega^2)/\{\tr(\Omega)\}^2$.
	\fi
\end{lemma}	

\begin{lemma}\label{lemma:SM_Q}
	Suppose Conditions \ref{c1}, \ref{c2} and \ref{c3} with $a_0(p)\asymp p^{1-\delta}$ for some positive constant
	$\delta\leq 1/2$ hold. %In addition, assume $\log n =O(p^{\gamma})$ for some positive $\gamma<\delta\alpha/(4\alpha+4)$. 
	Define a random $p\times p$ matrix $Q=n^{-1}\sum_{i=1}^nR_i^{-1}W_iW_i^{\top}$ and let $Q_{j\ell}$ be the $(j,\ell)$th element of $Q$.
	Then, \\
	(i)
	$|Q_{j\ell}|\lesssim \zeta_{1}p^{-1}|\omega_{j\ell}|+O_p(\zeta_{1}n^{-1/2}p^{-1}+\zeta_{1}p^{-7/6}+\zeta_{1}p^{-1-\delta/2})$.  \\
	(ii) $Q_{j\ell}=Q_{0, j\ell}+O_p(\zeta_{1}p^{-7/6}+\zeta_{1}p^{-1-\delta/2})$, where $Q_{0, j\ell}$ is the $(j,\ell)$th element of  $$Q_{0}=n^{-1}p^{-1/2}\sum_{i=1}^{n}\nu_{i}^{-1}\{\Gamma S(U_{i})\}\{\Gamma S(U_{i})\}^{\top}.
	$$ In addition, $Q_{0}$ satisfies $$\tr[\mE(Q_{0}^2)-\{\mE(Q_{0})\}^2] = O(n^{-1}p^{-1}).$$
\end{lemma}

\begin{lemma}\label{lemma:05}
	Suppose Conditions \ref{c1}, \ref{c2} and \ref{c3} with $a_0(p)\asymp p^{1-\delta}$ for some positive constant
	$\delta\leq 1/2$ hold.
	Then, \\
	(i) $\mE\{(\zeta_{1}^{-1}W_{i,j})^4\} \lesssim \bar{M}^2$ and $\mE\{(\zeta_{1}^{-1}W_{i,j})^2\} \gtrsim \underline{m}$ for all $i=1,\ldots,n$ and $j=1,\ldots,p$. \\
	(ii) $\|\zeta_{1}^{-1}W_{i,j}\|_{\psi_{\alpha}}\lesssim \bB$ for all $i=1,\ldots,n$ and $j=1,\ldots,p$. \\
	(iii) $\mE(W_{i,j}^2)=p^{-1}\omega_{jj}+O(p^{-1-\delta/2})$ for $j=1,\ldots,p$ and $\mE(W_{i,j}^2)=p^{-1}\omega_{j\ell}+O(p^{-1-\delta/2})$ for $1\leq j\neq \ell\leq p$. \\
	(iv) if $\log p = o(n^{1/3})$,  $$\left|n^{-1/2}\sum_{i=1}^n\zeta_{1}^{-1}W_i\right|_{\infty}=O_{p}\{\log^{1/2} (np)\}
	{\rm ~~and~~} \left|n^{-1}\sum_{i=1}^n(\zeta_{1}^{-1}W_i)^2\right|_{\infty}=O_{p}(1)\,.$$
\end{lemma}

\begin{lemma}\label{lem:SM_boot_br}
	Suppose the conditions of Theorem \ref{theo:boot} hold, then %Denote $\tilde{W}_i=(X_1-\hat{\btheta}_n)/\|X_i-\hat{\btheta}_n\|$ for $i=1,\ldots,n$. Then, 
	\begin{align}\label{eq:tildetheta}
		n^{1/2}\tilde{\btheta}_{n}=n^{-1/2}\zeta_{1}^{-1}\sum_{i=1}^n Z_i W_i+{\tilde C}_n\,,
	\end{align}
	where $|{\tilde C}_n|_{\infty}=O_p\{ n^{-1/4}\log^{1/2} (np)+p^{-(1/6 \wedge \delta/2)}\log^{1/2} (np)\}$.
\end{lemma}

The following lemma is Nazarov's inequality, and its proof can be found in \cite{Cher2017}.

\begin{lemma}[Nazarov's inequality]\label{lem:SM_Naza}
	Let $Y_0=(Y_{0,1}, \ldots, Y_{0,p})^{\top}$ be a centered Gaussian random vector in $\mR^p$ and $\mE(Y_{0,j}^2)\geq b$ for all $j=1 ,\ldots, p$
	and some constant $b>0$,  then for every $y\in \mR^p$ and $a>0$, 
	\[
	\P(Y_0\leq y+a)-\P(Y_0\leq y)\lesssim a\log^{1/2}(p)\,.
	\]
\end{lemma}

\subsection{Proof of main results}

\begin{proof}[\textit{Proof of Lemma \ref{lem:Br}}]
	As $\btheta$ is a location parameter, we assume $\btheta=0$ without loss of generality. Then, $W_i = S(X_i) = \|X_i\|^{-1}X_i%\|\epsilon_i\|^{-1}\epsilon_i=R_i^{-1}\epsilon_i 
	= \|\Gamma U_i\|^{-1}\Gamma U_i$ for $i=1,\ldots,n$.
	\iffalse
	\begin{eqnarray}
		\notag W_i = S(X_i) = \frac{X_i}{\|X_i\|}=\frac{\epsilon_i}{||\epsilon_i||}=\frac{\epsilon_i}{R_i}\,.
	\end{eqnarray}
	\fi
	The sample spatial median $\hat{\btheta}_{n}$ satisfies
	\[
	\sum_{i=1}^nS(X_i-\hat{\btheta}_{n})=\sum_{i=1}^{n}\frac{X_i-\hat{\btheta}_{n}}{\|X_i-\hat{\btheta}_{n}\|}=\sum_{i=1}^{n}\frac{W_i-R_i^{-1}\hat{\btheta}_{n}}{\|W_i-R_{i}^{-1}\hat{\btheta}_{n}\|}=0\,,
	\]
	which is  is equivalent to
	\[
	n^{-1}\sum_{i=1}^n(W_i-R_i^{-1}\hat{\btheta}_{n})(1-2R_i^{-1}W_i^{\top}\hat{\btheta}_{n}+R_i^{-2}\|\hat{\btheta}_{n}\|^2)^{-1/2}=0\,
	\]
	as $W_i^{\top}W_i=1$. 
	
	Under Condition \ref{c2}, $\zeta_{k}=\mE(R_{i}^{-k})=O(p^{-k/2})$ for $k=1,2,3,4$. In addition, Lemma \ref{lemma:SM_Q} indicates that 
	$Q_{j\ell}=Q_{0, j\ell}+O_p(\zeta_{1}p^{-7/6}+\zeta_{1}p^{-1-\delta/2})$, where $Q_{0, j\ell}$ is the $(j,\ell)$th element of  $Q_{0}=n^{-1}p^{-1/2}\sum_{i=1}^{n}\nu_{i}\{\Gamma S(U_{i})\}\{\Gamma S(U_{i})\}^{\top}\,.
	$ In addition, $Q_{0}$ satisfies $\tr[\mE(Q_{0}^2)-\{\mE(Q_{0})\}^2] = O(n^{-1}p^{-1}).$
	Thus, from the similar procedure as in the proof of Lemma 1.2 of \citet{Cheng2019}, we can show that $$\|\hat{\btheta}_{n}\|=O_p(\zeta_{1}^{-1}n^{-1/2}).$$ 
	Then, for $i=1,\ldots,n$, we have $|R_i^{-1}W_i^{T}\hat{\btheta}_{n}|
	\leq R_i^{-1}\|\hat{\btheta}_{n}\| = O_p(n^{-1/2})$ and $R_i^{-2}||\hat{\btheta}_{n}||^2= O_p(n^{-1})$.
	By the first-order Taylor expansion, the above equation can be rewritten as
	\begin{eqnarray}\label{eq:SM_lemma1_01}
		& & n^{-1}\sum_{i=1}^n (W_i-R_i^{-1}\hat{\btheta}_{n})(1+R_i^{-1}W_i^{\top}\hat{\btheta}_{n}-2^{-1}R_i^{-2}\|\hat{\btheta}_{n}\|^2+\delta_{1i} )=0\,,
	\end{eqnarray}
	where  
	$
	\delta_{1i}=O_p\{(R_i^{-1}W_i^{\top}\hat{\btheta}_{n}-2^{-1}R_i^{-2}\|\hat{\btheta}_{n}\|^2)^2\}=O_{p}(n^{-1}).
	$
	By Markov's inequality, for any $\varepsilon>0$,
	\begin{eqnarray}
		\notag \P\left(\max_{1\leq i\leq n}R_{i}^{-1}\geq \varepsilon \zeta_{1}n^{1/4}\right) & = & \P\left(\max_{1\leq i\leq n}R_{i}^{-4}\geq \varepsilon^4 \zeta_{1}^4n\right) \\
		\notag & \leq & \mE\left(\max_{1\leq i\leq n}R_{i}^{-4}\right)/(\varepsilon^4 \zeta_{1}^4n) 
		\leq  n\mE(R_{i}^{-4})/(\varepsilon^4 \zeta_{1}^4n) \lesssim \varepsilon^{-4}\,,
	\end{eqnarray}
	where the last inequality is due to Condition \ref{c2}. Thus, $\max_{1\leq i \leq n}R_{i}^{-2}=O_{p}(\zeta_{1}^2n^{1/2})$, and consequently, 
	$\max_{1\leq i\leq n}\delta_{1i} = O_{p}(\|\hat{\btheta}_n\|^2\max_{1\leq i \leq n}R_{i}^{-2})=O_p(n^{-1/2})$.
	Rewrite \eqref{eq:SM_lemma1_01} as
	\begin{eqnarray}
		\notag & & n^{-1}\sum_{i=1}^n (1-2^{-1}R_i^{-2}\|\hat{\btheta}_{n}\|^2+\delta_{1i})W_i + n^{-1}\sum_{i=1}^nR_i^{-1}(W_i^{\top}\hat{\btheta}_{n})W_i \\
		\notag &  = & n^{-1}\sum_{i=1}^nR_i^{-1}(1-2^{-1}R_i^{-2}\|\hat{\btheta}_{n}\|^2+\delta_{1i})\hat{\btheta}_{n}+n^{-1}\sum_{i=1}^nR_i^{-2}(W_i^{\top}\hat{\btheta}_{n})\hat{\btheta}_{n}\,,
	\end{eqnarray}
	which implies
	\begin{eqnarray}\label{eq:s}
		& & n^{-1}\sum_{i=1}^n (1-2^{-1}R_i^{-2}\|\hat{\btheta}_{n}\|^2+\delta_{1i})W_i+n^{-1}\sum_{i=1}^nR_i^{-1}(W_i^{\top}\hat{\btheta}_{n})W_i \\
		\notag & = & n^{-1}\sum_{i=1}^nR_i^{-1}(1+\delta_{1i}+\delta_{2i})\hat{\btheta}_{n}\,,
	\end{eqnarray}
	where $\delta_{2i}=R_i^{-1}W_{i}^{\top}\hat{\btheta}_n-2^{-1}R_i^{-2}\|\hat{\btheta}_{n}\|^2 = O_{p}(\delta_{1i}^{1/2})$ satisfies $\max_{1\leq i \leq n}\delta_{2i}=O_{p}(n^{-1/4})$.
	It is straightforward to check that $n^{-1}\sum_{i=1}^nR_i^{-1}(W_i^{\top}\hat{\btheta}_{n})W_i = n^{-1}\sum_{i=1}^nR_i^{-1}W_iW_i^{\top}\hat{\btheta}_{n}=Q\hat{\btheta}_{n}$.
	From Lemma \ref{lemma:SM_Q}, 
	$$|Q_{j\ell}|\lesssim \zeta_{1}p^{-1}|\omega_{j\ell}|+O_p(\zeta_{1}n^{-1/2}p^{-1}+\zeta_{1}p^{-7/6}+\zeta_{1}p^{-1-\delta/2}),$$ and this implies that
	$$|Q\hat{\btheta}_{n}|_{\infty}\leq \|Q\|_{1}||\hat{\btheta}_{n}|_{\infty}\lesssim \zeta_{1}p^{-1}\|\Omega\|_1|\hat{\btheta}_{n}|_{\infty}+O_p(\zeta_{1}n^{-1/2}+\zeta_{1}p^{-1/6}+\zeta_{1}p^{-\delta/2})|\hat{\btheta}_{n}|_{\infty}.$$
	
	According to Lemma \ref{lemma:05}, we have that $|n^{-1}\sum_{i=1}^n\zeta_{1}^{-1}W_i|_{\infty}=O_p\{ n^{-1/2}\log^{1/2}(np)\}$.
	Then, $$\left|\zeta_{1}^{-1}n^{-1}\sum_{i=1}^n \delta_{1i}W_i\right|^2_{\infty}\leq \left|n^{-1}\sum_{i=1}^n(\zeta_{1}^{-1}W_{i})^2\right|_{\infty} \left(n^{-1}\sum_{i=1}^n\delta^2_{1i}\right)\lesssim O_{p}(n^{-2}).$$
	In addition, we have that 
	$|\zeta_{1}^{-1}n^{-1}\sum_{i=1}^n R_i^{-2}\|\hat{\btheta}_{n}\|^2W_i|_{\infty}\lesssim O_{p}(n^{-1})$.
	Regarding equation \eqref{eq:s} and the fact that $\zeta_{1}^{-1}n^{-1}\sum_{i=1}^nR_i^{-1}=1+O_p(n^{-1/2})$ , we obtain
	\begin{eqnarray}
		\notag  \hat{\btheta}_{n}|_{\infty} & \lesssim &  \left|\zeta_{1}^{-1}n^{-1}\sum_{i=1}^nW_i\right|_{\infty}+\zeta_{1}^{-1}|Q\hat{\btheta}_{n}|_{\infty} \\
		\notag & \lesssim & p^{-1}a_0(p)|\hat{\btheta}_{n}|_{\infty} + O_p(n^{-1/2}+p^{-(1/6 \wedge \delta/2)})|\hat{\btheta}_{n}|_{\infty} +O_p\{n^{-1/2}\log^{1/2} (np)\}\,.
	\end{eqnarray}
	Thus, we conclude that $|\hat{\btheta}_{n}|_{\infty}=O_p\{n^{-1/2}\log^{1/2} (np)\}$ as $a_0(p)\asymp p^{1-\delta}$.
	In addition, we have
	$|\zeta_{1}^{-1}Q\hat{\btheta}_{n}|_{\infty} =  O_p\{n^{-1/2}p^{-(1/6 \wedge \delta/2)}\log^{1/2}(np)+n^{-1}\log^{1/2}(np)\}$
	and
	$n^{-1}\sum_{i=1}^nR_i^{-1}(1+\delta_{1i}+\delta_{2i})=\zeta_{1}\{1+O_{p}(n^{-1/4})\}$.
	Finally, we can write
	\[
	n^{1/2}(\hat{\btheta}_{n}- \btheta)=n^{-1/2}\zeta_{1}^{-1}\sum_{i=1}^nW_i+C_n\,,
	\]
	where $C_n$ satisfies $|C_n|_{\infty}=O_p\{ n^{-1/4}\log^{1/2}(np)+p^{-(1/6 \wedge \delta/2)}\log^{1/2}(np)\}.$
\end{proof}

\begin{proof}[\textit{Proof of Theorem \ref{theo1}}]
	%Let $G_1,\ldots,G_n$ be a sequence of i.i.d.~$p$-dimensional random vectors from $N(0, I_p)$, then $U_i=G_i/\|G_i\|$ for $i=1,\ldots,n$.
	%Recall that $W_i = \Gamma U_i/\|\Gamma U_i\|$ %and $U_i=G_i/\|G_i\|$, we have $W_i=\Gamma G_i/\|\Gamma G_i\|$ 
	%for $i=1,\ldots,n$.
	%Let $\mathcal{A}_{2i} = \{(1-\epsilon)\tr(\Omega)\leq \|\Gamma U_i\|^2\leq (1+\epsilon)\tr(\Omega)\}$ for a fixed $0<\epsilon<1$, then
	%$$ \P(\cap_{i=1}^{n}\mathcal{A}_{2i}) \geq 1-nc_1\exp(-c_2p^{\delta\alpha/(4\alpha+4)}) $$
	%according to Lemma \ref{lem:norm}. 	
	%From the Lemma \ref{lem:Br}, we have $W_i=\Gamma g_i/ ||\Gamma g_i||$, and ${\mathcal A}=\{0.5 \tr(\Omega)\leq ||\Gamma g_i||^2\leq 1.5 \tr(\Omega) \}$ occurs with probability at least $1-p^{-\delta}$.
	Let $L_{n,p}=n^{-1/4}\log^{1/2}(np)+p^{-(1/6 \wedge \delta/2)}\log^{1/2}(np)$. Then, for any sequence $\eta_n\rightarrow \infty$ and any $t\in\mathbb{R}^{p}$,
	\begin{eqnarray}
		\notag & & \P\left\{n^{1/2}(\hat{\btheta}_{n}-\btheta)\leq t\right\} 
		=  \P\left(n^{-1/2}\zeta_{1}^{-1}\sum_{i=1}^nW_i+C_n\leq t\right) \\
		\notag & \leq & \P\left(n^{-1/2}\zeta_{1}^{-1}\sum_{i=1}^nW_i\leq t+\eta_nL_{n,p}\right) + \P(|C_n|_{\infty}>\eta_nL_{n,p})\,.
		%\\
		%\notag & \leq & \P\left(n^{-1/2}\zeta_{1}^{-1}\sum_{i=1}^nW_i\leq t+\eta_nL_{n,p}\right)+o(1).
		%\\
		%\notag & \leq & \P\left(n^{-1/2}\zeta_{1}^{-1}\sum_{i=1}^nW_i\leq t+\eta_nL_{n,p}, {\mathcal A}_a\right)+\P({\mathcal A}_a^c)+o(1)\,.
	\end{eqnarray}
	According to Lemma \ref{lemma:05}, $\mE\{(\zeta_{1}^{-1}W_{i,j})^4\} \lesssim \bar{M}^2$, $\mE\{(\zeta_{1}^{-1}W_{i,j})^2\} \gtrsim \underline{m}$, and $\|\zeta_{1}^{-1}W_{i,j}\|_{\psi_{\alpha}}\lesssim \bB$ for all $i=1,\ldots,n$ and $j=1,\ldots,p$. According to the Gaussian approximation for independent partial sums in \citet{Koike2019}, let $G\sim N(0, \zeta_{1}^{-2}\B)$ with $\B=\mE(W_{1}W_{1}^{\top})$, we have
	\begin{eqnarray}
		\notag & & \P\left(n^{-1/2}\zeta_{1}^{-1}\sum_{i=1}^nW_i\leq t+\eta_nL_{n,p}\right) 
		\leq  \P(G\leq t+\eta_n L_{np}) + O\left(\left\{n^{-1}\log^5(np)\right\}^{1/6}\right) \\
		\notag & \leq & \P(G\leq t) + O\{\eta_n L_{np}\log^{1/2}(p)\} + O\left(\left\{n^{-1}\log^5(np)\right\}^{1/6}\right)\,,
	\end{eqnarray}
	where the last inequality is from Nazarov's inequality in Lemma \ref{lem:SM_Naza}. 
	It is also worth noting that the order $O\left(\left\{n^{-1}\log^5(np)\right\}^{1/6}\right)$ %obtained in \citet{Koike2019} 
	is improved to 
	$O\left(\left\{n^{-1}\log^5(np)\right\}^{1/4}\right)$ in \citet{Cher2019}. Thus,
	\begin{eqnarray}
		\notag & & \P\left\{n^{1/2}(\hat{\btheta}_{n}-\btheta)\leq t\right\} \\
		\notag & \leq & \P(G\leq t) + O\{\eta_n L_{np}\log^{1/2}(p)\} + O\left(\left\{n^{-1}\log^5(np)\right\}^{1/6}\right) \\
		\notag & & + \P(|C_n|_{\infty}>\eta_nL_{n,p})\,.
	\end{eqnarray}
	On the other hand, we also have
	\begin{eqnarray}
		\notag & & \P\left\{n^{1/2}(\hat{\btheta}_{n}-\btheta)\leq t\right\} \\
		\notag & \geq & \P(G\leq t)-O\{\eta_n L_{np}\log^{1/2}(p)\}-O\left(\left\{n^{-1}\log^5(np)\right\}^{1/6}\right) \\
		\notag & & -\P(|C_n|_{\infty}>\eta_nL_{n,p})\,,
	\end{eqnarray}
	where $\P(|C_n|_{\infty}>\eta_nL_{n,p})\to0$ as $n\to\infty$ according to Lemma \ref{lem:Br}.
	
	Then, %under Condition \ref{c3} with $a_{0}(p)\asymp p^{1-\delta}$, 
	if
	$\log p=o(n^{1/5})$ and $\log n=o(p^{1/3 \wedge \delta})$,
	with sufficiently slow $\eta_n\rightarrow \infty$, we have
	\[
	\sup_{t \in \mathbb{R}^p}\left|\P\{n^{1/2}(\hat{\btheta}_{n}-\btheta)\leq t\}- \P(G\leq t)\right|\rightarrow 0\,.
	\]
	We obtain immediately from Corollary 5.1 in \cite{Cher2017} that 
	\[
	\rho_n(\mathcal{A}^{\mathrm{re}})= \sup_{A \in {\mathcal A}^{\mathrm{re}}}\left|\P\{n^{1/2}(\hat{\btheta}_{n}-\btheta) \in A \}-\P\left(G \in A\right)\right| \rightarrow 0\,,
	\]
	which leads to the conclusion of this theorem.
\end{proof}

\begin{proof}[\textit{Proof of Theorem \ref{theo:boot}}]
	Let $\tilde{X}_{i}=X_i-\hat{\btheta}_{n}$ and $\tilde{R}_{i}=\|\tilde{X}_i\|$ for $i=1,\ldots,n$. According to Lemma \ref{lem:SM_boot_br}, 
	\begin{align}
		\notag n^{1/2}\tilde{\btheta}_{n}=n^{-1/2}\zeta_{1}^{-1}\sum_{i=1}^n Z_i W_i+{\tilde C}_n\,,
	\end{align}
	where ${\tilde C}_n$ satisfies $|{\tilde C}_n|_{\infty}=O_p\{n^{-1/4}\log^{1/2} (np)+p^{-(1/6 \wedge \delta/2)}\log^{1/2} (np)\}$. 
	
	\iffalse
	\begin{eqnarray}
		\notag \mathrm{Cov}^{*}\left\{n^{-1/2}\zeta_{1}^{-1}\sum_{i=1}^n Z_iW_i+{\tilde C}_n\right\} = \zeta_{1}^{-1}\hat{\B} +\hat{\btheta}_{n}\hat{\btheta}_{n}^{\top}+o_p\{n^{-1}\log(np)\}.
	\end{eqnarray}
	Denote $\B_{j\ell}$ and $\hat{\B}_{j\ell}$ be the $(j,\ell)$th element of $\B$ and $\hat{\B}$, respectively. In addition, 
	define
	\begin{eqnarray}
		\notag \Delta_n = \max_{1\leq j,\ell\leq p}\left|\zeta_{1}^{-2}\hat{\B}_{j\ell}-\hat{\btheta}\hat{\btheta}^{\top}+o_p\{n^{-1}\log(np)\}-\zeta_{1}^{-2}\B_{j\ell}\right|,
	\end{eqnarray}
	then with the fact $|\hat{\btheta}|_{\infty}=O_p(n^{-1/2}\log^{1/2}(np))$,  we have
	\begin{eqnarray}
		\notag \Delta_n & \leq & \Delta_{n1} +|\hat{\btheta}\hat{\btheta}^{\top}|_{\infty}+o_p\{n^{-1}\log(np)\}  \lesssim \Delta_{n1}+|\hat{\btheta}|^2_{\infty} ,
	\end{eqnarray}
	where
	\begin{eqnarray}
		\notag \Delta_{n1} =\max_{1\leq j,\ell\leq p}|\zeta_{1}^{-2}\hat{\B}_{j\ell}-\zeta_{1}^{-2}\B_{j\ell}| = \max_{1\leq j,\ell\leq p}\left|n^{-1}\zeta_{1}^{-2}\sum_{i=1}^n\left\{W_{i,j}W_{i,\ell}-\mE(W_{i,j}W_{i,\ell})\right\}\right|.
	\end{eqnarray}
	\fi
	
	Denote $\bar{W}_n=n^{-1}\sum_{i=1}^{n}W_i$ and rewrite
	\begin{eqnarray}
		\notag n^{1/2}\tilde{\btheta}_{n}=n^{-1/2}\zeta_{1}^{-1}\sum_{i=1}^n Z_i (W_i-\bar{W}_n) +\left(n^{-1/2}\zeta_{1}^{-1}\sum_{i=1}^n Z_i\right) \bar{W}_n+{\tilde C}_n,
	\end{eqnarray}
	where 
	\begin{eqnarray}
		\notag \left|\left(n^{-1/2}\zeta_{1}^{-1}\sum_{i=1}^n Z_i\right) \bar{W}_n\right|_{\infty} \leq \zeta_{1}^{-1}\left|n^{-1/2}\sum_{i=1}^n Z_i\right|\left|\bar{W}_n\right|_{\infty} \lesssim n^{-1/2}\log^{1/2}(np)
	\end{eqnarray}
	according to Lemma \ref{lemma:05} (iii). 
	
	It is clear that $\mE^{*}\left\{n^{-1/2}\zeta_{1}^{-1}\sum_{i=1}^n Z_i (W_i-\bar{W}_n)\right\}=0$.
	Let $\hat{\B}=n^{-1}\sum_{i=1}^n W_i W_i^{\top}$, then
	\begin{eqnarray}
		\notag \mathrm{Var}^{*}\left\{n^{-1/2}\zeta_{1}^{-1}\sum_{i=1}^n Z_i (W_i-\bar{W}_n)\right\} = \zeta_{1}^{-1}\hat{\B} - \zeta_{1}^{-2}\bar{W}_n\bar{W}_n^{\top}.
	\end{eqnarray}
	Denote $\B_{j\ell}$ and $\hat{\B}_{j\ell}$ be the $(j,\ell)$th element of $\B$ and $\hat{\B}$, respectively. In addition, denote $\bar{W}_{n,j}$ as the $j$th element of $\bar{W}_n$.
	Define
	\begin{eqnarray}
		\notag \Delta_n = \max_{1\leq j,\ell\leq p}\left|\zeta_{1}^{-2}\hat{\B}_{j\ell}-\zeta_{1}^{-2}\bar{W}_{n,j}\bar{W}_{n,\ell}-\zeta_{1}^{-2}\B_{j\ell}\right|,
	\end{eqnarray}
	then
	\begin{eqnarray}
		\notag \Delta_n & \leq & \Delta_{n1} + \max_{1\leq j,\ell\leq p}\left|\zeta_{1}^{-2}\bar{W}_{n,j}\bar{W}_{n,\ell}\right| \lesssim \Delta_{n1} + n^{-1}\log(np),
	\end{eqnarray}
	where
	\begin{eqnarray}
		\notag \Delta_{n1} & = & \max_{1\leq j,\ell\leq p}|\zeta_{1}^{-2}\hat{\B}_{j\ell}-\zeta_{1}^{-2}\B_{j\ell}| 
		=  \max_{1\leq j,\ell\leq p}\left|n^{-1}\zeta_{1}^{-2}\sum_{i=1}^n\left\{W_{i,j}W_{i,\ell}-\mE(W_{i,j}W_{i,\ell})\right\}\right|.
	\end{eqnarray}
	%\iffalse
	%$\Delta_n=\max_{1\leq j,\ell\leq p}|\zeta_{1}^{-2}\hat{\B}_{j\ell}-\zeta_{1}^{-2}\B_{j\ell}|$, we know that $\Delta_n\leq \Delta_{n1}+|\mathbb{C}_{1n}|_{\infty}$, where
	%\[
	%\Delta_{n1}=\max_{1\leq j,\ell\leq p}\left|n^{-1}\zeta_{1}^{-2}\sum_{i=1}^n\left\{W_{i,j}W_{i,\ell}-\mE(W_{i,j}W_{i,\ell})\right\}\right|\,.
	%\]
	%\fi
	From the properties of the $\psi_{\alpha}$ norm, %  in \cite{Koike2019}, 
	it holds that
	\begin{eqnarray}
		\notag \left\|\max_{1\leq i\leq n; 1\leq j,\ell\leq p}|\zeta_{1}^{-2}W_{i,j}W_{i,\ell}|\right\|_{\psi_{\alpha/2}} & \lesssim & \left\|\max_{1\leq i \leq n,1\leq j\leq p }|\zeta_{1}^{-2}W_{i,j}|^2\right\|_{\psi_{\alpha/2}} \\
		\notag & = & \zeta_{1}^{-2}\left\|\max_{1\leq i\leq n,1\leq j\leq p}|W_{i,j}|\right\|_{\psi_{\alpha}}^2 
		\lesssim  \log^{2} (np)\,.
	\end{eqnarray}
	
	Let $J_n=\max_{1\leq i\leq n; 1\leq j,\ell\leq p}\zeta_{1}^{-2}|W_{i,j}W_{i,\ell}-\mE(W_{i,j}W_{i,\ell})|$,  and 
	\begin{eqnarray}
		\notag \sigma_n^2 & = & \max_{1\leq j,\ell\leq p} \zeta_{1}^{-2}\sum_{i=1}^n\mE\{W_{i,j}W_{i,\ell}-\mE(W_{i,j}W_{i,\ell})\}^2 \\
		\notag & \lesssim &
		\max_{1\leq j,\ell\leq p} \zeta_{1}^{-2}\sum_{i=1}^n\mE\{|W_{i,j}W_{i,\ell}|^2\}\lesssim n\,.
	\end{eqnarray}
	It also follows that 
	\begin{eqnarray}
		\notag  \|J_{n}\|_{\psi_{\alpha/2}}  \lesssim   \zeta_{1}^{-2}\left\|\max_{1\leq i\leq n; 1\leq j,\ell\leq p}|W_{ij,}W_{i,\ell}|\right\|_{\psi_{\alpha/2}}+\max_{1\leq i\leq n; 1\leq j,\ell\leq p}\zeta_{1}^{-2}\mE(|W_{i,j}W_{i,\ell}|) 
		\lesssim  \log^{2} (np)\,.
	\end{eqnarray}
	By Lemma E.1 in \cite{Cher2017}, it holds that
	\begin{eqnarray}
		\notag \mE(\Delta_{n1}) & \lesssim &  n^{-1}\left[\sigma_n\log^{1/2} (p)+\{\mE(J_n^2)\}^{1/2}\log p\right] \\
		\notag & \lesssim & n^{-1}\left\{n^{1/2}\log^{1/2} (p)+\log^{1/\alpha+1}(np)\right\} \\
		\notag & \lesssim & n^{-1/2}\log^{1/2}(np)\,.
	\end{eqnarray}
	Then applying Lemma E.2 in \cite{Cher2017} with $\eta=1$ and $\beta=\alpha/2$, we obtain that
	\begin{align*}
		\P( \Delta_{n1} \geq 2 \mE(\Delta_{n})+t)
		\lesssim \exp\left(-C_1nt^2\right)+3\exp\left\{-C_2\{tn\log^{-{2/\alpha}}(np)\}^{\alpha/2}\right\}\,.
	\end{align*} 
	Thus, there exist a constant $C_1$ depends on $\delta$ such that
	\[
	\P\left\{\Delta_{n1}>C_1n^{-1/2}\log^{1/2}(np)\right\} \lesssim p^{-\delta}\rightarrow 0\,.
	\]
	
	From the multiplier bootstrap theorem and Gaussian comparison in \citet{Cher2017} and \citet{Koike2019}, % as well as the Gaussian comparison in \cite{Cher2015}, 
	\begin{eqnarray}
		\notag & &  \sup_{t\in\mathbb{R}^{p}}\left|\P^{*}\left\{n^{-1/2}\zeta_{1}^{-1}\sum_{i=1}^n Z_i (W_i-\bar{W}_n)\leq t\right\}- \P(G\leq t)\right| \\
		\notag & \lesssim & \Delta_n^{1/2}\log(p)+\{n^{-1}\log^{5}(np)\}^{1/4}\,,
	\end{eqnarray}
	on $\{\Delta_{n}\lesssim n^{-1/2}\log^{1/2}(np)\}$, which occurs with probability $1-p^{-\delta}$. 
	
	Finally, similar to the proof of Theorem \ref{theo1}, we can show that under Conditions \ref{c2} and \ref{c3} with $a_{0}(p)\asymp p^{1-\delta}$, if 
	$\log p =o(n^{1/5})$ and $\log n =o(p^{1/3 \wedge \delta})$,
	we have
	\[
	\sup_{A \in {\mathcal A}^{\mathrm{re}}}\left|\P\{n^{1/2}(\hat{\btheta}_{n}-\btheta)\in A\}- \P^{*}(n^{1/2}\tilde{\btheta}_{n} \in A)\right|\to 0
	\]
	in probability,
	which completes the proof of this theorem.
\end{proof}

%\subsection{Proof of Theorems \ref{theo:SCIs}--\ref{theo:boottwo}}\label{sec:SM_proof}
\begin{proof}[\textit{Proof of Theorem \ref{theo:SCIs}}]
	Theorems \ref{theo1} and \ref{theo:boot} indicates that there exists a positive sequence $\beta_{n,p}\rightarrow 0$ as $n,p\rightarrow \infty$ such that
	\[
	\sup_{t\in {\mathbb R}}\left|\P(n^{1/2}|\hat{\btheta}_{n}-\btheta|_{\infty}\leq t)- \P(|G|_{\infty}\leq t)\right| \leq \beta_{n,p}/2\,
	\]
	and
	\[
	\sup_{t\in {\mathbb R}}\left|\P(n^{1/2}|\hat{\btheta}_{n}-\btheta|_{\infty}\leq t)- \P^{*}(n^{1/2}|\tilde{\btheta}_{n}|_{\infty}\leq t)\right| \leq \beta_{n,p}\,
	\]
	with probability approaching one when $n\rightarrow \infty$. %Setting $T_n=n^{1/2}|\hat{\btheta}_{n}-\btheta|_{\infty}$, and 
	Letting $q_{1-\alpha}$ be the $(1-\alpha)$th quantile of $n^{1/2}|\hat{\btheta}_{n}-\btheta|_{\infty}$, that is, $q_{1-\alpha}=\inf\{u \in \mR: \P(n^{1/2}|\hat{\btheta}_{n}-\btheta|_{\infty}>u) \leq\alpha\}$. Then,
	\[
	\P^{*}(n^{1/2}|\tilde{\btheta}_{n}|_{\infty}\leq q_{1-\alpha+\beta_{n,p}})\geq \P(n^{1/2}|\hat{\btheta}_{n}-\btheta|_{\infty}\leq q_{1-\alpha+\beta_{n,p}})-\beta_{n,p}\geq 1-\alpha\,,
	\]
	with probability approaching one as $n\to\infty$.
	On the other hand, it holds with the same probability that
	\begin{eqnarray}
		\notag & & \P^{*}(n^{1/2}|\tilde{\btheta}_{n}|_{\infty} \leq q_{1-\alpha-3\beta_{n,p}}) \\
		\notag & \leq & \P(n^{1/2}|\hat{\btheta}_{n}-\btheta|_{\infty}\leq q_{1-\alpha-3\beta_{n,p}})+\beta_{n,p} \\
		\notag & = & \P(n^{1/2}|\hat{\btheta}_{n}-\btheta|_{\infty}\leq q_{1-\alpha-3\beta_{n,p}}-n^{-1/6})+\beta_{n,p} \\
		\notag & & + \P(n^{1/2}|\hat{\btheta}_{n}-\btheta|_{\infty}\leq q_{1-\alpha-3\beta_{n,p}}) \\
		\notag & & - \P(n^{1/2}|\hat{\btheta}_{n}-\btheta|_{\infty}\leq q_{1-\alpha-3\beta_{n,p}}-n^{-1/6}) \\
		\notag & < &
		1-\alpha-2\beta_{n,p} + \P(n^{1/2}|\hat{\btheta}_{n}-\btheta|_{\infty}\leq q_{1-\alpha-3\beta_{n,p}}) \\
		\notag & &  - \P(n^{1/2}|\hat{\btheta}_{n}-\btheta|_{\infty}\leq q_{1-\alpha-3\beta_{n,p}}-n^{-1/6})\,,
	\end{eqnarray}
	where $\P(n^{1/2}|\hat{\btheta}_{n}-\btheta|_{\infty}\leq q_{1-\alpha-3\beta_{n,p}}) - \P(n^{1/2}|\hat{\btheta}_{n}-\btheta|_{\infty}\leq q_{1-\alpha-3\beta_{n,p}}-n^{-1/6})$ can be bounded by
	\begin{eqnarray}
		\notag & & \left|\P(n^{1/2}|\hat{\btheta}_{n}-\btheta|_{\infty}\leq q_{1-\alpha-3\beta_{n,p}}) \right. \\
		\notag & & \left. - \P(n^{1/2}|\hat{\btheta}_{n}-\btheta|_{\infty}\leq q_{1-\alpha-3\beta_{n,p}}-n^{-1/6})\right| \\
		\notag & \leq & \left|\P(|G|_{\infty}\leq q_{1-\alpha-3\beta_{n,p}}) - \P(|G|_{\infty}\leq q_{1-\alpha-3\beta_{n,p}}-n^{-1/6})\right| \\
		\notag & & + \left|\P(n^{1/2}|\hat{\btheta}_{n}-\btheta|_{\infty}\leq q_{1-\alpha-3\beta_{n,p}}) - \P(|G|_{\infty}\leq q_{1-\alpha-3\beta_{n,p}})\right| \\
		\notag & & + \left|\P(n^{1/2}|\hat{\btheta}_{n}-\btheta|_{\infty}\leq q_{1-\alpha-3\beta_{n,p}}-n^{-1/6})\right. \\
		\notag & & \left. ~~~~~~~~~~~~ - \P(|G|_{\infty}\leq q_{1-\alpha-3\beta_{n,p}}-n^{-1/6})\right| \\
		\notag & \leq & \left|\P(|G|_{\infty}\leq q_{1-\alpha-3\beta_{n,p}}) - \P(|G|_{\infty}\leq q_{1-\alpha-3\beta_{n,p}}-n^{-1/6})\right| + \beta_{n,p} \\
		\notag & \leq & C_1\left\{n^{-1}\log^5(np)\right\}^{1/6} + \beta_{n,p},
	\end{eqnarray}
	for some positive constant $C_1$, where the last inequality follows from the Nazarov's inequality. Choosing $C_1\left\{n^{-1}\log^5(np)\right\}^{1/6} \leq \beta_{n,p}$, 
	we obtain 
	\begin{eqnarray}
		\notag & & \P^{*}(n^{1/2}|\tilde{\btheta}_{n}|_{\infty} \leq q_{1-\alpha-3\beta_{n,p}}) < 1-\alpha
	\end{eqnarray}
	with probability approaching one.
	It follows that
	\[
	\P(q_{1-\alpha-3\beta_{n,p}}<q^{B}_{1-\alpha}\leq q_{1-\alpha+\beta_{n,p}}) \rightarrow 1,
	\text{~~as~} n, p\rightarrow \infty\,.
	\]
	Therefore, 
	\begin{eqnarray}\label{th3q}
		\notag & & \P(n^{1/2}|\hat{\btheta}_{n}-\btheta|_{\infty}>q^B_{1-\alpha}) \\
		\notag & \leq & \P(n^{1/2}|\hat{\btheta}_{n}-\btheta|_{\infty}>q_{1-\alpha-3\beta_{n,p}}) + \P(q^B_{1-\alpha}\leq q_{1-\alpha-3\beta_{n,p}}) \\
		& \leq & \alpha + 3\beta_{n,p} + o(1)
	\end{eqnarray}
	and
	\begin{eqnarray}
		\notag & &  \P(n^{1/2}|\hat{\btheta}_{n}-\btheta|_{\infty}>q^B_{1-\alpha}) \\
		\notag & \geq & \P(n^{1/2}|\hat{\btheta}_{n}-\btheta|_{\infty}>q_{1-\alpha+\beta_{n,p}})- \P(q^B_{1-\alpha}> q_{1-\alpha+\beta_{n,p}}) \\
		\notag & \geq &  \P(n^{1/2}|\hat{\btheta}_{n}-\btheta|_{\infty}>q_{1-\alpha+\beta_{n,p}}-n^{-1/6})- o(1) \\
		\notag & & + \P(n^{1/2}|\hat{\btheta}_{n}-\btheta|_{\infty}>q_{1-\alpha+\beta_{n,p}}) \\
		\notag & & - \P(n^{1/2}|\hat{\btheta}_{n}-\btheta|_{\infty}>q_{1-\alpha+\beta_{n,p}}-n^{-1/6}) \\
		\notag & \geq & \alpha-2\beta_{n,p}-C_2\left\{n^{-1}\log^5(np)\right\}^{1/6}\geq \alpha-3\beta_{n,p}.
	\end{eqnarray}
	for some positive constant $C_2$, where the second last inequality follows from the Nazarov's inequality and the last inequality is from choosing $\beta_{n,p} \geq C_2\left\{n^{-1}\log^5(np)\right\}^{1/6}$.
	Finally, as $\beta_{n,p}\rightarrow 0$,
	\[
	%\sup_{\alpha \in (0,1)}
	\P(n^{1/2}|\hat{\btheta}_{n}-\btheta|_{\infty}\geq q^{B}_{1-\alpha})-\alpha \rightarrow 0,
	\]
	which completes the proof of this theorem.
\end{proof}

\begin{proof}[\textit{Proof of Theorem \ref{theo:power}}]
	Without loss of generality, we assume $\btheta_0=0$. Rewrite the test statistic as $T_n=n^{1/2}|\hat{\btheta}_{n}|_{\infty}$, and let
	${T}^{c}_n=n^{1/2}|\hat{\btheta}_{n}-\btheta|_{\infty}$, which has the same distribution of $T_n$ under $H_0$.
	Then, it holds that
	\[
	T_n\geq  n^{1/2}|\btheta|_{\infty}-{T}^{c}_n\,.
	\]
	Therefore, the power of the test based on $T_n$ satisfies
	\begin{eqnarray}
		\notag \P(T_n>q^{B}_{1-\alpha}\mid H_1) & \geq & \P(n^{1/2}|\btheta|_{\infty}-{T}^{c}_n \geq q^{B}_{1-\alpha} \mid H_1) \\
		\notag & = & \P({T}^{c}_n\leq  n^{1/2}|\btheta|_{\infty}-q^{B}_{1-\alpha}\mid H_1)
	\end{eqnarray} 
	
	Under the conditions of Theorem \ref{theo:boot}, there exists a positive sequence $\beta_{n,p}\rightarrow 0$ as $n,p\rightarrow \infty$, satisfies
	\begin{align}\label{th4eq1}
		\sup_{t\in {\mathbb R}}|\P(T_n^{c}>t\mid H_1)-\P(|G|_{\infty}>t\mid H_1)|\leq \beta_{n,p},
	\end{align}
	where $G\sim N(0, \zeta_{1}^{-2}\B)$.
	Letting $q_{1-\alpha}$ be the $(1-\alpha)$th quantile of $T_{n}^{C}$ and $q^{G}_{1-\alpha}$ be the $(1-\alpha)$th quantile of $|G|_{\infty}$.
	Choosing $t=q^{G}_{1-\alpha+2\beta_{n,p}}$ in equation (\ref{th4eq1}), we obtain that $|\P(T_n^{c}>q^{G}_{1-\alpha+2\beta_{n,p}}\mid H_1)-\alpha+2\beta_{n,p}|\leq \beta_{n,p}$ and $\P(T_n^{c}>q^{G}_{1-\alpha+2\beta_{n,p}}\mid H_1)\leq \alpha-{\beta}_{n,p}$,
	which implies that $q_{1-\alpha+\beta_{n,p}}\leq q^{G}_{1-\alpha+2\beta_{n,p}}$. 
	
	Note that $q^{B}_{1-\alpha}$ is the $(1-\alpha)$th quantile of $n^{1/2}|\tilde{\btheta}_{n}|_{\infty}$ conditional on $X_1,\ldots,X_n$. 
	By carrying out similar procedure as in the proof of equation \eqref{th3q}, we can show that
	\begin{align}
		\P({T}^{c}_n>  n^{1/2}|\btheta|_{\infty}-q^{B}_{1-\alpha}\mid H_1)\leq \P({T}^{c}_n>  n^{1/2}|\btheta|_{\infty}-q_{1-\alpha+\beta_{n,p}}\mid H_1)+o(1)\,.
	\end{align}
	It follows that
	\[
	\P({T}^{c}_n>  n^{1/2}|\btheta|_{\infty}-q^{B}_{1-\alpha}\mid H_1)\leq \P({T}^{c}_n>  n^{1/2}|\btheta|_{\infty}-q^{G}_{1-\alpha+2\beta_{n,p}}\mid H_1)+o(1)\,.
	\]
	For $|G|_{\infty}$, we know that $\||G|_{\infty}\|_{\psi_2}\lesssim \log^{1/2}(np)$. In addition, for any $t>0$,
	\[
	\P(|G|_{\infty}>t) \leq 2\exp\{-C_1(t/\||G|_{\infty}\|_{\psi_2})^2\}\leq 2\exp\{-C_2t^2\log^{-1}(np)\}\,.
	\]
	%for some positive constants $C_1$ and $C_2$.
	Choosing $t=C_2^{-1/2}\log^{1/2}(2/(\alpha-2\beta_{n,p}))\log^{1/2}(np)$, we arrive at
	\begin{eqnarray}
		\notag \P(|G|_{\infty}>C_2^{-1/2}\log^{1/2}(2/(\alpha-2\beta_{n,p}))\log^{1/2}(np))\leq \alpha-2\beta_{n,p}\,,
	\end{eqnarray}
	which leads to 
	$$q^{G}_{1-\alpha+2\beta_{n,p}}\leq C_2^{-1/2}\log^{1/2}(2/(\alpha-2\beta_{n,p}))\log^{1/2}(np).$$
	Then, if $|\btheta|_{\infty}\geq  Cn^{-1/2}\log^{1/2}(\alpha^{-1})\log^{1/2}(np)$ for a large enough constant $C$, it holds with sufficiently large $C_3$ that
	\begin{eqnarray}
		\notag & & \P(T_n>q^{B}_{1-\alpha}\mid H_1) \\
		\notag & \geq & \P({T}^{c}_n \leq  n^{1/2}|\btheta|_{\infty}-q^{G}_{1-\alpha+2\beta_{n,p}}\mid H_1)+o(1) \\
		\notag & \geq & \P\{T_n^{c}\leq C_3\log^{1/2}(np)\log^{1/2}(\alpha^{-1})\mid H_1\} + o(1) \\
		\notag & \geq & \P\{|G|_{\infty}\leq C_3\log^{1/2}(np)\log^{1/2}(\alpha^{-1})\mid H_1\} - \beta_{n,p} + o(1) \\
		\notag & \geq & 1- 2\alpha^{C_2C_3^2} - \beta_{n,p} + o(1).
	\end{eqnarray}
	We complete the proof of this theorem.
\end{proof}

\begin{proof}[\textit{Proof of Theorem \ref{theo:FDR}}]
	Recall that $\hat{\zeta}_{1}=n^{-1}\sum_{i=1}^{n}\|X_i-\hat{\btheta}_n\|^{-1}$.
	It has been shown in the proof of Lemma \ref{lem:SM_boot_br} that
	\begin{eqnarray}
		\notag \|X_i-\hat{\btheta}_n\|^{-1} = R_{i}^{-1}\left(1+R_{i}^{-1}W_{i}^{\top}\hat{\btheta}_n-2^{-1}R_{i}^{-2}\|\hat{\btheta}_n\|^2+\tilde{\delta}_{1i}\right),
	\end{eqnarray}		
	where $\tilde{\delta}_{1i}$ satisfies $\tilde{\delta}_{1i}=O_{p}(n^{-1})$ and $\max_{1\leq i \leq n}\tilde{\delta}_{1i}=O_{p}(n^{-1/2})$. Thus,
	\begin{eqnarray}
		\notag \hat{\zeta}_{1} & = & n^{-1}\sum_{i=1}^{n}R_{i}^{-1}\left(1+R_{i}^{-1}W_{i}^{\top}\hat{\btheta}_n-2^{-1}R_{i}^{-2}\|\hat{\btheta}_n\|^2+\tilde{\delta}_{1i}\right) \\
		\notag & = & n^{-1}\sum_{i=1}^{n}R_{i}^{-1}(1+\tilde{\delta}_{3i}),
	\end{eqnarray}
	where $\tilde{\delta}_{3i}$ satisfies $\tilde{\delta}_{3i}=O_{p}(n^{-1/2})$ and $\max_{1\leq i \leq n}\tilde{\delta}_{3i}=O_{p}(n^{-1/4})$. By the fact that $n^{-1}\sum_{i=1}^{n}R_{i}^{-1} = \zeta_{1} + O_{p}(\zeta_{1}n^{-1/2})$, we conclude that
	\begin{eqnarray}
		\notag \hat{\zeta}_1/\zeta_{1}-1=O_p(n^{-1/2}).
	\end{eqnarray}
	
	Let $\tilde{W}_i = (X_i-\hat{\btheta}_n)/\|X_i-\hat{\btheta}_n\|$ for $i=1,\ldots,n$. From the proof of Lemma \ref{lem:SM_boot_br}, 
	\begin{eqnarray}
		\notag \tilde{W}_i = (W_i-R_{i}^{-1}\hat{\btheta}_n)(1+\tilde{\delta}_{2i})=W_i+W_i\tilde{\delta}_{2i}-R_{i}^{-1}\hat{\btheta}_n(1+\tilde{\delta}_{2i}),
	\end{eqnarray}
	where $\tilde{\delta}_{2i}$ satisfies $\tilde{\delta}_{2i}=O_{p}(n^{-1/2})$ and $\max_{1\leq i \leq n}\tilde{\delta}_{2i}=O_{p}(n^{-1/4})$. Let $\tilde{W}_{i,j}$ be the $j$th component of $\tilde{W}_{i}$, then
	\begin{eqnarray}
		\notag \hat{\B}_{jj} & = & n^{-1}\sum_{i=1}^{n}\tilde{W}_{i,j}^2 \\
		\notag & = & n^{-1}\sum_{i=1}^{n}W_{i,j}^2\{1+O_{p}(\tilde{\delta}_{2i})\} + n^{-1}\sum_{i=1}^{n}R_{i}^{-1}W_{i,j}\hat{\theta}_{n,j}\{1+O_{p}(\tilde{\delta}_{2i})\}  \\
		\notag & &   + n^{-1}\sum_{i=1}^{n}R_{i}^{-2}\hat{\theta}_{n,j}^2\{1+O_{p}(\tilde{\delta}_{2i})\},
	\end{eqnarray}
	where $\max_{1\leq j \leq p}n^{-1}\sum_{i=1}^{n}W_{i,j}^2\{1+O_{p}(\tilde{\delta}_{2i})\}/\B_{jj} = 1+ O_{p}(n^{-1/4})$,
	\begin{eqnarray}
		\notag & & \max_{1\leq j \leq p}\left|n^{-1}\sum_{i=1}^{n}R_{i}^{-2}\hat{\theta}_{n,j}^2\{1+O_{p}(\tilde{\delta}_{2i})\}\right\| \\
		\notag & \lesssim & \left|n^{-1}\sum_{i=1}^{n}R_{i}^{-2}\right|\max_{1\leq j \leq p}|\hat{\theta}_{n,j}^2| \\
		\notag & = & O_{p}\{\zeta_{1}^{2}n^{-1}\log^{1/2}(np)\}
	\end{eqnarray}
	and
	\begin{eqnarray}
		\notag & &  \max_{1\leq j \leq p}\left|n^{-1}\sum_{i=1}^{n}R_{i}^{-1}W_{i,j}\hat{\theta}_{n,j}\{1+O_{p}(\tilde{\delta}_{2i})\}\right| \\
		\notag & \lesssim & \left(n^{-1}\sum_{i=1}^{n}R_{i}^{-2}\right)^{1/2}\max_{1\leq j \leq p}\left\{\left(n^{-1}\sum_{i=1}^{n}W_{i,j}^2\right)^{1/2}\right\}\max_{1\leq j \leq p}|\hat{\theta}_{n,j}| \\
		\notag & = & O_{p}\{\zeta_{1}p^{-1/2}n^{-1/2}\log^{1/2}(np)\}.
	\end{eqnarray}
	It follows that 
	\begin{eqnarray}
		\notag \max_{1\leq j \leq p}\hat{\B}_{jj}/\B_{jj} = 1+O_{p}\{n^{-1/4}\log^{1/2}(np)\}.
	\end{eqnarray}
	Thus, Condition A (ii) of \citet{Bell2018} is satisfied by $s_{n,j}$.
	It is clear that Condition A (i) of \citet{Bell2018} is satisfied by the remainder term $C_{n}$.
	Hence, from Theorem 2.4 in \cite{Bell2018}, for any $1\leq j\leq p$, if
	$\log p=o(n^{1/5})$ and $\log n=o(p^{1/3\wedge\delta})$,
	we have
	\begin{eqnarray}
		\notag \sup_{0 \leq x\leq 2^{1/2}\log^{1/2}(np)}\left|\P \left\{n^{1/2}(\hat{\theta}_{n,j}-\theta_{j})/s_{n,j}>x\right\}-\{1-\Phi(x)\}\right|\rightarrow 0\,.
	\end{eqnarray} 
	
	%Let $G_j^{*}(t)=\P(|T_{ij}^{*}|\geq t)$ be the conditional distribution of  $T_{ij}^{*}$ given $X$, for $1\leq i\leq n$ and $1\leq j\leq p$. Based on the Equation (\ref{eq:tildetheta}), and using the classical wild bootstrap theorems, we have
	%\begin{align}\label{eq: eqP}
	%	\frac{G_j^{*}(t)}{\P(|n^{1/2}(\hat{\theta}_{j}-\theta_j)/s_n|\geq t)}=\frac{G_j^{*}(t)}{\P(|T_j-n^{1/2}(\theta_{0j}-\theta_j)/s_n|\geq t)}=1+o_p(1)\,,
	%\end{align}
	%uniformly in $j$ with $0\leq t\leq o(n^{1/4})$.
	
	Let ${\bar T}_j=n^{-1/2}\sum_{i=1}^nW_{i,j}/\{n^{-1}\sum_{i=1}^nW^2_{i,j}-(n^{-1}\sum_{i=1}^nW_{i,j})^2\}^{1/2}$. Based on Equation (13) of \cite{Liu2014}, for any sequence $d_n\rightarrow \infty$ and $d_n=o(p)$ as $n\rightarrow \infty$, with Condition \ref{c5},
	\[
	\sup_{0\leq t\leq \mathcal{G}_{\kappa}^{-1}(d_n/p)}\left|\frac{\sum_{j\in \H_0}\I\{|{\bar T}_j|\geq t\}}{p_0\mathcal{G}_{\kappa}(t)}-1\right|=o_p(1)\,,
	\]
	where $\mathcal{G}_{\kappa}(t)$ is some function such that $\mathcal{G}_{\kappa}(t)\geq \mathcal{G}(t)=2\{1-\Phi(t)\}$ for all $t\in \mR$, and
	$\mathcal{G}_{\kappa}(t)=\mathcal{G}(t)\{1+o(1)\}$ uniformly over $0\leq t\leq 2^{1/2}\log^{1/2}(p)$. Then, with enough large $n$, as long as $|\H|\rightarrow \infty$ and $|\H|> 2/\alpha$,  we have
	$$
	\alpha|\H|/p\geq 2/p=2\exp\{-(2^{1/2}\log^{1/2}p)^2/2\}\geq 2\{1-\Phi(2^{1/2}\log^{1/2}p)\}=\mathcal{G}(2^{1/2}\log^{1/2}p)\,.
	$$
	It follows that $\mathcal{G}^{-1}(\alpha|\H|/p)\leq 2^{1/2}\log^{1/2}p$, and consequently,
	\[
	\sup_{0\leq t\leq \mathcal{G}^{-1}(\alpha|\H|/p)}\bigg|\frac{\sum_{j\in \H_0}\I\{|{\bar T}_j|\geq t\}}{p_0\mathcal{G}(t)}-1\bigg|=o_p(1)\,.
	\]
	%then it holds that with  (\ref{eq: eqP}) as long as $\log p=o(n^{1/2})$, 
	%\[
	%\sup_{0\leq t\leq G^{-1}(\alpha|\H|/p)}\bigg|\frac{\sum_{j\in \H_0}\I\{|{\bar T}_j|\geq t\}}{p_0G_j^{*}(t)}-1\bigg|=o_p(1)\,,
	%\]
	Let $T^{\prime}_j=n^{1/2}(\hat{\theta}_{n,j}-\theta_j)/s_{n,j}$, we obtain $\max_{1\leq j\leq p}|T^{\prime}_j-{\bar T}_j|=o_p\{\log^{-1/2}(p)\}$ with some careful calculations. With similar procedure to Page 84 of \cite{Bell2018}, it holds that
	\begin{align}\label{eq:pT}
		\sup_{0\leq t\leq \mathcal{G}^{-1}(\alpha|\H|/p)}\bigg|\frac{\sum_{j\in \H_0}\I\{|{T}^{\prime}_j|\geq t\}}{p_0\mathcal{G}(t)}-1\bigg|=o_p(1)\,.
	\end{align}
	
	The B-H method with $P_1, \ldots, P_p$ is equivalent to the following procedure: reject $H_{0j}$, if only if $P_j\leq \hat{t}_0$,  where 
	\[
	\hat{t}_0=\sup\bigg\{0\leq t\leq 1: t\leq \frac{\alpha\max\{\sum_{j=1}^p\I(P_{j}\leq t)\}}{p}\bigg\}\,.
	\]
	Then we have $\hat{t}_0=\frac{\alpha\max\{\sum_{j=1}^p\I(P_{j}\leq \hat{t}_0),1\} }{p}$,  and $\alpha |\H|/p\geq \mathcal{G}(2^{1/2}\log^{1/2}p)$. Set $t=\mathcal{G}^{-1}(\alpha |\H|/p)$,  then $t\leq 2^{1/2}\log^{1/2}p$
	with probability tends to 1. Thus, we have
	\begin{eqnarray}
		\notag \mathcal{G}(t)=\frac{\alpha |\H|}{p} & \leq & \frac{\alpha \max \{\sum_{j=1}^p\I(|T_i|\geq 2^{1/2} \log^{1/2}p), 1\}}{p} \\
		\notag & \leq & \frac{\alpha \max \{\sum_{j=1}^p\I(|T_i|\geq t), 1\}}{p}\,, 
	\end{eqnarray}
	where the second inequality implied by (B.29) of \cite{Bell2018}. It implies that $\P(\hat{t}_0 \geq \alpha |\H|/p)\rightarrow 1$
	with $\hat{t}_0=\mathcal{G}(\hat{t})$, and together with (\ref{eq:pT}), we have
	\[
	\frac{\sum_{j\in \H_0}\I\{|{T}^{\prime}_j|\geq \hat{t}\}}{p_0\mathcal{G}(\hat{t})}= \frac{\sum_{j \in \H_0}\I(|T_j|\geq \hat{t})}{p_0\mathcal{G}(\hat{t})}\rightarrow 1\,,
	\]
	which  is equivalent to
	\[
	\frac{\sum_{j\in \H_0}\I(P_j\leq \hat{t}_0)}{p_0\hat{t}_0}\rightarrow 1\,.
	\]
	Finally,
	\[
	\mathrm {FDR}_M=\frac{\sum_{j\in \H_0}\I(P_j\leq \hat{t}_0)}{\max\{\sum_{j=1}^p\I(P_{j}\leq \hat{t}_0),1\}}=
	\frac{\sum_{j\in \H_0}\I(P_j\leq \hat{t}_0)}{p\hat{t}_0/\alpha}\rightarrow \frac{\alpha p_0}{p}\,
	\]
	as $n\to\infty$, which completes the proof of this theorem. 
\end{proof}

\section*{Appendix B: Proof of preliminary lemmas}\label{sec:SM_tech}
%\renewcommand\thesection{B}

%\subsection{Preliminary lemmas}\label{sec:appendix_lem}

\renewcommand\thelemma{B\arabic{lemma}}
%\setcounter{lemma}{0}

%\subsection{Proofs of preliminary lemmas}\label{sec:SM_lemmas}

In this section, we present proofs of preliminary lemmas in Section A1 of Appendix A.

\begin{proof}[Proof of Lemma \ref{lem:norm}]
	As the components of $U_{1}$ are independent and standardized, simple calculations yield $\mE(\|U_{1}\|^2) = p$ 
	and
	\begin{eqnarray}
		\notag \mE(\|\Gamma U_{1}\|^2) = \mE(U_{1}^{\top}\Gamma^{\top}\Gamma U_{1}) = \tr\{\Gamma^{\top}\Gamma\mE(U_{1}U_{1}^{\top})\} = \tr(\Omega)\,. 
	\end{eqnarray}
	Under Condition \ref{c1}, the components of $U_1=(U_{1,1},\ldots,U_{1,p})^{\top}$ are independent sub-exponential random variables such that $\max_{1\leq j \leq p}\|U_{1,j}\|_{\psi_{\alpha}}\leq c_0$.
	Applying the concentration inequality in the proof of Lemma S2.1 in \citep{Wang2015}, for every $t\geq0$,
	\begin{eqnarray}\label{eq:lemma_02_01}
		& \P\left(\left|\|U_1\|^2-p\right|\geq t\right) \leq C_1\exp\left\{-C_2 \left(p^{-1}t^2\right)^{\alpha/(4\alpha+4)}\right\}\,. &
	\end{eqnarray}
	and
	\begin{eqnarray}\label{eq:lemma_02_02}
		& \P\left\{\left|\|\Gamma U_1\|^2-\tr(\Omega)\right|\geq t\right\}\leq C_1\exp\left[-C_2 \left\{ \frac{t^2}{\tr(\Omega^2)}\right\}^{\alpha/(4\alpha+4)}\right]\,. &
	\end{eqnarray}
	%where $C_1$ and $C_2$ are positive constants.
	For any fixed $0<\epsilon<1$, let 
	\begin{eqnarray}
		\notag \mathcal{A}_1=\{p-\epsilon p^{(1+\delta)/2}\leq \|U_1\|^2\leq p+\epsilon p^{(1+\delta)/2}\}
	\end{eqnarray}
	and
	\begin{eqnarray}
		\notag \mathcal{A}_2=\{(1-\epsilon)\tr(\Omega)\leq \|\Gamma U_1\|^2\leq (1+\epsilon)\tr(\Omega)\}.
	\end{eqnarray}
	Taking $t=\epsilon p^{(1+\delta)/2}$ in \eqref{eq:lemma_02_01} and $t=\epsilon\tr(\Omega)$ in \eqref{eq:lemma_02_02}, we have
	\begin{eqnarray}
		\notag \P(\mathcal{A}_1) \geq 1-C_1\exp\left\{-C_2 (\epsilon^2p^{\delta})^{\alpha/(4\alpha+4)}\right\}\,
	\end{eqnarray}
	and
	\begin{eqnarray}
		\notag \P(\mathcal{A}_2)\geq 1-C_1\exp\left[-C_2  \left\{\frac{\epsilon^2\tr^2(\Omega)}{\tr(\Omega^2)}\right\}^{\alpha/(4\alpha+4)}\right]\,.
	\end{eqnarray}
	Under Condition \ref{c3},
	\begin{eqnarray}
		\notag \tr(\Omega^2) = \sum_{j=1}^{p}\sum_{\ell=1}^{p}\omega_{j\ell}^2 \leq \bar{M}p\max_{1\leq \ell\leq p}\sum_{j=1}^{p}|\omega_{j\ell}|\leq \bar{M}pa_0(p)\,.
	\end{eqnarray}
	Since $\tr(\Omega)=p$ and $a_0(p)\asymp p^{1-\delta}$, we conclude that 
	\begin{eqnarray}
		\notag \frac{\tr^2(\Omega)}{\tr(\Omega^2)}\geq \frac{p^2}{\bar{M}pa_0(p)} \asymp p^{\delta}\,. 
	\end{eqnarray}
	Consequently, for some positive constants $c_1$ and $c_2$, we get that
	\begin{eqnarray}
		\notag \P(\mathcal{A}_1) \geq 1-c_1\exp\{-c_2p^{\delta\alpha/(4\alpha+4)}\}\,.
	\end{eqnarray}
	and
	\begin{eqnarray}
		\notag \P(\mathcal{A}_2) \geq 1-c_1\exp\{-c_2p^{\delta\alpha/(4\alpha+4)}\}\,
	\end{eqnarray}
	for sufficient large $p$.
	\iffalse
	It is observed that $\|\Gamma S(U_1)\|=\|\Gamma U_1\|/\|U_1\|$. Let 
	\begin{eqnarray}
		\notag \mathcal{A}_{3} = \{1-\epsilon \leq \|\Gamma S(U_1)\|^2\leq 1+\epsilon \}\,,
	\end{eqnarray}
	then for sufficient large $p$,
	\begin{eqnarray}
		\notag & & 1-\P(\mathcal{A}_3) \\
		\notag & \leq & \P\left\{\|\Gamma S(U_1)\|< 1-\epsilon \text{~or~} \|\Gamma S(U_1)\| > 1+\epsilon, \mathcal{A}_1\right\} + \P(\mathcal{A}_1^{c}) \\
		\notag & = & \P\left\{\|\Gamma U_1\|< (1-\epsilon)\|U_1\| \text{~or~} \|\Gamma U_1\| > (1+\epsilon)\|U_1\|, \mathcal{A}_1\right\} + \P(\mathcal{A}_1^{c}) \\
		\notag & \leq & \P\left(\|\Gamma U_1\|< (1-\epsilon)\left\{p+\epsilon p^{(1+\delta)/2}\right\} \text{~or~} \|\Gamma U_1\| > (1+\epsilon)\left\{p-\epsilon p^{(1-\delta)/2}\right\}\right) \\
		\notag & & + \P(\mathcal{A}_1^{c}) \\
		\notag & \leq & c_1\exp\{-c_2p^{\delta\alpha/(4\alpha+4)}\}\,.
	\end{eqnarray}
	\fi
	Thus, we finish the proof of this lemma.	
\end{proof}

%Lemma \ref{lem:norm} shows that $p^{-1}\|\Gamma U_1\|^2$ and $p^{-1}\|U_1\|^2$ %, and $\|\Gamma S(U_1)\|^2$ 
%are bounded in probability with large enough $p$.

\begin{proof}[Proof of Lemma \ref{lemma:SM_moments}]
	(i) As the components of $U_i=(U_{i,1},\ldots,U_{i,p})^{\top}$ are i.i.d.~standardized sub-exponential random variables, simple algebra yields
	\begin{eqnarray}
		\notag \mE(\|U_i\|^4) & = & \mE\left\{\left(\sum_{j=1}^{p}U_{i,j}^2\right)^2\right\} \\
		\notag & = & \sum_{j=1}^{p}\mE(U_{i,j}^4) + \sum_{1\leq j_1\neq j_2\leq p}\mE(U_{i,j_1}^2)\mE(U_{i,j_2}^2) \\
		\notag & = & p\mE(U_{i,j}^4) + p(p-1)\,
	\end{eqnarray}	
	and
	\begin{eqnarray}
		\notag \mE(\|U_i\|^6) & = & \sum_{j=1}^{p}\mE(U_{i,j}^6) + 3\sum_{1\leq j_1\neq j_2\leq p}\mE(U_{i,j_1}^4)\mE(U_{i,j_2}^2) \\
		\notag & & + \sum_{1\leq j_1\neq j_2\neq j_3\leq p}\mE(U_{i,j_1}^2)\mE(U_{i,j_2}^2)\mE(U_{i,j_3}^2)\\
		\notag & = & p\mE(U_{i,j}^6) + 3p(p-1)\mE(U_{i,j}^4) + p(p-1)(p-2)\,.
	\end{eqnarray}
	
	In addition,
	\begin{eqnarray}
		\notag \mE(\|U_i\|^8) %& = & \mE\left\{\left(\sum_{j=1}^{p}U_{i,j}^2\right)^4\right\} \\
		\notag & = & \sum_{j=1}^{p}\mE(U_{i,j}^8) + 4\sum_{1\leq j_1\neq j_2\leq p}\mE(U_{i,j_1}^6)\mE(U_{i,j_2}^2) \\
		\notag & & + 3\sum_{1\leq j_1\neq j_2\leq p}\mE(U_{i,j_1}^4)\mE(U_{i,j_2}^4) \\
		\notag & & + 6\sum_{1\leq j_1\neq j_2\neq j_3\leq p}\mE(U_{i,j_1}^4)\mE(U_{i,j_2}^2)\mE(U_{i,j_3}^2)\\
		\notag & & + \sum_{1\leq j_1\neq j_2\neq j_3\neq j_4\leq p}\mE(U_{i,j_1}^2)\mE(U_{i,j_2}^2)\mE(U_{i,j_3}^2)\mE(U_{i,j_4}^2) \\
		\notag & = & p\mE(U_{i,j}^8) + 4p(p-1)\mE(U_{i,j_1}^6) + 3p(p-1)\{\mE(U_{i,j_1}^4)\}^2 \\
		\notag & & + 3p(p-1)\mE(U_{i,j}^4) + p(p-1)(p-2)(p-3)\,.
	\end{eqnarray}
	The result of $\mE(\|U_i\|^{2k}) = p^k +  O(p^{k-1})$ for any positive integer $k$ can be checked by
	\begin{eqnarray}
		\notag \mE(\|U_i\|^{2k}) & = & \sum_{1\leq j_1\neq \cdots \neq j_k\leq p}\mE(U_{i,j_1}^2)\times \cdots\times \mE(U_{i,j_k}^2)\{1+O(p^{-1})\} \\
		\notag & = & p^k +  O(p^{k-1})\,.
	\end{eqnarray}
	Moreover, by the fact that $\{1+u-(u-1)^2\}/2\leq u^{1/2}\leq (1+u)/2$ for all $u\geq 0$, we can get that $\mE(\|U\|^{k})=p^{k/2}+O(p^{k/2-1})$ for all positive integer $k$.
	
	(ii) Write $\Lambda_{j\ell}=\sum_{j_1=1}^{p}\Gamma_{j_1j}\Gamma_{j_1\ell}$ as the $(j,\ell)$th element of $\Gamma^{\top}\Gamma$, then
	\begin{eqnarray}
		\notag \mE(\|\Gamma U_i\|^4) %& = & \mE\{(U_{i}^{\top}\Gamma^{\top}\Gamma U_{i})^2\} \\
		\notag & = & \mE\left\{\left(\sum_{j=1}^{p}\sum_{\ell=1}^{p}\Lambda_{j\ell}U_{i,j}U_{i,\ell}\right)^2\right\} \\
		\notag & = & \sum_{j=1}^{p}\Lambda_{jj}^2\mE(U_{i,j}^4) + 2\sum_{1\leq j_1\neq j_2\leq p}\Lambda_{j_1j_2}^2\mE(U_{i,j_1}^2)\mE(U_{i,j_2}^2) \\
		\notag & & + \sum_{1\leq j_1\neq j_2\leq p}\Lambda_{j_1j_1}\Lambda_{j_2j_2}\mE(U_{i,j_1}^2)\mE(U_{i,j_2}^2) \\
		\notag & = & \mE(U_{i,j}^4)\sum_{j=1}^{p}\Lambda_{jj}^2 + 2\sum_{1\leq j_1\neq j_2\leq p}\Lambda_{j_1j_2}^2 + \sum_{1\leq j_1\neq j_2\leq p}\Lambda_{j_1j_1}\Lambda_{j_2j_2} \\
		\notag & = & \left(\sum_{j=1}^{p}\Lambda_{jj}\right)^2 +  \{\mE(U_{i,j}^4)-1\}\sum_{j=1}^{p}\Lambda_{jj}^2 + 2\sum_{1\leq j_1\neq j_2\leq p}\Lambda_{j_1j_2}^2 \\
		%\notag & = & \{\tr(\Omega)\}^2 +  \{\mE(U_{i,j}^4)-1\}\sum_{j=1}^{p}\Lambda_{jj}^2 + 2\sum_{1\leq j_1\neq j_2\leq p}\Lambda_{j_1j_2}^2 \\
		\notag & = & \{\tr(\Omega)\}^2+O\{\tr(\Omega^2)\} %\\
		%\notag & = & p^2+O(p^{2-\delta})\,
	\end{eqnarray}
	as $\sum_{j=1}^{p}\Lambda_{jj}^2 + \sum_{1\leq j_1\neq j_2\leq p}\Lambda_{j_1j_2}^2 = \sum_{j=1}^{p}\sum_{\ell=1}^{p}\Lambda_{j\ell}^2=\tr(\Omega^2)$
	and $\tr(\Omega^2)\lesssim p^{2-\delta}$
	based on Condition \ref{c3}.
	Similarly, we can show that
	\begin{eqnarray}
		\notag \mE(\|\Gamma U_i\|^6) %& = & \mE\left\{\left(\sum_{j=1}^{p}\sum_{\ell=1}^{p}\Lambda_{j\ell}U_{i,j}U_{i,\ell}\right)^3\right\} \\
		\notag & = & \sum_{1\leq j_1\neq j_2\neq j_3\leq p}(\Lambda_{j_1j_1}\Lambda_{j_2j_2}\Lambda_{j_3j_3}+\Lambda_{j_1j_2}^2\Lambda_{j_3j_3} \\
		\notag & & ~~~~~~~~~~~~~~~~~~~~~~~~~+\Lambda_{j_1j_2}\Lambda_{j_1j_3}\Lambda_{j_2j_3})\mE(U_{i,j_1}^2)\mE(U_{i,j_2}^2)\mE(U_{i,j_3}^2) \{1+O(p^{-1})\}\\
		\notag & = & p^3 + O(p^{3-\delta})\,
	\end{eqnarray}
	and $\mE(\|\Gamma U_i\|^{12})=p^6 + O(p^{6-\delta})$.
	\iffalse
	and
	\begin{eqnarray}
		\notag & & \mE(\|\Gamma U_i\|^8) \\
		%\notag & = & \mE\left\{\left(\sum_{j=1}^{p}\sum_{\ell=1}^{p}\Lambda_{j\ell}U_{i,j}U_{i,\ell}\right)^4\right\} \\
		\notag & \asymp & \sum_{1\leq j_1\neq j_2\neq j_3\neq j_4\leq p}\Lambda_{j_1j_1}\Lambda_{j_2j_2}\Lambda_{j_3j_3}\Lambda_{j_4j_4}\mE(U_{i,j_1}^2)\mE(U_{i,j_2}^2)\mE(U_{i,j_3}^2)\mE(U_{i,j_4}^2) \\
		\notag & \asymp & \{\tr(\Omega)\}^4\,.
	\end{eqnarray}
	\fi
	
	Similar to the proof of part (i), the result $\mE(\|\Gamma U_{i}\|) = p^{1/2} + O(p^{1/2-\delta})$ and $\mE(\|\Gamma U_{i}\|^{3}) = p^{3/2} + O(p^{3/2-\delta})$ are directly consequences of $\mE(\|\Gamma U_{i}\|^2) = p$, $\mE(\|\Gamma U_{i}\|^4) = p^2+O(p^{2-\delta})$, $\mE(\|\Gamma U_i\|^6)=p^3 + O(p^{3-\delta})$, $\mE(\|\Gamma U_i\|^{12})=p^6 + O(p^{6-\delta})$ and $\{1+u-(u-1)^2\}/2\leq u^{1/2}\leq (1+u)/2$ for all $u\geq 0$.
	
	(iii) Now we consider $\mE\{\|\Gamma S(U_i)\|^2\}$.
	For $i=1,\ldots,n$, let $$\mathcal{A}_{1i} =\{p-\epsilon p^{(1+\delta)/2}\leq \|U_i\|^2\leq p+\epsilon p^{(1+\delta)/2}\}$$ for a fixed $0<\epsilon<1$. According to Lemma \ref{lem:norm} and the fact that $\|\Gamma U_{i}\|^2\leq \tr(\Omega)\|U_{i}\|^2$,
	\begin{eqnarray}
		\notag \mE\{\|\Gamma S(U_i)\|^2\} & = & \mE(\|\Gamma U_{i}\|^2 \|U_{i}\|^{-2})  \\
		\notag & = & p^{-1}\mE\{\|\Gamma U_i\|^2\} + \mE\left\{\|\Gamma U_{i}\|^2\left(\|U_{i}\|^{-2}-p^{-1}\right)\right\} \\
		\notag & = & 1 + \mE\left\{\|\Gamma U_{i}\|^2\left(\|U_{i}\|^{-2}-p^{-1}\right)\right\}\,,
	\end{eqnarray}
	where
	\begin{eqnarray}
		\notag & & \mE\left\{\|\Gamma U_{i}\|^2\left(\|U_{i}\|^{-2}-p^{-1}\right)\right\} \\
		\notag & \leq & p^{-1}\mE\left(\|\Gamma U_{i}\|^2\|U_{i}\|^{-2}\left|\|U_{i}\|^2-p\right|\right) \\
		\notag & = & p^{-1}\mE\left\{\|\Gamma U_{i}\|^2\|U_{i}\|^{-2}\left|\|U_{i}\|^2-p\right|\mI(\mathcal{A}_{1i})\right\}\\
		\notag & & + p^{-1}\mE\left\{\|\Gamma U_{i}\|^2\|U_{i}\|^{-2}\left|\|U_{i}\|^2-p\right|\mI(\mathcal{A}_{1i}^{c})\right\} \\
		\notag & \leq & p^{-1}\{p-\epsilon p^{(1+\delta)/2}\}^{-1}\mE\left(\|\Gamma U_{i}\|^2\left|\|U_{i}\|^2-p\right|\right) \\
		\notag & & + p^{-1}\tr(\Omega)\mE\left\{\left|\|U_{i}\|^2-p\right|\mI(\mathcal{A}_{1i}^{c})\right\} \\
		\notag & \leq & p^{-1}\{p-\epsilon p^{(1+\delta)/2}\}^{-1}\left\{\mE(\|\Gamma U_{i}\|^4)\right\}^{1/2}\left\{\mE(\left|\|U_{i}\|^2-p\right|^2)\right\}^{1/2} \\
		\notag & & + \left\{\mE(\left|\|U_{i}\|^2-p\right|^2)\right\}^{1/2}\left\{\P(\mathcal{A}_{1i}^{c})\right\}^{1/2} \\
		\notag & \leq & p^{-1}(p-\epsilon p^{1-\delta})^{-1}\{p^2+O(p^{2-\delta})\}^{1/2}\times O(p^{1/2}) \\
		\notag & & + O(p^{1/2}) \times c_1^{1/2}\exp\{-c_2p^{\delta\alpha/(4\alpha+4)}/2\} \\
		\notag & = & O(p^{-1/2})\,.
	\end{eqnarray}
	It follows that $\mE\{\|\Gamma S(U_i)\|^2\} = 1 + O(p^{-1/2})$.
	
	Similarly, the last result follows from
	\begin{eqnarray}
		\notag \mE\{\|\Gamma S(U_i)\|^4\} & = & p^{-2}\mE\{\|\Gamma U_i\|^4\} + \mE\left\{\|\Gamma U_{i}\|^4\left(\|U_{i}\|^{-4}-p^{-2}\right)\right\} \\
		\notag & = & 1 + O(p^{-\delta}) + \mE\left\{\|\Gamma U_{i}\|^4\left(\|U_{i}\|^{-4}-p^{-2}\right)\right\}\,,
	\end{eqnarray}
	where
	\begin{eqnarray}
		\notag & & \mE\left\{\|\Gamma U_{i}\|^4\left(\|U_{i}\|^{-4}-p^{-2}\right)\right\} \\
		\notag & \leq & p^{-2}\mE\left(\|\Gamma U_{i}\|^4\|U_{i}\|^{-4}\left|\|U_{i}\|^4-p^2\right|\right) \\
		\notag & = & p^{-2}\mE\left\{\|\Gamma U_{i}\|^4\|U_{i}\|^{-4}\left|\|U_{i}\|^4-p^2\right|\mI(\mathcal{A}_{1i})\right\}\\
		\notag & & + p^{-2}\mE\left\{\|\Gamma U_{i}\|^4\|U_{i}\|^{-4}\left|\|U_{i}\|^4-p^2\right|\mI(\mathcal{A}_{1i}^{c})\right\} \\
		\notag & \leq & p^{-2}(p-\epsilon p^{1-\delta})^{-2}\mE\left(\|\Gamma U_{i}\|^4\left|\|U_{i}\|^4-p^2\right|\right) \\
		\notag & & + p^{-2}\{\tr(\Omega)\}^2\mE\left\{\left|\|U_{i}\|^4-p^2\right|\mI(\mathcal{A}_{1i}^{c})\right\} \\
		\notag & \leq & p^{-2}(p-\epsilon p^{1-\delta})^{-2}\left\{\mE(\|\Gamma U_{i}\|^6)\right\}^{2/3}\left\{\mE(\left|\|U_{i}\|^4-p^2\right|^3)\right\}^{1/3} \\
		\notag & & + \left\{\mE(\left|\|U_{i}\|^4-p^2\right|^2)\right\}^{1/2}\left\{\P(\mathcal{A}_{1i}^{c})\right\}^{1/2} \\
		\notag & \leq & p^{-2}(p-\epsilon p^{1-\delta})^{-2}\times O(p^{2})\times O(p^{3/2}) \\
		\notag & & + O(p^{3/2}) \times c_1^{1/2}\exp\{-c_2p^{\delta\alpha/(4\alpha+4)}/2\} \\
		\notag & = & O(p^{-1/3})\,.
	\end{eqnarray}
	
	(iv) as $\nu_{i}$ and $S(U_{i})$ are independent,
	\begin{eqnarray}
		\notag & & \mE(\nu_{i}^{-1})\mE\{\|\Gamma S(U_{i})\|^{-1}\} \\
		\notag & = & \mE(\nu_{i}^{-1}\|\Gamma U_{i}\|^{-1}\|U_{i}\|) \\
		\notag & = & \mE(R_{i}^{-1}\|U_{i}\|) \\
		\notag & = & \mE\{R_{i}^{-1}\|U_{i}\|\mI(\mathcal{A}_{1i})\} + \mE\{R_{i}^{-1}\|U_{i}\|\mI(\mathcal{A}_{1i}^{c})\} \\
		\notag & \leq & \{p+\epsilon p^{(1+\delta)/2}\}^{1/2}\mE\{R_{i}^{-1}\mI(\mathcal{A}_{1i})\} + \{\mE(R_i^{-4})\}^{1/4}\{\mE\|U_{i}\|^{4}\}^{1/4} \{\P(\mathcal{A}_{1i}^{c})\}^{1/2} \\
		\notag & \lesssim & \{p+\epsilon p^{(1+\delta)/2}\}^{1/2}\mE(R_i^{-1}) + \zeta_{4}^{1/4}\times p^{1/2} \times c_1^{1/2}\exp\{-c_2p^{\delta\alpha/(4\alpha+4)}/2\} \\
		\notag & \lesssim & \zeta_{1}p^{1/2}\,,
	\end{eqnarray}
	and
	\begin{eqnarray}
		\notag & & \mE(\nu_{i}^{-2})\mE\{\|\Gamma S(U_{i})\|^{-2}\} \\
		%\notag & = & \mE(\nu_{i}^{-2}\|\Gamma U_{i}\|^{-2}\|U_{i}\|^{2}) \\
		%\notag & = & \mE(R_{i}^{-2}\|U_{i}\|^{2}) \\
		\notag & = & \mE\{R_{i}^{-2}\|U_{i}\|^2\mI(\mathcal{A}_{1i})\} + \mE\{R_{i}^{-2}\|U_{i}\|^2\mI(\mathcal{A}_{1i}^{c})\} \\
		\notag & \leq & \{p+\epsilon p^{(1+\delta)/2}\}\mE\{R_{i}^{-2}\mI(\mathcal{A}_{1i})\} + \{\mE(R_i^{-4})\}^{1/2}\{\mE\|U_{i}\|^{6}\}^{1/3} \{\P(\mathcal{A}_{1i}^{c})\}^{1/6} \\
		\notag & \lesssim & \{p+\epsilon p^{(1+\delta)/2}\}\mE(R_i^{-2}) + \zeta_{4}^{1/2}\times p \times c_1^{1/6}\exp\{-c_2p^{\delta\alpha/(4\alpha+4)}/6\} \\
		\notag & \lesssim & \zeta_2p\,,
	\end{eqnarray}
	%where the second last inequality uses H\"{o}lder's inequality.
	In addition, we also have
	\begin{eqnarray}
		\notag & & \mE(\nu_{i}^{-3})\mE\{\|\Gamma S(U_{i})\|^{-3}\} \\
		%\notag & = & \mE(\nu_{i}^{-2}\|\Gamma U_{i}\|^{-2}\|U_{i}\|^{2}) \\
		%\notag & = & \mE(R_{i}^{-3}\|U_{i}\|^{3}) \\
		\notag & = & \mE\{R_{i}^{-3}\|U_{i}\|^3\mI(\mathcal{A}_{1i})\} + \mE\{R_{i}^{-3}\|U_{i}\|^3\mI(\mathcal{A}_{1i}^{c})\} \\
		\notag & \leq & \{p+\epsilon p^{(1+\delta)/2}\}^{3/2}\mE\{R_{i}^{-3}\mI(\mathcal{A}_{1i})\} + \{\mE(R_i^{-4})\}^{3/4}\{\mE\|U_{i}\|^{18}\}^{1/6} \{\P(\mathcal{A}_{1i}^{c})\}^{1/12} \\
		\notag & \lesssim & \{p+\epsilon p^{(1+\delta)/2}\}^{3/2}\mE(R_i^{-3}) + \zeta_{4}^{3/4}\times p^{3/2} \times c_1^{1/12}\exp\{-c_2p^{\delta\alpha/(4\alpha+4)}/12\} \\
		\notag & \lesssim & \zeta_3p^{3/2}\,.
	\end{eqnarray}
	
	By Cauchy-Schwarz inequality and Jensen's inequality, we can show that
	\begin{eqnarray}
		\notag & [\mE\{\|\Gamma S(U_i)\|^{-1}\}]^{-1} \leq \mE\{\|\Gamma S(U_i)\|\} \leq [\mE\{\|\Gamma S(U_i)\|^2\}]^{1/2} = 1 + O(p^{-1/2})\,, & \\
		\notag & [\mE\{\|\Gamma S(U_i)\|^{-2}\}]^{-1} \leq \mE\{\|\Gamma S(U_i)\|^2\} = 1 + O(p^{-1/2})\,, &
	\end{eqnarray}
	and
	\begin{eqnarray}
		\notag [\mE\{\|\Gamma S(U_i)\|^{-3}\}]^{-1} \leq \mE\{\|\Gamma S(U_i)\|^3\} \leq [\mE\{\|\Gamma S(U_i)\|^4\}]^{3/4} = 1 + O(p^{-1/3})\,.
	\end{eqnarray}
	Then, the results of this part follows immediately.
	We finish the proof of this lemma.
\end{proof}

\iffalse
By the inequality of arithmetic and geometric means, 
\begin{eqnarray}
	\notag p\|U_{i}\|^{-2} = \frac{p}{U_{i,1}^2+\cdots+U_{i,p}^2} \leq \frac{1}{\left(\prod_{j=1}^{p}U_{i,j}^2\right)^{1/p}} = \prod_{j=1}^{p}|U_{i,j}|^{-2/p}\,,
\end{eqnarray}
from which we conclude that
\begin{eqnarray}
	\notag \mE(p\|U_{i}\|^{-2}) \leq \prod_{j=1}^{p}\mE(|U_{i,j}|^{-2/p}) \leq \prod_{j=1}^{p} \{\mE(U_{i,j}^{-2})\}^{1/p} = \mE(U_{i,j}^{-2}).
\end{eqnarray}
In addition,
\begin{eqnarray}
	\notag \mE(p^{k/2}\|U_{i}\|^{-k}) \leq \prod_{j=1}^{p}\mE(|U_{i,j}|^{-k/p}) \leq \prod_{j=1}^{p} \{\mE(|U_{i,j}|^{-k})\}^{1/p} = \mE(|U_{i,j}|^{-k})
\end{eqnarray}
for any positive integer $k\geq 1$.

By Cauchy-Schwarz inequality,
\begin{eqnarray}
	\notag \mE(\|U_{i}\|^{-k})\mE(\|U_{i}\|^{k})\geq 1\,,
\end{eqnarray}
from which we know 
\begin{eqnarray}
	\notag \mE(\|U_{i}\|^{-k}) \geq \{\mE(\|U_{i}\|^{k})\}^{-1} \gtrsim p^{-k/2}\,.
\end{eqnarray}
\fi

\begin{proof}[Proof of Lemma \ref{lemma:SM_Q}]
	(i) For $i=1,\ldots,n$, let $\mathcal{A}_{2i}=\{(1-\epsilon)\tr(\Omega)\leq \|\Gamma U_i\|^2\leq (1+\epsilon)\tr(\Omega)\}$ for a fixed $0<\epsilon<1$.
	Recall that $\Gamma_j$ is the $j$th row of $\Gamma$ and $W_{i,j}=\Gamma_{j} U_{i}/\|\Gamma U_{i}\|$, then
	\begin{eqnarray}
		\notag Q_{j\ell} & = & n^{-1} \sum_{i=1}^nR_i^{-1} W_{i,j}W_{i, \ell} = n^{-1} \sum_{i=1}^n\nu_{i}^{-1}(\Gamma_j U_i)(\Gamma_{\ell} U_i)\|\Gamma U_{i}\|^{-3} \\
		\notag & = & n^{-1}p^{-3/2}\sum_{i=1}^n\nu_{i}^{-1}(\Gamma_j U_i)(\Gamma_{\ell} U_i) \\
		\notag & & + n^{-1} \sum_{i=1}^n\nu_{i}^{-1}(\Gamma_j U_i)(\Gamma_{\ell} U_i)\left(\|\Gamma U_{i}\|^{-3}-p^{-2/3}\right)\,,
	\end{eqnarray}
	where the last term satisfies
	\begin{eqnarray}
		\notag & & \left|\mE\left\{n^{-1} \sum_{i=1}^n\nu_{i}^{-1}(\Gamma_j U_i)(\Gamma_{\ell} U_i)\left(\|\Gamma U_{i}\|^{-3}-p^{-2/3}\right)\right\}\right| \\
		\notag & \leq & p^{-3/2}\mE\left\{\nu_{i}^{-1}\left|(\Gamma_j U_i)(\Gamma_{\ell} U_i)\right|\|\Gamma U_{i}\|^{-3}\left|\|\Gamma U_{i}\|^{3}-p^{3/2}\right|\right\} \\
		\notag & = & p^{-3/2}\mE\left\{R_{i}^{-1}\left|(\Gamma_j U_i)(\Gamma_{\ell} U_i)\right|\|\Gamma U_{i}\|^{-2}\left|\|\Gamma U_{i}\|^{3}-p^{3/2}\right|\right\} \\
		\notag & = & p^{-3/2}\mE\left\{R_{i}^{-1}\left|(\Gamma_j U_i)(\Gamma_{\ell} U_i)\right|\|\Gamma U_{i}\|^{-2}\left|\|\Gamma U_{i}\|^{3}-p^{3/2}\right|\mI(\mathcal{A}_{2i})\right\} \\
		\notag & & + p^{-3/2}\mE\left\{R_{i}^{-1}\left|(\Gamma_j U_i)(\Gamma_{\ell} U_i)\right|\|\Gamma U_{i}\|^{-2}\left|\|\Gamma U_{i}\|^{3}-p^{3/2}\right|\mI(\mathcal{A}_{2i}^{c})\right\} \\
		\notag & \leq & (1-\epsilon)^{-1}p^{-5/2}\mE\left\{R_{i}^{-1}\left|(\Gamma_j U_i)(\Gamma_{\ell} U_i)\right|\left|\|\Gamma U_{i}\|^{3}-p^{3/2}\right|\mI(\mathcal{A}_{2i})\right\} \\
		\notag & & + p^{-3/2}\mE\left\{R_{i}^{-1}\left|\|\Gamma U_{i}\|^{3}-p^{3/2}\right|\mI(\mathcal{A}_{2i}^{c})\right\} \\
		\notag & \lesssim & p^{-5/2}\{\mE(R_{i}^{-4})\}^{1/4}\left[\mE\left\{\left|(\Gamma_j U_i)(\Gamma_{\ell} U_i)\right|^4\right\}\right]^{1/4}\left\{\mE\left(\left|\|\Gamma U_{i}\|^{3}-p^{3/2}\right|^{2}\right)\right\}^{1/2} \\
		\notag & & + p^{-3/2}\{\mE(R_{i}^{-4})\}^{1/4}\left\{\mE\left(\left|\|\Gamma U_{i}\|^{3}-p^{3/2}\right|^{2}\right)\right\}^{1/2}\{\P(\mathcal{A}_{2i}^{c})\}^{1/4} \\
		\notag & \lesssim & \zeta_{1}p^{-1-\delta/2}\,.
	\end{eqnarray}
	It follows that
	\begin{eqnarray}
		\notag Q_{j\ell} = n^{-1}p^{-3/2}\sum_{i=1}^n\nu_{i}^{-1}(\Gamma_j U_i)(\Gamma_{\ell} U_i) + O_{p}(\zeta_{1}p^{-1-\delta/2})\,.
	\end{eqnarray}
	
	For $i=1,\ldots,n$, let $\mathcal{A}_{1i} = \{p-\epsilon p^{(1+\delta)/2}\leq \|U_{i}\|^2\leq p+\epsilon p^{(1+\delta)/2}\}$ for a fixed $0<\epsilon<1$. According to Lemma \ref{lem:norm},
	\begin{eqnarray}
		\notag & & \mE\left[\left\{\Gamma_j S(U_i)S(U_i)^{\top}\Gamma_{\ell}^{\top}\right\}^2\right] \\
		\notag & = & \mE\left\{\|U_{i}\|^{-4}(\Gamma_j U_i U_i^{\top}\Gamma_{\ell}^{\top})^2\right\} \\
		\notag & = & \mE\left\{\|U_{i}\|^{-4}(\Gamma_j U_i 	U_i^{\top}\Gamma_{\ell}^{\top})^2\mI(\mathcal{A}_{1i})\right\} +  \mE\left\{\|U_{i}\|^{-4}(\Gamma_j U_i U_i^{\top}\Gamma_{\ell}^{\top})^2\mI(\mathcal{A}_{1i}^{c})\right\} \\
		\notag & \lesssim & \{p-\epsilon p^{(1+\delta)/2}\}^{-2}\mE\left\{(\Gamma_j U_i U_i^{\top}\Gamma_{\ell}^{\top})^2\right\} + p^2 \P(\mathcal{A}_{1i}^{c}) \\
		\notag & \lesssim & \{p-\epsilon p^{(1+\delta)/2}\}^{-2} + p^2\times c_1\exp\{-c_2p^{\delta\alpha/(4\alpha+4)}\} \\
		\notag & \lesssim & p^{-2}\,.
	\end{eqnarray}	
	Then, we can show that
	\begin{eqnarray}
		\notag & & n^{-1}p^{-3/2}\sum_{i=1}^n\nu_{i}^{-1}(\Gamma_j U_i)(\Gamma_{\ell} U_i) \\
		\notag & = & n^{-1} p^{-1/2}\sum_{i=1}^n \nu_i^{-1}\Gamma_j S(U_i)S(U_i)^{\top}\Gamma_{\ell}^{\top}+O_p(\zeta_{1}p^{-7/6})\,,
	\end{eqnarray}
	where the last equality is indicated by
	\begin{eqnarray}
		\notag & & \mE|p^{-3/2}\nu_i^{-1}\Gamma_j S(U_i)S(U_i)^{\top}\Gamma_{\ell}^{\top}(\|U_i\|^{2}-p)| \\
		\notag & \lesssim &  p^{-3/2}\{\mE(\nu_{i}^{-3})\}^{1/3}\left(\mE\left[\left\{\Gamma_j S(U_i)S(U_i)^{\top}\Gamma_{\ell}^{\top}\right\}^2\right]\right)^{1/2}\left[\mE\left\{(\|U_i\|^{2}-p)^6\right\}\right]^{1/6} \\
		\notag & \lesssim & \zeta_{1}p^{-7/6}\,.
	\end{eqnarray}
	
	Thus, we obtain that
	\begin{eqnarray}
		\notag Q_{j\ell} = n^{-1} p^{-1/2}\sum_{i=1}^n \nu_i^{-1}\Gamma_j S(U_i)S(U_i)^{\top}\Gamma_{\ell}^{\top} + O_{p}(\zeta_{1}p^{-7/6}+\zeta_{1}p^{-1-\delta/2})\,.
	\end{eqnarray}
	
	As $\nu_{i}$ and $S(U_{i})$ are independent with each other, we have
	\begin{eqnarray}
		\notag & & \mE\left\{n^{-1} p^{-1/2}\sum_{i=1}^n \nu_i^{-1}\Gamma_j S(U_i)S(U_i)^{\top}\Gamma_{\ell}^{\top}\right\} \\
		\notag & = & p^{-1/2}\mE(\nu_i^{-1})\mE\left\{\Gamma_j S(U_i)S(U_i)^{\top}\Gamma_{\ell}^{\top}\right\}\,,
	\end{eqnarray}
	where $\mE(\nu_i^{-1}) \lesssim p^{1/2}\zeta_{1}$
	from Lemma \ref{lemma:SM_moments}.
	
	According to Lemma \ref{lem:norm} and regarding that $\Gamma_{j}\Gamma_{\ell}^{\top}=\omega_{j\ell}$,
	\begin{eqnarray}
		\notag & & \mE\left\{\Gamma_j S(U_i)S(U_i)^{\top}\Gamma_{\ell}^{\top}\right\} \\
		\notag & = & \mE\left(\Gamma_j U_iU_i^{\top}\Gamma_{\ell}^{\top}\|U_{i}\|^{-2}\right) \\
		\notag & = & p^{-1}\mE\left(\Gamma_j U_iU_i^{\top}\Gamma_{\ell}^{\top}\right) + \mE\left\{\Gamma_j U_iU_i^{\top}\Gamma_{\ell}^{\top}\left(\|U_i\|^{-2}-p^{-1}\right)\right\}\\
		\notag & = & p^{-1}\omega_{j\ell}+\mE\left\{\Gamma_j U_iU_i^{\top}\Gamma_{\ell}^{\top}\left(\|U_i\|^{-2}-p^{-1}\right)\right\} \\
		\notag & \leq & p^{-1}|\omega_{j\ell}| + \mE\left(\left|\Gamma_j U_iU_i^{\top}\Gamma_{\ell}^{\top}\right|\left|\|U_i\|^{-2}-p^{-1}\right|\right) \\
		\notag & = & p^{-1}|\omega_{j\ell}| + p^{-1}\mE\left(\left|\Gamma_j U_iU_i^{\top}\Gamma_{\ell}^{\top}\right|\|U_{i}\|^{-2}\left|\|U_{i}\|^2-p\right|\right) \\
		\notag & = & p^{-1}|\omega_{j\ell}| + p^{-1}\mE\left\{\left|\Gamma_j U_iU_i^{\top}\Gamma_{\ell}^{\top}\right|\|U_{i}\|^{-2}\left|\|U_{i}\|^2-p\right|\mI(\mathcal{A}_{1i})\right\} \\
		\notag & & + p^{-1}\mE\left\{\left|\Gamma_j U_iU_i^{\top}\Gamma_{\ell}^{\top}\right|\|U_{i}\|^{-2}\left|\|U_{i}\|^2-p\right|\mI(\mathcal{A}_{1i}^{c})\right\} \\
		\notag & \lesssim & p^{-1}|\omega_{j\ell}| + \{p^2-\epsilon p^{(3+\delta)/2}\}^{-1}\mE\left(\left|\Gamma_j U_iU_i^{\top}\Gamma_{\ell}^{\top}\right|\left|\|U_{i}\|^2-p\right|\right) \\
		\notag & & + \mE\left\{\left|\|U_{i}\|^2-p\right|\mI(\mathcal{A}_{1i}^{c})\right\} \\
		\notag & \leq & p^{-1}|\omega_{j\ell}| + \{p^2-\epsilon p^{(3+\delta)/2}\}^{-1}\left[\mE\left\{\left(\Gamma_j U_iU_i^{\top}\Gamma_{\ell}^{\top}\right)^2\right\}\right]^{1/2}\left[\mE\left\{\left(\|U_{i}\|^2-p\right)^2\right\}\right]^{1/2} \\
		\notag & & + \left[\mE\left\{\left(\|U_{i}\|^2-p\right)^2\right\}\right]^{1/2}\left\{\P(\mathcal{A}_{1i}^{c})\right\}^{1/2} \\
		\notag & \leq & p^{-1}|\omega_{j\ell}| + O(p^{-3/2}) + O(p^{1/2})\times c_1^{1/2}\exp\{-c_2p^{\delta\alpha/(4\alpha+4)}/2\} \\
		\notag & \lesssim & p^{-1}|\omega_{j\ell}| + O(p^{-3/2})\,,
	\end{eqnarray}
	where the second last inequality is due to
	\begin{eqnarray}
		\notag \mE\left\{\left(\|U_{i}\|^2-p\right)^2\right\} & = & \mE(\|U_{i}\|^4-2p\|U_{i}\|^2+p^2) \\
		\notag & = & p\mE(U_{i,j}^4) + p(p-1) - 2p^2 + p^2 \\
		\notag & = & O(p).
	\end{eqnarray}
	Thus, it follows that
	\begin{eqnarray}
		\notag \mE\left\{n^{-1} p^{-1/2}\sum_{i=1}^n \nu_i^{-1}\Gamma_j S(U_i)S(U_i)^{\top}\Gamma_{\ell}^{\top}\right\} \lesssim \zeta_{1}p^{-1}|\omega_{jj}| + O(\zeta_{1}p^{-3/2})\,.
	\end{eqnarray}
	
	Furthermore, as $\mE(\nu_{i}^{-2}) \lesssim p\zeta_{2}$,
	we can conclude that
	\begin{eqnarray}
		\notag & & \mathrm{Var}\left\{n^{-1} p^{-1/2}\sum_{i=1}^n \nu_i^{-1}\Gamma_j S(U_i)S(U_i)^{\top}\Gamma_{\ell}^{\top}\right\} \\
		\notag & = & n^{-1}p^{-1}\mE(\nu_{i}^{-2})\mE\left[\left\{\Gamma_j S(U_i)S(U_i)^{\top}\Gamma_{\ell}^{\top}\right\}^2\right] \\
		\notag & & - n^{-1}p^{-1}\{\mE(\nu_{i}^{-1})\}^{2}\left[\mE\left\{\Gamma_j S(U_i)S(U_i)^{\top}\Gamma_{\ell}^{\top}\right\}\right]^2 \\
		\notag & \lesssim & \zeta_{1}^{2}n^{-1}p^{-2}.
	\end{eqnarray}
	It follows from the Chebychev's inequality that
	\begin{eqnarray}
		\notag \left|n^{-1} p^{-1/2}\sum_{i=1}^n \nu_i^{-1}\Gamma_j S(U_i)S(U_i)^{\top}\Gamma_{\ell}^{\top}\right|\lesssim \zeta_{1}p^{-1}|\omega_{j\ell}|+O_p(\zeta_{1}n^{-1/2}p^{-1}+\zeta_{1}p^{-3/2})\,.
	\end{eqnarray}
	Finally, we arrive at $|Q_{j,\ell}|\lesssim\zeta_{1}p^{-1}|\omega_{j\ell}|+O_p(\zeta_{1}n^{-1/2}p^{-1}+\zeta_{1}p^{-7/6}+\zeta_{1}p^{-1-\delta/2})$.

	(ii)
	From the proof of part (i), we know that
	$
	Q_{j\ell}=Q_{0, j\ell}+O_p(\zeta_{1}p^{-7/6}+\zeta_{1}p^{-1-\delta/2})\,,
	$	
	where $Q_{0,j\ell}$ is the $(j,\ell)$th component of the random matrix $Q_{0}=n^{-1}p^{-1/2}\sum_{i=1}^{n}\nu_{i}^{-1}\{\Gamma S(U_{i})\}\{\Gamma S(U_{i})\}^{\top}$.
	In addition, $\mE\left\{\Gamma_j S(U_i)S(U_i)^{\top}\Gamma_{\ell}^{\top}\right\} \lesssim p^{-1}|\omega_{j\ell}| + O(p^{-3/2})$. It follows that
	\begin{eqnarray}
		\notag & & \tr\left\{\left(\mE\left[\{\Gamma S(U_{i})\}\{\Gamma S(U_{i})\}^{\top}\right]\right)^2\right\} \\
		\notag & = & \sum_{j=1}^{p}\sum_{\ell=1}^{p}\left[\mE\left\{\Gamma_j S(U_i)S(U_i)^{\top}\Gamma_{\ell}^{\top}\right\}\right]^2 \\
		\notag & \lesssim & p^{-2}\sum_{j=1}^{p}\sum_{\ell=1}^{p}|\omega_{j\ell}|^2 + p^{-5/2}\sum_{j=1}^{p}\sum_{\ell=1}^{p}|\omega_{j\ell}| + p^{-1} \\
		\notag & \lesssim & p^{-1}a_{0}(p) + p^{-3/2}a_{0}(p) + p^{-1} \\
		\notag & \lesssim & p^{-\delta}\,.
	\end{eqnarray}
	This implies that
	\begin{eqnarray}
		\notag \tr[\{\mE(Q_{0})\}^2] & = & p^{-1}\{\mE(\nu_{i}^{-1})\}^2\tr\left\{\left(\mE\left[\{\Gamma S(U_{i})\}\{\Gamma S(U_{i})\}^{\top}\right]\right)^2\right\} \\
		\notag & \lesssim & p^{-1-\delta}
	\end{eqnarray}
	and
	\begin{eqnarray}
		\notag & & \mE\{\tr(Q_{0}^{2})\} \\
		\notag & = & n^{-1}p^{-1}\tr\left(\mE\left[\nu_{i}^{-2}\{\Gamma S(U_{i})\}\{\Gamma S(U_{i})\}^{\top}\{\Gamma S(U_{i})\}\{\Gamma S(U_{i})\}^{\top}\right]\right) \\
		\notag & & + (1-n^{-1})p^{-1}\tr\left\{\left(\mE\left[\nu_{i}^{-1}\{\Gamma S(U_{i})\}\{\Gamma S(U_{i})\}^{\top}\right]\right)^2\right\} \\
		\notag & = & n^{-1}p^{-1}\mE(\nu_{i}^{-2})\mE\left\{\|\Gamma S(U_{i})\|^4\right\} \\
		\notag & & + (1-n^{-1})p^{-1}\{\mE(\nu_{i}^{-1})\}^2\tr\left\{\left(\mE\left[\{\Gamma S(U_{i})\}\{\Gamma S(U_{i})\}^{\top}\right]\right)^2\right\} \\
		\notag & = & O(n^{-1}p^{-1}) + \tr[\{\mE(Q_{0})\}^2](1-n^{-1})\,.
	\end{eqnarray}
	Thus, we have
	\begin{eqnarray}
		\notag \tr[\mE(Q_{0}^2)-\{\mE(Q_{0})\}^2] = O(n^{-1}p^{-1})\,.
	\end{eqnarray}
	We complete the proof of this lemma.
\end{proof}

\begin{proof}[Proof of Lemma \ref{lemma:05}]
	Recall that $\Gamma_{j}$ is the $j$th row of $\Gamma$, and denote $\Gamma_{j\ell}$ to be the $(j,\ell)$th element of $\Gamma$, then $$\Gamma_{j}U_{i}=\sum_{\ell=1}^{p}\Gamma_{j\ell}U_{i,\ell}.$$ It is noticed that $\omega_{j\ell}=\sum_{j_1=1}^{p}\Gamma_{jj_1}\Gamma_{\ell j_1}$, then
	\begin{eqnarray}
		\notag \mathrm{Var}(\Gamma_{j}U_{i}) = \sum_{\ell=1}^{p}\Gamma_{j\ell}^2=\omega_{jj}
	\end{eqnarray} 
	and
	\begin{eqnarray}
		\notag & & \mE\{(\Gamma_{j}U_{i})^4\} \\
		\notag & = & \mE\left\{\left(\sum_{\ell=1}^{p}\Gamma_{j\ell}U_{i,\ell}\right)^4\right\} \\
		\notag & = & \sum_{\ell=1}^{p}\Gamma_{j\ell}^4\mE(U_{i,\ell}^4) + 6\sum_{1\leq \ell_1\neq \ell_2\leq p}\Gamma_{j\ell_1}^2\Gamma_{j\ell_2}^2\mE(U_{i,\ell_1}^2)\mE(U_{i,\ell_2}^2) \\
		\notag & \lesssim & \omega_{jj}^2\,.
	\end{eqnarray}
	
	(i) For $i=1,\ldots,n$, let $\mathcal{A}_{2i} = \{(1-\epsilon)\tr(\Omega) \leq \|\Gamma U_i\|^2\leq (1+\epsilon)\tr(\Omega)\}$ for a fixed $0<\epsilon<1$, then
	\begin{eqnarray}
		\notag \P(\mathcal{A}_{2i}) \geq 1-c_1\exp\{-c_2p^{\delta\alpha/(4\alpha+4)}\}\,
	\end{eqnarray}
	according to the proof of Lemma \ref{lem:norm}. 
	It follows that
	\begin{eqnarray}
		\notag \mE(W_{i,j}^4) & = & \mE\{\|\Gamma U_{i}\|^{-4}(\Gamma_j U_i)^4\} \\
		\notag & = & \mE\{\|\Gamma U_{i}\|^{-4}(\Gamma_j U_i)^4\mI(\mathcal{A}_{2i})\} + \mE\{\|\Gamma U_{i}\|^{-4}(\Gamma_j U_i)^4\mI(\mathcal{A}_{2i}^{c})\} \\
		\notag & \leq & \{(1-\epsilon)\tr(\Omega)\}^{-2}\mE\{(\Gamma_j U_i)^4\} + \P(\mathcal{A}_{2i}^{c}) \\
		\notag & \lesssim & \omega_{jj}^2\{(1-\epsilon)\tr(\Omega)\}^{-2} + c_1\exp\{-c_2p^{\delta\alpha/(4\alpha+4)}\} \\
		\notag & \lesssim & \omega_{jj}^2\{\tr(\Omega)\}^{-2}
	\end{eqnarray}
	and
	\begin{eqnarray}
		\notag \mE(W_{i,j}^2) & \geq & \mE\{\|\Gamma U_{i}\|^{-2}(\Gamma_j U_i)^2\mI(\mathcal{A}_{2i})\} \\
		\notag & \geq & \{(1+\epsilon)\tr(\Omega)\}^{-1}\mE\{(\Gamma_j U_i)^2\mI(\mathcal{A}_{2i})\} \\
		\notag & = & \{(1+\epsilon)\tr(\Omega)\}^{-1}\mE\{(\Gamma_j U_i)^2\} - \{(1+\epsilon)\tr(\Omega)\}^{-1}\mE\{(\Gamma_j U_i)^2\mI(\mathcal{A}_{2i}^{c})\} \\
		\notag & \geq & \{(1+\epsilon)\tr(\Omega)\}^{-1}\mE\{(\Gamma_j U_i)^2\} - \{(1+\epsilon)\tr(\Omega)\}^{-1}[\mE\{(\Gamma_j U_i)^4\}]^{1/2}\{\P(\mathcal{A}_{2i}^{c})\}^{1/2} \\
		\notag & \gtrsim & \omega_{jj}\{(1+\epsilon)\tr(\Omega)\}^{-1} - \{(1+\epsilon)\tr(\Omega)\}^{-1}\times \omega_{jj} \times c_1^{1/2}\exp\{-c_2p^{\delta\alpha/(4\alpha+4)}/2\} \\
		\notag & \gtrsim & \omega_{jj}\{\tr(\Omega)\}^{-1},
	\end{eqnarray}
	from which we conclude that 
	\begin{eqnarray}
		\notag \mE\{(\zeta_{1}^{-1}W_{i,j})^4\} \lesssim \zeta_{1}^{-4}p^{-2}\omega_{jj}^2 \lesssim \bar{M}^2
	\end{eqnarray}
	and
	\begin{eqnarray}
		\notag \mE\{(\zeta_{1}^{-1}W_{i,j})^2\}\gtrsim \zeta_{1}^{-2}p^{-1}\omega_{jj} \gtrsim \underline{m}.
	\end{eqnarray}

	(ii) Similar to the proof of part (i), for any $\varrho\geq 1$,
	\begin{eqnarray}
		\notag \mE\left\{|\zeta_{1}^{-1}W_{i,j}|^{\varrho}\right\} & = & \mE\left\{|\zeta_{1}^{-1}W_{i,j}|^{\varrho}\mI(\mathcal{A}_{1i})\right\}  + \mE\left\{|\zeta_{1}^{-1}W_{i,j}|^{\varrho}\mI(\mathcal{A}_{1i}^c)\right\} \\
		\notag & \lesssim & \zeta_{1}^{-\varrho}\{\tr(\Omega)\}^{-\varrho/2}\mE\{|\Gamma_{j} U_i|^{\varrho}\} + \zeta_{1}^{-\varrho}\P(\mathcal{A}_{1i}^c) \\
		\notag & \lesssim & \mE\{|\Gamma_{j} U_i|^{\varrho}\} + p^{\varrho/2}\exp\{-c_2p^{\delta\alpha/(4\alpha+4)}\}.
	\end{eqnarray}
	Since $\max_{1\leq j \leq p}\|U_{i,j}\|_{\psi_{\alpha}}\leq c_0$ for some constant $c_0$, we have $\|\Gamma_j U_i\|_{\psi_{\alpha}}\lesssim c_0$ according to Lemma B.4 in \citet{Koike2019}. Then, we known that $\mE\{|\Gamma_{j} U_i|^{\varrho}\}\lesssim \varrho^{\varrho/\alpha}$ for any $\varrho\geq 1$ by the equivalent sub-exponential properties \citep{Koike2019}. Therefore,
	\begin{eqnarray}
		\notag  \mE\left\{|\zeta_{1}^{-1}W_{i,j}|^{\varrho}\right\} & \lesssim & \varrho^{\varrho/\alpha}
	\end{eqnarray}
	for any $\varrho\geq1$ for sufficient large $p$, which indicates that $\zeta_{1}^{-1}W_{i,j}$ is sub-exponential, and thus $\|\zeta_{1}^{-1}W_{i,j}\|_{\psi_{\alpha}}\lesssim \bB$.
	
	(iii) By simple algebra,
	\begin{eqnarray}
		\notag \mE(W_{i,j}^2) & = & p^{-1}\mE\{(\Gamma_{j}U_{i})^2\} + \mE\{(\Gamma_{j}U_{i})^2(\|\Gamma U_{i}\|^{-2}-p^{-1})\} \\
		\notag & = & p^{-1}\omega_{jj} + \mE\{(\Gamma_{j}U_{i})^2(\|\Gamma U_{i}\|^{-2}-p^{-1})\}\,,
	\end{eqnarray}
	where $\mE\{(\Gamma_{j}U_{i})^2(\|\Gamma U_{i}\|^{-2}-p^{-1})\}$ satisfies
	\begin{eqnarray}
		\notag & & \left|\mE\{(\Gamma_{j}U_{i})^2(\|\Gamma U_{i}\|^{-2}-p^{-1})\}\right| \\
		\notag & \leq & p^{-1}\mE\{(\Gamma_{j}U_{i})^2\|\Gamma U_{i}\|^{-2}|\|\Gamma_{j}U_{i}\|^2-p|\} \\
		\notag & = & p^{-1}\mE\{(\Gamma_{j}U_{i})^2\|\Gamma U_{i}\|^{-2}|\|\Gamma_{j}U_{i}\|^2-p|\mI(\mathcal{A}_{2i})\} \\
		\notag & & + p^{-1}\mE\{(\Gamma_{j}U_{i})^2\|\Gamma U_{i}\|^{-2}|\|\Gamma_{j}U_{i}\|^2-p|\mI(\mathcal{A}_{2i}^{c})\} \\
		\notag & \leq & p^{-1}\{(1-\epsilon)\tr(\Omega)\}^{-1}\mE\{(\Gamma_{j}U_{i})^2|\|\Gamma_{j}U_{i}\|^2-p|\} + p^{-1}\mE\{\|\Gamma_{j}U_{i}\|^2-p|\mI(\mathcal{A}_{2i}^{c})\} \\
		\notag & \leq & p^{-2}(1-\epsilon)^{-1}[\mE\{(\Gamma_{j}U_{i})^4\}]^{1/2}\{\mE(|\|\Gamma_{j}U_{i}\|^2-p|^2)\}^{1/2} \\
		\notag & & + p^{-1}\{\mE(|\|\Gamma_{j}U_{i}\|^2-p|^2)\}^{1/2}\{\P(\mathcal{A}_{2i}^{c})\}^{1/2} \\
		\notag & \lesssim & p^{-2}\times p^{1-\delta/2} + p^{-1}\times p^{1-\delta/2}\times c_1^{1/2}\exp\{-c_2p^{\delta\alpha/(4\alpha+4)}/2\} \\
		\notag & \lesssim & p^{-1-\delta/2}.
	\end{eqnarray}
	
	In addition, for $1\leq j\neq \ell\leq p$, we have
	\begin{eqnarray}
		\notag \mE(W_{i,j}W_{i,\ell}) & = & p^{-1}\mE\{(\Gamma_{j}U_{i})(\Gamma_{\ell}U_{i})\} + \mE\{(\Gamma_{j}U_{i})(\Gamma_{\ell}U_{i})(\|\Gamma U_{i}\|^{-2}-p^{-1})\} \\
		\notag & = & p^{-1}\omega_{j\ell} + \mE\{(\Gamma_{j}U_{i})(\Gamma_{\ell}U_{i})(\|\Gamma U_{i}\|^{-2}-p^{-1})\}\,,
	\end{eqnarray}
	where $\mE\{(\Gamma_{j}U_{i})(\Gamma_{\ell}U_{i})(\|\Gamma U_{i}\|^{-2}-p^{-1})\}$ satisfies
	\begin{eqnarray}
		\notag & & \left|\mE\{(\Gamma_{j}U_{i})(\Gamma_{\ell}U_{i})(\|\Gamma U_{i}\|^{-2}-p^{-1})\}\right| \\
		\notag & \leq & p^{-1}\mE\{|(\Gamma_{j}U_{i})(\Gamma_{\ell}U_{i})|\|\Gamma U_{i}\|^{-2}|\|\Gamma_{j}U_{i}\|^2-p|\} \\
		\notag & = & p^{-1}\mE\{|(\Gamma_{j}U_{i})(\Gamma_{\ell}U_{i})|\|\Gamma U_{i}\|^{-2}|\|\Gamma_{j}U_{i}\|^2-p|\mI(\mathcal{A}_{2i})\} \\
		\notag & & + p^{-1}\mE\{|(\Gamma_{j}U_{i})(\Gamma_{\ell}U_{i})|\|\Gamma U_{i}\|^{-2}|\|\Gamma_{j}U_{i}\|^2-p|\mI(\mathcal{A}_{2i}^{c})\} \\
		\notag & \leq & p^{-1}\{(1-\epsilon)\tr(\Omega)\}^{-1}\mE\{|(\Gamma_{j}U_{i})(\Gamma_{\ell}U_{i})||\|\Gamma_{j}U_{i}\|^2-p|\} + p^{-1}\mE\{\|\Gamma_{j}U_{i}\|^2-p|\mI(\mathcal{A}_{2i}^{c})\} \\
		\notag & \leq & p^{-2}(1-\epsilon)^{-1}[\mE\{|(\Gamma_{j}U_{i})(\Gamma_{\ell}U_{i})|^2\}]^{1/2}\{\mE(|\|\Gamma_{j}U_{i}\|^2-p|^2)\}^{1/2} \\
		\notag & & + p^{-1}\{\mE(|\|\Gamma_{j}U_{i}\|^2-p|^2)\}^{1/2}\{\P(\mathcal{A}_{2i}^{c})\}^{1/2} \\
		\notag & \lesssim & p^{-2}\times p^{1-\delta/2} + p^{-1}\times p^{1-\delta/2}\times c_1^{1/2}\exp\{-c_2p^{\delta\alpha/(4\alpha+4)}/2\} \\
		\notag & \lesssim & p^{-1-\delta/2}.
	\end{eqnarray}

	(iv) According to part (ii), $\zeta_{1}^{-1}W_1,\ldots,\zeta_{1}^{-1}W_n$ are i.i.d.~$p$-dimensional random vectors satisfies $\|\zeta_{1}^{-1}W_{i,j}\|_{\psi_{\alpha}}\lesssim \bar{B}$ for all $i=1,\ldots,n$ and $j=1,\ldots,p$. By Lemma 2.2.2 of \citet{VanderVaart1996}, 
	\begin{eqnarray}
		\notag \left\|\max_{1\leq i\leq n}\max_{1\leq j\leq p}|\zeta_{1}^{-1}W_{i,j}|\right\|_{\psi_{\alpha}}\lesssim \log^{1/\alpha}(np)\,.
	\end{eqnarray}
	Similar to the proof of part (i), we can show that 
	\begin{eqnarray}
		\notag \mE\{(\zeta_{1}^{-1}W_{i,j})^2\} & = & \zeta_{1}^{-2}\mE\{\|\Gamma U_{i}\|^{-2}(\Gamma_j U_i)^2\mI(\mathcal{A}_{1i})\} \\
		\notag & & + \zeta_{1}^{-2}\mE\{\|\Gamma U_{i}\|^{-4}(\Gamma_j U_i)^4\mI(\mathcal{A}_{1i}^{c})\} \\
		\notag & \leq & \zeta_{1}^{-2}\{(1+\epsilon)\tr(\Omega)\}^{-1}\mE\{(\Gamma_j U_i)^2\} + \zeta_{1}^{-2}\mE\{\mI(\mathcal{A}_{1i}^{c})\} \\
		\notag & \leq & \zeta_{1}^{-2}\omega_{jj}\{(1+\epsilon)\tr(\Omega)\}^{-1} + \zeta_{1}^{-2}c_1\exp\{-c_2p^{\delta/(4+4\alpha)}\} \\
		\notag & = & \zeta_{1}^{-2}\omega_{jj}\{(1+\epsilon)\tr(\Omega)\}^{-1}\{1+o(1)\}\,.
	\end{eqnarray}
	It follows that
	\begin{eqnarray}
		\notag \max_{1\leq j \leq p}\sum_{i=1}^{n}\mE\{(\zeta_{1}^{-1}W_{i,j})^2\} & \lesssim & \max_{1\leq j \leq p}\sum_{i=1}^{n}\zeta_{1}^{-2}\omega_{jj}\{(1+\epsilon)\tr(\Omega)\}^{-1} 
		\notag  \lesssim  n\max_{1\leq j \leq p}\omega_{jj} \leq \bar{M}n\,,
	\end{eqnarray}
	Applying Lemma E.1 of \citet{Cher2017}, it holds that with $\alpha\geq 1$ and $n^{-1/2} \log^{3/2} (np) \lesssim 1$,
	\begin{eqnarray}
		\notag \mE\left(\left|n^{-1/2}\sum_{i=1}^n\zeta_{1}^{-1}W_{i}\right|_{\infty}\right) & \lesssim & n^{-1/2} \{n^{1/2}\log^{1/2}(p)+\log^{1/\alpha}(np)\log(p)\} \\
		\notag & \lesssim & \log^{1/2} (np)\,.
	\end{eqnarray}
	From the properties of the $\psi_{\alpha}$ norm, %  in \cite{Koike2019}, 
	it holds that
	\begin{eqnarray}
		\notag \left\|\max_{1\leq i \leq n,1\leq j\leq p }|\zeta_{1}^{-1}W_{i,j}|^2\right\|_{\psi_{\alpha/2}}  \lesssim  \log^2(np).
	\end{eqnarray}
	According to Lemma E.3 of \citet{Cher2017}, we have that
	\begin{eqnarray}
		\notag \mE\left(\left|n^{-1}\sum_{i=1}^n(\zeta_{1}^{-1}W_{i})^2\right|_{\infty}\right) & \lesssim & n^{-1} \{\bar{M}n+\log^{2}(np)\log(p)\} 
		\lesssim  \bar{M}\,.
	\end{eqnarray} 
	We finish the proof of this lemma.
\end{proof}

\begin{proof}[Proof of Lemma \ref{lem:SM_boot_br}]
	Let $\tilde{X}_{i}=X_i-\hat{\btheta}_{n}$ and $\tilde{R}_{i}=\|\tilde{X}_i\|$ for $i=1,\ldots,n$.
	According to the proof of Lemma \ref{lem:Br}, $\|\hat{\btheta}_{n}\|=O_p(\zeta_{1}^{-1}n^{-1/2})$ and $\max_{1\leq i \leq n}R_{i}^{-1}=O_{p}(\zeta_{1}n^{1/4})$. Then $R_{i}^{-1}\|\hat{\btheta}_{n}\|$ satisfies
	\begin{eqnarray}
		\notag R_{i}^{-1}\|\hat{\btheta}_{n}\|=O_{p}(n^{-1/2}) \text{~~and~~} \max_{1\leq i \leq n}R_{i}^{-1}\|\hat{\btheta}_{n}\|=O_{p}(n^{-1/4})\,.
	\end{eqnarray}
	As $\tilde{R}_{i}^{-1}=R_{i}^{-1}\|W_i-R_{i}^{-1}\hat{\btheta}_n\|^{-1}=R_{i}^{-1}\left(1-2R_{i}^{-1}W_{i}^{\top}\hat{\btheta}_{n}+R_{i}^{-2}\|\hat{\btheta}_{n}\|^2\right)^{-1/2}$, 
	by Taylor expansion, 
	\begin{eqnarray}
		\notag \tilde{R}_{i}^{-1} = R_{i}^{-1}\left(1+R_{i}^{-1}W_{i}^{\top}\hat{\btheta}_{n}-2^{-1}R_{i}^{-2}\|\hat{\btheta}_n\|^2+\tilde{\delta}_{1i}\right),
	\end{eqnarray}
	where $\tilde{\delta}_{1i}$ satisfies $\tilde{\delta}_{1i} = O_{p}(n^{-1})$ and $\max_{1\leq i \leq n}\tilde{\delta}_{1i}=O_{p}(n^{-1/2})$. It follows that
	\begin{eqnarray}
		\notag \tilde{R}_{i}^{-1} = R_{i}^{-1}(1 + \tilde{\delta}_{2i})\,,
	\end{eqnarray}
	where $\tilde{\delta}_{2i}=R_{i}^{-1}W_{i}^{\top}\hat{\btheta}_{n}-2^{-1}R_{i}^{-2}\|\hat{\btheta}_n\|^2+\tilde{\delta}_{1i}$ satisfies $\tilde{\delta}_{2i} = O_{p}(n^{-1/2})$ and $\max_{1\leq i \leq n}\tilde{\delta}_{2i}=O_{p}(n^{-1/4})$. Thus, 
	\begin{eqnarray}
		\notag \tilde{R}_{i}^{-1}=O_{p}(\zeta_{1}) \text{~~and~~} \max_{1\leq i \leq n}\tilde{R}_{i}^{-1}=O_{p}(\zeta_{1}n^{1/4})\,.
	\end{eqnarray}
	
	Denote $\tilde{W}_{i} = \tilde{X}_{i}/\|\tilde{X}_{i}\|$ for $i=1,\ldots,n$. Then,
	\begin{eqnarray}
		\notag \tilde{W}_{i} & = & \tilde{R}_{i}^{-1}(X_i-\hat{\btheta}_n) \\
		\notag & = & R_{i}^{-1}(X_i-\hat{\btheta}_n)(1 + \tilde{\delta}_{2i}) \\
		\notag & = & (W_{i}-R_{i}^{-1}\hat{\btheta}_n)(1 + \tilde{\delta}_{2i})\,.
	\end{eqnarray}
	
	We first show that $\|\tilde{\btheta}_n\| = O_{p}(\zeta_{1}^{-1}n^{-1/2})$. It is noticed that ${\tilde \btheta}_n$ minimizes
	\begin{eqnarray}%\label{btheta}
		\notag L_{n}^{*}(\bbeta) = \sum_{i=1}^n\|Z_i\tilde{X}_{i}-\bbeta\|\,,
	\end{eqnarray}
	which is a strictly convex function of $\bbeta$. Thus, if we can show that $L_{n}^{*}(\bbeta)$ has a $\zeta_{1}n^{1/2}$-consistent local
	minimizer, then this local minimizer must be a $\zeta_{1}n^{1/2}$-consistent global minimizer of $L_{n}^{*}(\bbeta)$. The existence of a $\zeta_{1}n^{1/2}$-consistent local minimizer is implied by
	the fact that for an arbitrarily small $\varepsilon>0$, there exists a constant $C_0$, which does not depend on $n$ and $p$, such that
	\begin{eqnarray}\label{eq:SM_lem_br_01}
		\liminf_{n}\P\left\{\inf_{\bq\in \mathbb{R}^p,~\|\bq\|=C_0}~ L_{n}^{*}(\zeta_{1}^{-1}n^{-1/2}\bq)>L_{n}^{*}(0)\right\}>1-\varepsilon,
	\end{eqnarray} 
	Since $|Z_i|=1$, we rewrite $\|Z_i{\tilde X}_i-\zeta_{1}^{-1}n^{-1/2}\bq \|$ as
	\begin{eqnarray}
		\notag & & 	\|Z_i{\tilde X}_i-\zeta_{1}^{-1}n^{-1/2}\bq \| \\
		\notag & = & \tilde{R}_{i} \left(1-2\zeta_{1}^{-1}n^{-1/2}\tilde{R}_{i}^{-1}Z_i\bq^{\top}\W_i+\zeta_{1}^{-2}n^{-1}\tilde{R}_{i}^{-2}\|\bq\|^2\right)^{1/2}\,.
	\end{eqnarray}
	As $|\zeta_{1}^{-1}n^{-1/2}\tilde{R}_{i}^{-1}Z_i\bq^{T}\W_i|=O_p(n^{-1/2})$ and $\zeta_{1}^{-2}n^{-1}\tilde{R}_{1i}^{-2}\|\bq\|^2=O_p(n^{-1})$, by Taylor expansion, we obtain that
	\begin{eqnarray}
		\notag & & \|Z_i\tilde{X}_i-\zeta_{1}^{-1}n^{-1/2}\bq \| \\
		\notag & = & \tilde{R}_i -\zeta_{1}^{-1}n^{-1/2}Z_i\bq^{\top}\W_i+2^{-1}\zeta_{1}^{-2}n^{-1}\tilde{R}_{i}^{-1}\|\bq\|^2 \\
		\notag & & -2^{-1}\zeta_{1}^{-2}n^{-1}\tilde{R}_{i}^{-1}\bq^{\top}\W_i\W_i^{\top}\bq+O_p(\zeta_{1}^{-1}n^{-3/2})\,.
	\end{eqnarray}
	Then,
	\begin{eqnarray}\label{eq:compare}
		\notag & & \zeta_{1}\left\{L_{n}^{*}(\zeta_{1}^{-1}n^{-1/2}\bq)-L_{n}^{*}(0)\right\} \\
		\notag & = & \zeta_{1}\sum_{i=1}^{n}\left(\|Z_i\tilde{X}_i-\zeta_{1}^{-1}n^{-1/2}\bq \|-\|\tilde{X}_i\|\right)\\
		\notag & = & -n^{-1/2}\bq^{\top}\left(\sum_{i=1}^{n}Z_i\W_i\right) + 2^{-1}\zeta_{1}^{-1}n^{-1}\|\bq\|^2\sum_{i=1}^{n}\tilde{R}_{i}^{-1} \\
		& & -2^{-1}\zeta_{1}^{-1}n^{-1}\bq^{\top}\left(\sum_{i=1}^{n}\tilde{R}_{i}\W_i\W_i^{\top}\right)
		\bq+O_p(n^{-1/2})\,.
	\end{eqnarray}
	
	As $\mE^{*}\left(n^{-1/2}\sum_{i=1}^{n}Z_i\W_i\right) = 0$ 
	and $$\mE^{*}\left(\left\|n^{-1/2}\sum_{i=1}^{n}Z_i\W_i\right\|^2\right) =  n^{-1}\sum_{i=1}^{n}\tilde{W}_{i}^{\top}\tilde{W}_{i}  = 1,$$
	we obtain that
	\begin{eqnarray}
		\notag \left|n^{-1/2}\bq^{\top}\sum_{i=1}^{n}Z_i\W_i\right| \leq \|\bq\|\left\|n^{-1/2}\sum_{i=1}^{n}Z_i\W_i\right\| = O_{p}(\|\bq\|)\,.
	\end{eqnarray}
	In the meanwhile, as $\zeta_{1}^{-1}n^{-1}\sum_{i=1}^{n}R_{i}^{-1}=1+O_{p}(n^{-1/2})$, we have
	\begin{eqnarray}
		\notag \zeta_{1}^{-1}n^{-1}\|\bq\|^2\sum_{i=1}^{n}\tilde{R}_{i}^{-1} & = & \zeta_{1}^{-1}n^{-1}\|\bq\|^2\sum_{i=1}^{n}R_{i}^{-1}(1+\tilde{\delta}_{2i}) \\
		\notag & = & \|\bq\|^2\{1+O_{p}(n^{-1/4})\}\,.
	\end{eqnarray}
	Simple algebra yields
	\begin{eqnarray}
		\notag & & n^{-1}\sum_{i=1}^{n}\tilde{R}_{i}^{-1}\W_i\W_i^{\top} \\
		\notag & = & n^{-1}\sum_{i=1}^{n}R_{i}^{-1}(W_i-R_{i}^{-1}\hat{\btheta}_n)(W_i-R_{i}^{-1}\hat{\btheta}_n)^{\top}(1+\tilde{\delta}_{2i}) \\
		\notag & = & n^{-1}\sum_{i=1}^{n}R_{i}W_iW_i^{\top}(1+\tilde{\delta}_{2i}) - 2n^{-1}\sum_{i=1}^{n}R_{i}^{-2}W_{i}\hat{\btheta}_{n}^{\top} (1+\tilde{\delta}_{2i}) \\
		\notag & & + n^{-1}\sum_{i=1}R_{i}^{-3}\hat{\btheta}_n\hat{\btheta}_n^{\top}(1+\tilde{\delta}_{2i})\,.
	\end{eqnarray}
	Similar to the proof in \citet{Cheng2019} and utilizing the results on $Q=n^{-1}\sum_{i=1}^{n}R_{i}^{-1}W_{i}W_{i}^{-1}$ in Lemma \ref{lemma:SM_Q}, we can show that %$\mE\left(n^{-1}\bq^{\top}\sum_{i=1}^{n}R_{i}W_iW_i^{\top}\bq\right)\lesssim \zeta_{1}p^{-\delta}$, and $\mathrm{Var}\left(n^{-1}\bq^{\top}\sum_{i=1}^{n}R_{i}W_iW_i^{\top}\bq\right)=O(\zeta_{1}^2n^{-1}p^{-\delta}+\zeta_{1}^2n^{-1})$, which lead to
	$n^{-1}\bq^{\top}\sum_{i=1}^{n}R_{i}W_iW_i^{\top}\bq(1+\tilde{\delta}_{2i}) = O_{p}(\zeta_{1}n^{-1/2}+\zeta_{1}p^{-(1/6 \wedge \delta/2)})$.
	In addition,
	as $$n^{-1}\sum_{i=1}^{n}R_{i}^{-2}\bq^{\top}W_{i}\leq n^{-1}\sum_{i=1}^{n}R_{i}^{-2}\|\bq\|\|W_i\|=\|\bq\|n^{-1}\sum_{i=1}^{n}R_{i}^{-2}=O_{p}(\zeta_{1}^{2})$$
	and
	$n^{-1}\sum_{i=1}^{n}R_{i}^{-3}=\zeta_{3}\{1+o_{p}(1)\}$,
	we have
	\begin{eqnarray}
		\notag & & n^{-1}\bq^{\top}\sum_{i=1}^{n}R_{i}^{-2}W_{i}\hat{\btheta}_{n}^{\top}\bq(1+\tilde{\delta}_{2i}) \\
		\notag & = & n^{-1}\sum_{i=1}^{n}R_{i}^{-2}\bq^{\top}W_{i}(1+\tilde{\delta}_{2i})(\hat{\btheta}_{n}^{\top}\bq) = O_{p}(\zeta_{1}n^{-1/2})\,.
	\end{eqnarray}
	and
	\begin{eqnarray}
		\notag n^{-1}\bq^{\top}\sum_{i=1}R_{i}^{-3}\hat{\btheta}_n\hat{\btheta}_n^{\top}\bq(1+\tilde{\delta}_{2i}) = n^{-1}\sum_{i=1}R_{i}^{-3}(1+\tilde{\delta}_{2i})\|\bq^{\top}\hat{\btheta}_n\|^2 = O_{p}(\zeta_{1}n^{-1})\,.
	\end{eqnarray}
	
	Thus, we obtain
	\begin{eqnarray}
		\notag & &  2^{-1}\zeta_{1}^{-1}n^{-1}\|\bq\|^2\sum_{i=1}^{n}\tilde{R}_{i}^{-1} +2^{-1}\zeta_{1}^{-1}n^{-1}\bq^{\top}\left(\sum_{i=1}^{n}\tilde{R}_{i}\W_i\W_i^{\top}\right)
		\bq \\
		\notag & = & 2^{-1}\|\bq\|^2 + O_{p}(n^{-1/4}+p^{-\delta})\,.
	\end{eqnarray}
	Choosing a sufficient large constant $C_0$, the second term dominates the first term in \eqref{eq:compare} and thus $\zeta_{1}\left\{L_{n}^{*}(\zeta_{1}^{-1}n^{-1/2}\bq)-L_{n}^{*}(0)\right\}>0$. Hence, we have $\|\tilde{\btheta}_n\|=O_p(\zeta_{1}^{-1}n^{-1/2})$.
	
	Denote $\Theta_{i}=Z_i\hat{\btheta}_n+\tilde{\btheta}_n$ for $i=1,\ldots,n$. Then $$\max_{1\leq i \leq n}\|\Theta_{i}\| \leq \|\hat{\btheta}_n\|+\|\tilde{\btheta}_n\|=O_{p}(\zeta_{1}^{-1}n^{-1/2}).$$ Recall that $\tilde{\btheta}_n$ satisfies
	\begin{eqnarray}
		\notag \sum_{i=1}^{n}\frac{Z_i\tilde{X}_i-\tilde{\btheta}_n}{\|Z_i\tilde{X}_i-\tilde{\btheta}_n\|} = \sum_{i=1}^{n}\frac{Z_iW_i - R_{i}^{-1}\Theta_{i}}{\|Z_iW_i - R_{i}^{-1}\Theta_{i}\|} = 0\,,
	\end{eqnarray}
	which is equivalently to
	\begin{eqnarray}
		\notag n^{-1}\sum_{i=1}^{n}(Z_iW_i-R_{i}^{-1}\Theta_{i})\left(1-2Z_iR_i^{-1}W_{i}^{\top}\Theta_{i} + R_{i}^{-2}\|\Theta_{i}\|^2\right)^{-1/2} = 0\,,
	\end{eqnarray}
	where $|R_{i}^{-1}W_{i}^{\top}\Theta_{i}|=O_{p}(n^{-1/2})$, $R_{i}^{-2}\|\Theta_{i}\|^2=O_{p}(n^{-1})$,
	\begin{eqnarray}
		\notag \max_{1\leq i \leq n}|R_{i}^{-1}W_{i}^{\top}\Theta_{i}|=O_{p}(n^{-1/4}) \text{~~and~~} \max_{1\leq i \leq n}R_{i}^{-2}\|\Theta_{i}\|^2=O_{p}(n^{-1/2})\,.
	\end{eqnarray}
	Taylor expansion leads to
	\[
	n^{-1}\sum_{i=1}^{n}(Z_iW_i-R_{i}^{-1}\Theta_{i})(1+Z_iR_{i}^{-1}W_{i}^{\top}\Theta_{i}-2R_{i}^{-2}\|\Theta_{i}\|^2+\tilde{\delta}_{3i})=0
	\]
	where $\delta_{3i}=O_p\{(Z_iR_{i}^{-1}W_{i}^{\top}\Theta_{i}-R_{i}^{-2}\|\Theta_{i}\|^2)^2\}=O_p(n^{-1})$, and $\max_{1\leq i\leq n}\delta_{3i}=O_p(n^{-1/2})$. Then,
	\begin{eqnarray}
		\notag & & n^{-1}\sum_{i=1}^{n}Z_iW_i(1-2R_{i}^{-2}\|\Theta_{i}\|^2+\tilde{\delta}_{3i}) + n^{-1}\sum_{i=1}^{n}R_{i}^{-1}(W_{i}^{\top}\Theta_{i})W_{i} \\ 
		\notag & = & n^{-1}\sum_{i=1}^{n}Z_iW_i(1-2R_{i}^{-2}\|\Theta_{i}\|^2+\tilde{\delta}_{3i}) + n^{-1}\sum_{i=1}^{n}Z_iR_{i}^{-1}W_{i}W_{i}^{\top}\hat{\btheta}_n \\
		\notag & & + n^{-1}\sum_{i=1}^{n}R_{i}^{-1}W_{i}W_{i}^{\top}\tilde{\btheta}_n \\ 
		\notag & = & n^{-1}\sum_{i=1}^{n}R_{i}^{-1}\Theta_{i}(1+\tilde{\delta}_{3i} + \tilde{\delta}_{4i}) \\
		\notag & = & n^{-1}\sum_{i=1}^{n}R_{i}^{-1}\tilde{\btheta}_n(1+\tilde{\delta}_{3i} + \tilde{\delta}_{4i}) + n^{-1}\sum_{i=1}^{n}Z_iR_{i}^{-1}\hat{\btheta}_n(1+\tilde{\delta}_{3i} + \tilde{\delta}_{4i})\,,
	\end{eqnarray}
	where $\tilde{\delta}_{4i}=Z_iR_{i}^{-1}W_{i}^{\top}\Theta_{i}-2R_{i}^{-2}\|\Theta_{i}\|^2=O_{p}(\tilde{\delta}_{3i}^{1/2})$ satisfies $\max_{1\leq i \leq n}\tilde{\delta}_{4i}=O_{p}(n^{-1/4})$.
	
	The proof of Lemma \ref{lem:Br} implies $|\hat{\btheta}|_{\infty}=O_{p}\{n^{-1/2}\log^{1/2}(np)\}$. %and $\left|n^{-1}\sum_{i=1}^{n}R_{i}^{-1}W_{i}W_{i}^{\top}\hat{\btheta}_n\right|_{\infty}=O_{p}\{\zeta_{1}n^{-1/2}p^{-\delta}\log^{1/2}(np)+\zeta_{1}n^{-1}\log^{1/2}(np)\}$. 
	As $\mE^{*}\left(n^{-1}\sum_{i=1}^{n}Z_iR_{i}^{-1}\right)=0$ and $\mE^{*}\left\{\left(n^{-1}\sum_{i=1}^{n}Z_iR_{i}^{-1}\right)^2\right\}=n^{-2}\sum_{i=1}^{n}R_{i}^{-2}=O_{p}(n^{-1}\zeta_{2})$, we have
	$n^{-1}\sum_{i=1}^{n}Z_iR_{i}^{-1} = O_{p}(\zeta_{1}n^{-1/2})$.
	
	As $Z_i$ is bounded, it is straightforward to show that $|n^{-1/2}\sum_{i=1}^{n}Z_{i}W_{i}|_{\infty}=O_{p}\{p^{-1/2}\log^{1/2}(np)\}$ similar as in the proof of Lemma \ref{lemma:05} (iii).
	Thus, similar to the proof of Lemma \ref{lem:Br}, we obtain that
	\begin{eqnarray}
		\notag |\tilde{\btheta}|_{\infty}=O_{p}\{n^{-1/2}\log^{1/2}(np)\}
	\end{eqnarray}
	and
	\begin{eqnarray}
		\notag & & \left|n^{-1}\sum_{i=1}^{n}R_{i}^{-1}W_{i}W_{i}^{\top}\tilde{\btheta}_n\right|_{\infty} \\
		\notag & = & O_{p}\{\zeta_{1}n^{-1/2}p^{-(1/6 \wedge \delta/2)}\log^{1/2}(np)+\zeta_{1}n^{-1}\log^{1/2}(np)\}\,.
	\end{eqnarray}
	In the meanwhile, it holds that $|n^{-1}\sum_{i=1}^nR_i^{-1}|=\zeta_{1}+O_p(\zeta_{1}n^{-1/2})$. Finally, 
	\begin{align}\label{eq:tildetheta}
		n^{1/2}\tilde{\btheta}_{n}=n^{-1/2}\zeta_{1}^{-1}\sum_{i=1}^n Z_i W_i+{\tilde C}_n\,,
	\end{align}
	and ${\tilde C}_n$ is the remainder term satisfies 
	$$|{\tilde C}_n|_{\infty}=O_p\{n^{-1/4}\log^{1/2} (np)+ p^{-\delta-(1/6 \wedge \delta/2)}\log^{1/2} (np)\}\,.$$
	We finish the proof of this lemma.
\end{proof}

\section*{Appendix C: Additional simulation results}\label{sec:SM_simulation}

\renewcommand\thesection{C}
\setcounter{subsection}{0}

In this section, we report additional simulation results. Section C1 %\ref{sec:SM_SCIs} 
presents simulation results on SCIs for $\rho=0.2$ and $0.5$. Section C2 %\ref{sec:simulations_test} 
reports simulations on global tests  for high-dimensional location parameters. %Section \ref{sec:SM_simu_ARE} presents simulations on asymptotic relative efficiency of the sample spatial median compared to the sample mean.

\subsection{Addition simulation results on simultaneous confidence intervals}\label{sec:SM_SCIs}

Tables \ref{tabS1} reports the coverage probability and median length of the SCIs based on $\hat{\btheta}_n$ for $\rho=0.2$ and $0.5$, the results of the SCIs based on the sample mean $\bar{X}_n$ are presented in parentheses. 
We observe that the performance of the SCIs based on $\hat{\btheta}_n$ with $\rho=0.2$ and $0.5$ is similar to that of $\rho=0.0$ and $0.8$ in the main paper. The SCIs achieve satisfactory coverage probability, and it is much shorter than those based on $\bar{X}_n$ under the multivariate $t$-distribution, which is heavy-tailed.

\begin{table}[htp]
	\footnotesize
	\centering
	{
		\caption{Coverage probability (in $\%$) and median length of the SCIs based on $\hat{\btheta}_n$, the results of the SCIs based on $\bar{X}_n$ are in parentheses.}
		\label{tabS1}
		\setlength\tabcolsep{3pt}
		\resizebox{\textwidth}{!}{
			\begin{tabular}{@{}cccccccccccccc@{}}
				
				\hline
				&&&&\multicolumn{4}{c}{$\btheta=\btheta_1$}& &\multicolumn{4}{c}{$\btheta=\btheta_2$}\cr
				&&&&\multicolumn{2}{c}{Coverage probability}&\multicolumn{2}{c}{Median length}&&\multicolumn{2}{c}{Coverage probability}&\multicolumn{2}{c}{Median length}\cr
				
				Model & $\rho$ & $n$&$p$& 90\% & 95\% & 90\% & 95\%& &90\% & 95\% & 90\% & 95\%\cr
				
				\hline
				%  I & 0 &100&~$100$ & 89.6 (89.9) & 94.4 (94.4) & 0.65 (0.65) & 0.69 (0.69)
				% && 88.9 (88.8) & 94.1 (93.9) & 0.65 (0.65) & 0.69 (0.69) \cr
				
				%&&&$1000$& 89.5 (89.6) & 94.7 (94.4) & 0.77 (0.77) & 0.80 (0.80)
				%&& 89.5 (89.5) & 94.0 (94.0) & 0.77 (0.77) & 0.81 (0.80) \cr
				
				%&&200&~$100$& 89.8 (89.8) & 95.1 (95.1) & 0.46 (0.46) & 0.49 (0.49)
				%&& 88.6 (88.8) & 94.4 (94.7) & 0.46 (0.46) & 0.49 (0.49) \cr
				
				%&&&$1000$& 89.7 (89.7) & 94.4 (94.6) & 0.55 (0.55) & 0.57 (0.57)
				%&& 89.1 (89.2) & 94.7 (94.6) & 0.55 (0.55) & 0.57 (0.57) \cr
				
				%\cline{2-13}
				
				I& 0.2 &100&~$100$ & 89.8 (89.9) & 94.5 (94.5) & 0.65 (0.65) & 0.69 (0.69)
				&& 88.8 (88.7) & 94.4 (94.4) & 0.65 (0.65) & 0.69 (0.69) \cr
				
				&&&$1000$& 88.7 (88.7) & 94.5 (94.3) & 0.77 (0.77) & 0.80 (0.80)
				&& 90.0 (89.6) & 94.7 (94.8) & 0.77 (0.77) & 0.80 (0.80) \cr
				
				&&200&~$100$& 89.0 (88.9) & 94.3 (94.1) & 0.46 (0.46) & 0.49 (0.49)
				&& 88.8 (88.8) & 94.0 (94.2) & 0.46 (0.46) & 0.49 (0.49) \cr
				
				&&&$1000$& 89.8 (89.8) & 94.4 (94.4) & 0.55 (0.55) & 0.57 (0.57)
				&& 88.7 (89.2) & 94.6 (94.3) & 0.55 (0.55) & 0.57 (0.57) \cr
				
				\cline{2-13}
				
				& 0.5 &100&~$100$ & 89.6 (89.8) & 94.5 (94.4) & 0.65 (0.65) & 0.69 (0.69)
				&& 88.4 (88.8) & 94.0 (94.1) & 0.65 (0.65) & 0.69 (0.69) \cr
				
				&&&$1000$& 88.4 (88.4) & 94.3 (94.3) & 0.77 (0.77) & 0.80 (0.80)
				&& 87.4 (87.4) & 94.1 (94.2) & 0.77 (0.77) & 0.80 (0.80) \cr
				
				&&200&~$100$& 90.9 (90.9) & 95.1 (95.2) & 0.46 (0.46) & 0.49 (0.49)
				&& 89.7 (90.0) & 95.3 (95.0) & 0.46 (0.46) & 0.49 (0.49) \cr
				
				&&&$1000$& 89.0 (89.0) & 94.2 (94.3) & 0.55 (0.55) & 0.57 (0.57)
				&& 88.8 (88.6) & 94.3 (94.0) & 0.55 (0.55) & 0.57 (0.57) \cr
				
				%\cline{2-13}
				
				%& 0.8 &100&~$100$ & 89.1 (88.7) & 94.6 (94.6) & 0.64 (0.63) & 0.68 (0.67)
				%&& 88.4 (88.6) & 93.7 (94.1) & 0.64 (0.63) & 0.68 (0.67) \cr
				
				%&&&$1000$& 88.4 (88.4) & 93.8 (93.7) & 0.76 (0.76) & 0.80 (0.79)
				%&& 89.0 (89.2) & 94.6 (94.6) & 0.76 (0.76) & 0.80 (0.79) \cr
				
				%&&200&~$100$& 90.5 (90.1) & 95.2 (94.9) & 0.45 (0.45) & 0.48 (0.48)
				%&& 89.6 (89.6) & 94.0 (94.1) & 0.45 (0.45) & 0.48 (0.48) \cr
				
				%&&&$1000$& 90.4 (90.4) & 94.5 (94.4) & 0.54 (0.54) & 0.56 (0.56)
				%&& 88.4 (88.5) & 93.6 (93.8) & 0.54 (0.54) & 0.56 (0.56) \cr
				
				\hline
				
				% II & 0 &100&~$100$ & 89.7 (88.6) & 94.7 (93.7) & 0.71 (1.05) & 0.75 (1.11)
				%&& 88.8 (88.8) & 94.5 (94.2) & 0.71 (1.05) & 0.75 (1.11) \cr
				
				%&&&$1000$& 89.4 (91.0) & 95.8 (95.0) & 0.84 (1.25) & 0.88 (1.30) 
				%&& 89.1 (89.0) & 94.4 (94.5) & 0.84 (1.25) & 0.88 (1.31) \cr
				
				%&&200&~$100$& 88.6 (89.1) & 94.2 (95.1) & 0.50 (0.76) & 0.53 (0.81)
				%&& 89.5 (89.7) & 94.4 (94.8) & 0.50 (0.76) & 0.53 (0.80) \cr
				
				%&&&$1000$& 89.6 (88.7) & 94.8 (94.6) & 0.59 (0.90) & 0.62 (0.94)
				%&& 90.1 (89.5) & 94.8 (93.9) & 0.59 (0.90) & 0.62 (0.94) \cr
				
				II& 0.2 &100&~$100$ & 89.2 (88.8) & 94.8 (94.2) & 0.71 (1.05) & 0.75 (1.12)
				&& 88.4 (88.8) & 93.7 (94.3) & 0.71 (1.05) & 0.75 (1.11) \cr
				
				&&&$1000$& 89.0 (89.4) & 94.1 (94.8) & 0.84 (1.24) & 0.88 (1.30)
				&& 89.0 (88.9) & 94.4 (94.6) & 0.84 (1.24) & 0.88 (1.30) \cr
				
				&&200&~$100$& 90.7 (89.8) & 95.3 (94.7) & 0.50 (0.76) & 0.53 (0.80)
				&& 89.2 (89.7) & 94.0 (94.4) & 0.50 (0.76) & 0.53 (0.80) \cr
				
				&&&$1000$& 88.6 (89.5) & 94.2 (94.6) & 0.59 (0.90) & 0.62 (0.93)
				&& 89.0 (90.6) & 95.0 (95.1) & 0.59 (0.90) & 0.62 (0.94) \cr
				
				\cline{2-13}
				
				& 0.5 &100&~$100$ & 89.2 (87.9) & 93.6 (93.9) & 0.71 (1.05) & 0.75 (1.12)
				&& 89.4 (88.6) & 94.6 (94.1) & 0.71 (1.05) & 0.75 (1.11) \cr
				
				&&&$1000$& 89.2 (88.9) & 94.4 (94.2) & 0.84 (1.24) & 0.88 (1.30)
				&& 90.0 (89.4) & 94.7 (94.6) & 0.84 (1.25) & 0.88 (1.30) \cr
				
				&&200&~$100$& 89.4 (90.0) & 94.1 (94.6) & 0.50 (0.76) & 0.53 (0.80)
				&& 89.7 (88.6) & 95.0 (93.6) & 0.50 (0.76) & 0.53 (0.80) \cr
				
				&&&$1000$& 90.0 (89.9) & 95.6 (94.8) & 0.59 (0.90) & 0.62 (0.94)
				&& 88.8 (89.5) & 93.8 (94.4) & 0.59 (0.89) & 0.62 (0.93) \cr
				
				%\cline{2-13}
				
				%& 0.8 &100&~$100$ & 89.1 (90.7) & 94.4 (94.9) & 0.69 (1.02) & 0.74 (1.09)
				%&& 89.4 (89.7) & 94.2 (94.4) & 0.69 (1.02) & 0.74 (1.09) \cr
				
				%&&&$1000$& 89.3 (89.1) & 94.6 (94.4) & 0.83 (1.23) & 0.87 (1.29)
				%&& 89.8 (88.8) & 94.7 (94.4) & 0.83 (1.23) & 0.87 (1.29) \cr
				
				%&&200&~$100$& 87.6 (87.7) & 93.4 (93.6) & 0.49 (0.73) & 0.52 (0.78)
				%&& 90.3 (90.1) & 94.9 (95.2) & 0.49 (0.73) & 0.52 (0.78) \cr
				
				%&&&$1000$& 88.7 (89.7) & 94.7 (94.6) & 0.59 (0.88) & 0.61 (0.92)
				%&& 90.2 (90.8) & 94.7 (95.7) & 0.59 (0.89) & 0.61 (0.93) \cr
				
				\hline

				III&0.2&100&~$100$& 89.6 (89.5) & 95.0 (95.1) & 0.65 (0.66) & 0.69 (0.70)
				&& 89.4 (89.4) & 94.6 (94.6) & 0.65 (0.66) & 0.69 (0.70) \cr
				
				&&&$1000$ & 89.3 (88.8) & 94.5 (94.5) & 0.78 (0.78) & 0.82 (0.82)
				&& 90.3 (90.7) & 95.0 (94.9) & 0.78 (0.78) & 0.82 (0.82) \cr
				
				&&200&~$100$ & 89.2 (89.0) & 94.4 (94.4) & 0.46 (0.46) & 0.49 (0.49)
				&& 90.0 (89.6) & 95.1 (95.2) & 0.46 (0.46) & 0.49 (0.49) \cr
				
				&&&$1000$ & 89.7 (89.7) & 94.6 (94.8) & 0.55 (0.55) & 0.57 (0.58)
				&& 90.4 (90.6) & 95.0 (95.0) & 0.55 (0.55) & 0.57 (0.57) \cr
				
				\cline{2-13}
				
				&0.5&100&~$100$& 88.9 (89.3) & 94.0 (94.6) & 0.65 (0.65) & 0.69 (0.69)
				&& 88.0 (88.5) & 94.2 (94.0) & 0.65 (0.65) & 0.69 (0.69) \cr
				
				&&&$1000$ & 89.1 (89.2) & 94.3 (94.2) & 0.78 (0.78) & 0.81 (0.81)
				&& 89.2 (88.9) & 94.1 (94.0) & 0.78 (0.78) & 0.81 (0.81) \cr
				
				&&200&~$100$ & 89.6 (89.7) & 95.0 (94.4) & 0.46 (0.46) & 0.49 (0.49)
				&& 89.6 (89.7) & 94.9 (94.4) & 0.46 (0.46) & 0.49 (0.49) \cr
				
				&&&$1000$ & 89.0 (89.1) & 94.3 (94.4) & 0.55 (0.55) & 0.57 (0.57) 
				&& 89.3 (89.6) & 95.4 (95.0) & 0.55 (0.55) & 0.57 (0.57) \cr
				
				\hline 
				
			\end{tabular}
		}
	}
\end{table}

\subsection{Simulations on global tests for high-dimensional location parameters}\label{sec:simulations_test}

In this section, we report the performance of the test based on $T_n$ (Median test) for one-sample high-dimensional location parameters, and compare it with three alternative approaches: the test of \citet[CQ test]{Chen2010}; the test based on $T_{\mathrm{Mean}}$ (Mean test) and bootstrap approximation for $\bar{X}_n$; the test of \citet[WPL test]{Wang2015} based on $T_{\mathrm{WPL}}$.
We consider the same data generation models (I, II and III) as in Section \ref{sec:simulations}.
%To illustrate the advantage of the test based on $T_n$ in detecting sparse signals, 
For $\btheta$,
we set its first $\lfloor c_0\log p \rfloor$ components as non-zero, while the other elements are all zero. $c_0$ is chosen from $0.5$ and $1$. 
The magnitude of non-zero entries in $\btheta$ is $\kappa (\log p/n)^{1/2}$, where $\kappa$ is chosen from $0$ to $5$.
Note that $\kappa=0$ refers to the null hypothesis. 
We consider $n=50$ or $100$, and $p=100$ and $1000$ for each sample size.

\begin{figure}[htbp]
	\centerline{
		\includegraphics[width=11cm]{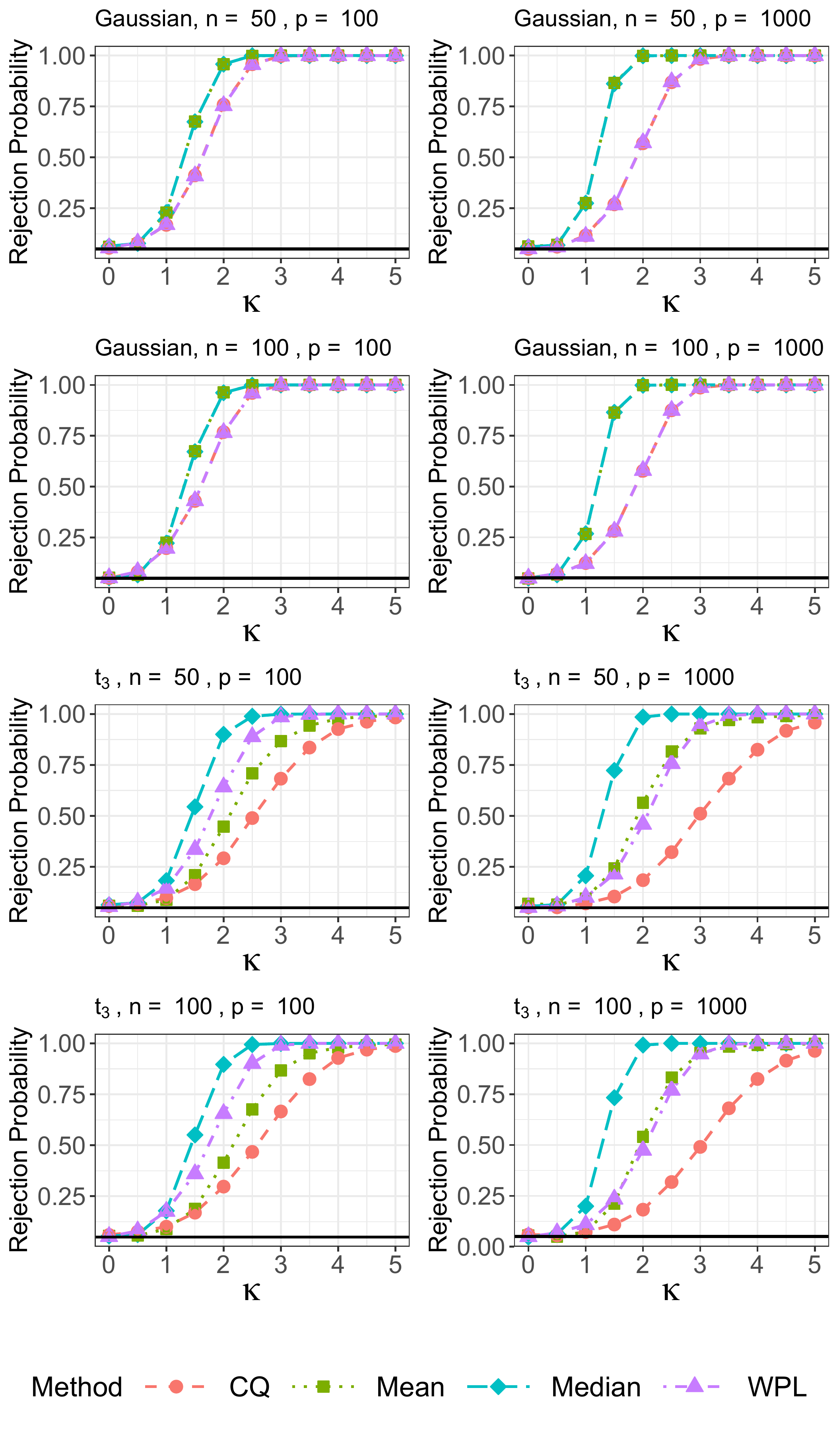}
	}
	\caption{Empirical size and power of the four tests (CQ, Mean, Median, WPL) for Models I and II with $c_0=0.5$ and $\rho=0$. The horizontal black solid line refers to the nominal $5\%$ significance level. ``Gaussian'' denotes the multivariate normal distribution, and $t_3$ denotes the multivariate $t$-distribution with $3$ degrees of freedom.}
	\label{Fig2}
\end{figure}

\begin{figure}[htbp]
	\centerline{
		\includegraphics[width=11cm]{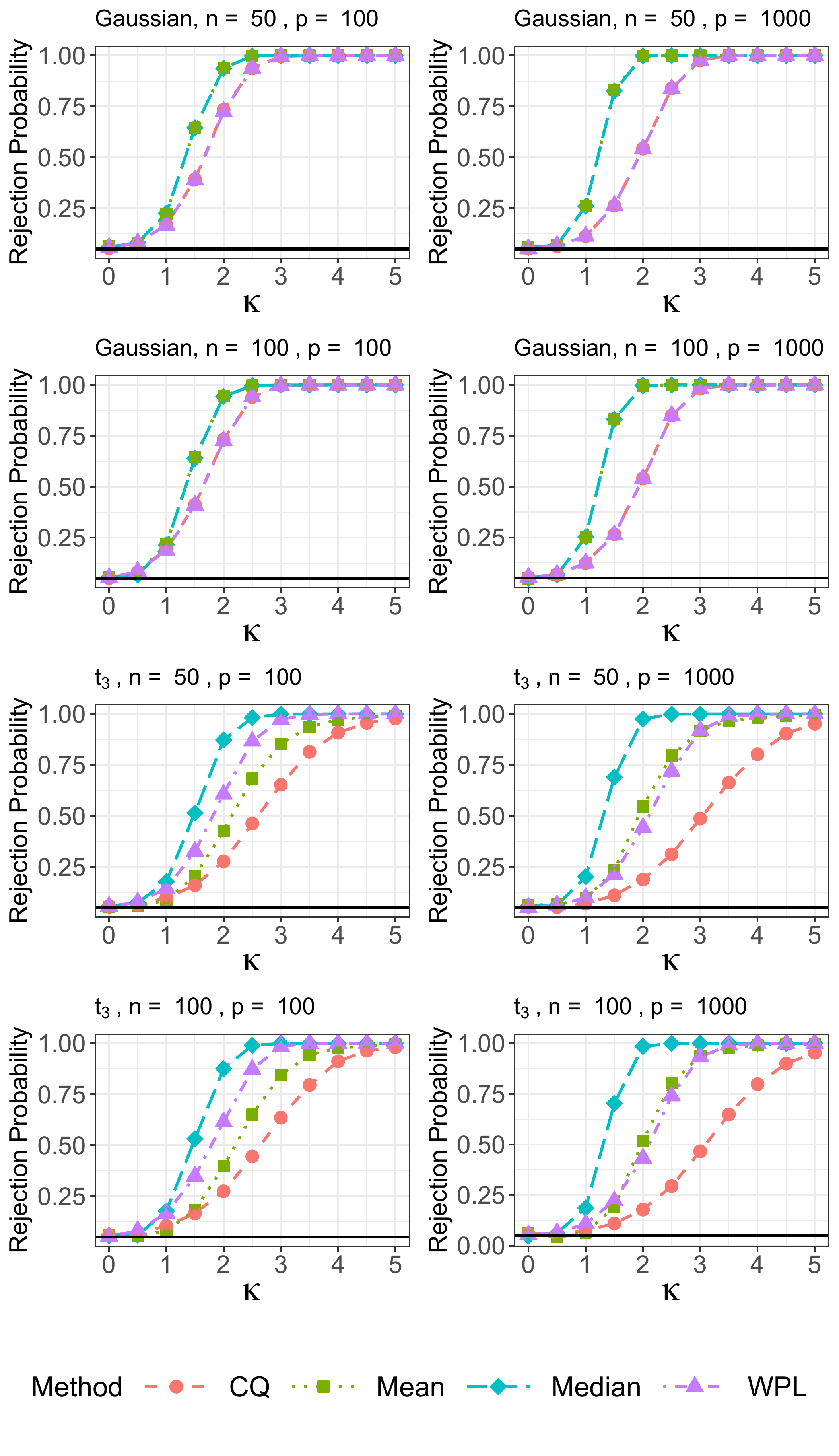}
	}
	\caption{Empirical size and power of the four tests (CQ, Mean, Median, WPL) for Models I and II with $c_0=0.5$ and $\rho=0.2$. The horizontal black solid line refers to the nominal $5\%$ significance level. ``Gaussian'' denotes the multivariate normal distribution, and $t_3$ denotes the multivariate $t$-distribution with $3$ degrees of freedom.}
	\label{FigS1}
\end{figure}

\begin{figure}[htbp]
	\centerline{
		\includegraphics[width=11cm]{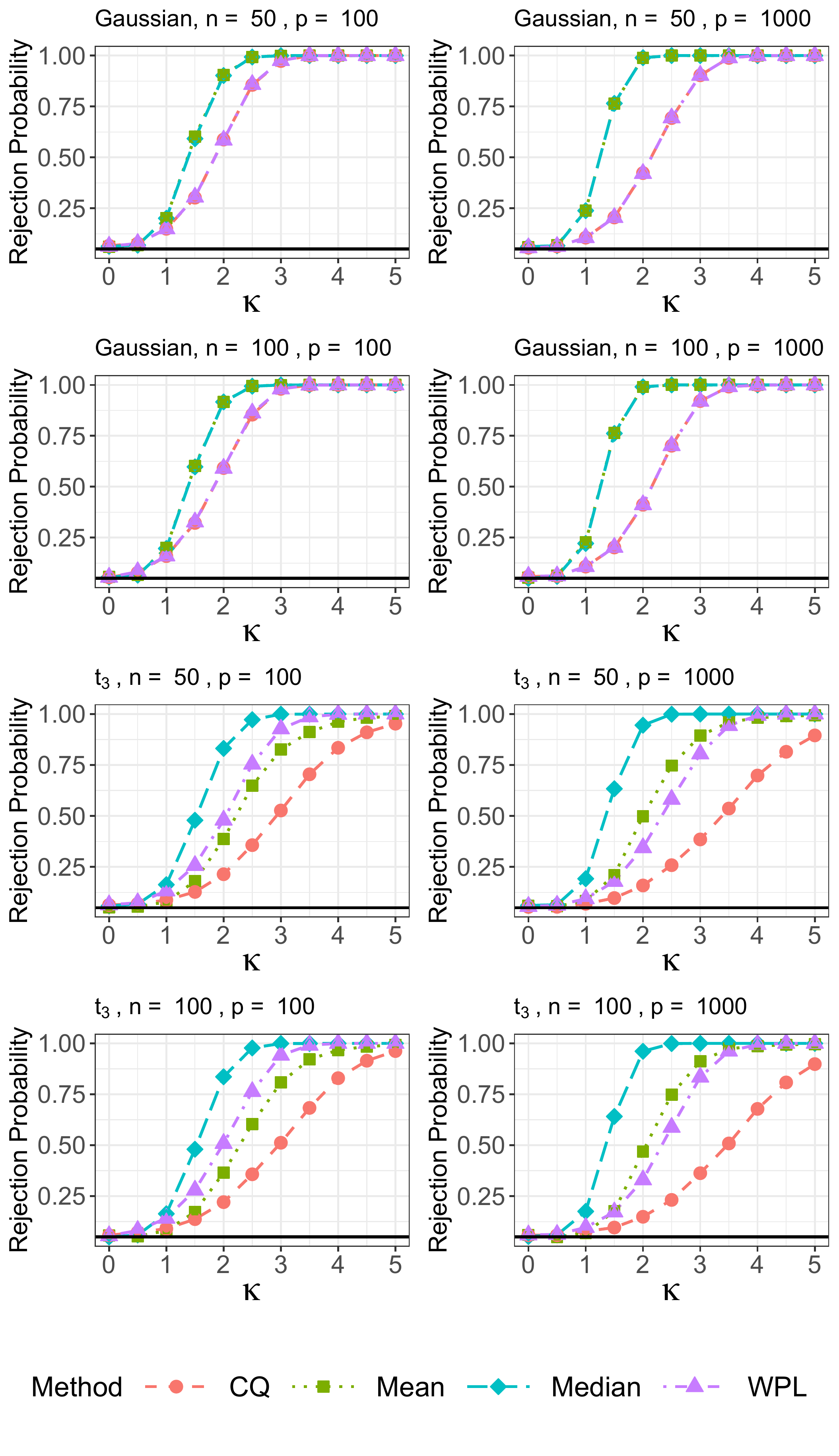}
	}
	\caption{Empirical size and power of the four tests (CQ, Mean, Median, WPL) for Models I and II with $c_0=0.5$ and $\rho=0.5$. The horizontal black solid line refers to the nominal $5\%$ significance level. ``Gaussian'' denotes the multivariate normal distribution, and $t_3$ denotes the multivariate $t$-distribution with $3$ degrees of freedom.}
	\label{FigS3}
\end{figure}

\begin{figure}[htbp]
	\centerline{
		\includegraphics[width=11cm]{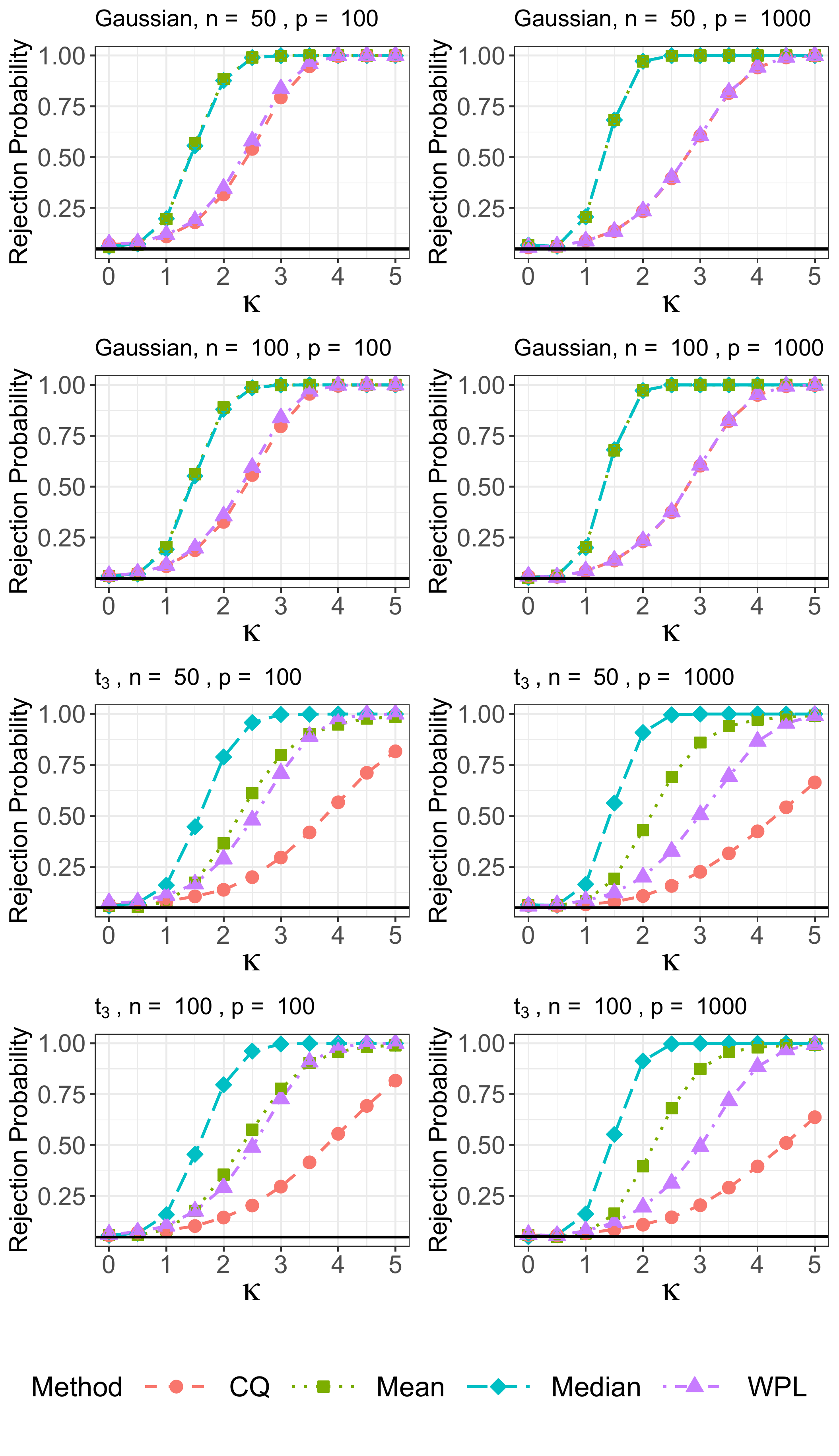}
	}
	\caption{Empirical size and power of the four tests (CQ, Mean, Median, WPL) for Models I and II with $c_0=0.5$ and $\rho=0.8$. The horizontal black solid line refers to the nominal $5\%$ significance level. ``Gaussian'' denotes the multivariate normal distribution, and $t_3$ denotes the multivariate $t$-distribution with $3$ degrees of freedom.}
	\label{Fig3}
\end{figure}

\begin{figure}[htbp]
	\centerline{
		\includegraphics[width=11cm]{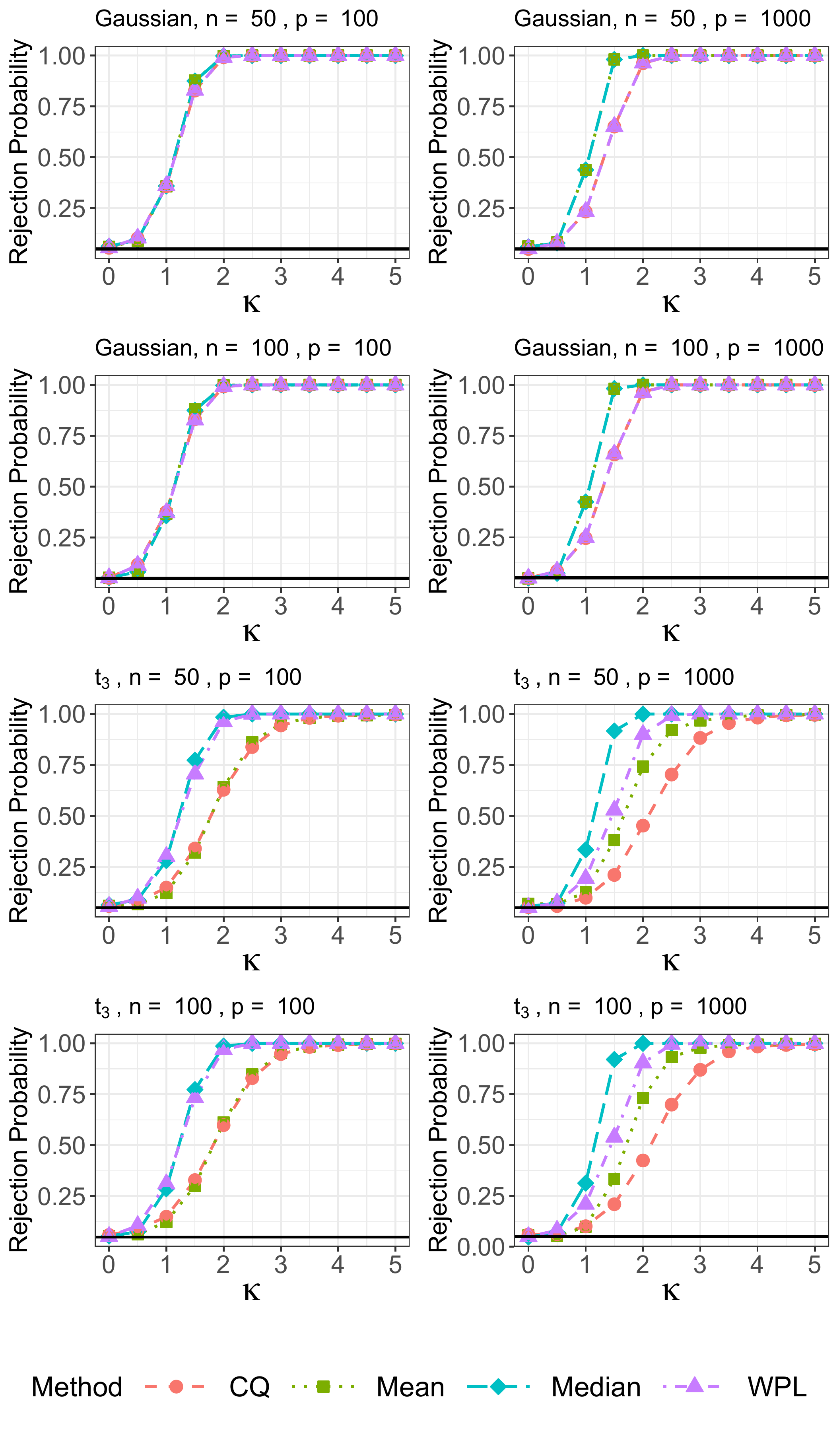}
	}
	\caption{Empirical size and power of the four tests (CQ, Mean, Median, WPL) for Models I and II with $c_0=1$ and $\rho=0$. The horizontal black line refers to the nominal $5\%$ significance level. ``Gaussian'' denotes the multivariate normal distribution, and $t_3$ denotes the multivariate $t$-distribution with $3$ degrees of freedom.}
	\label{FigS5}
\end{figure}

\begin{figure}[htbp]
	\centerline{
		\includegraphics[width=11cm]{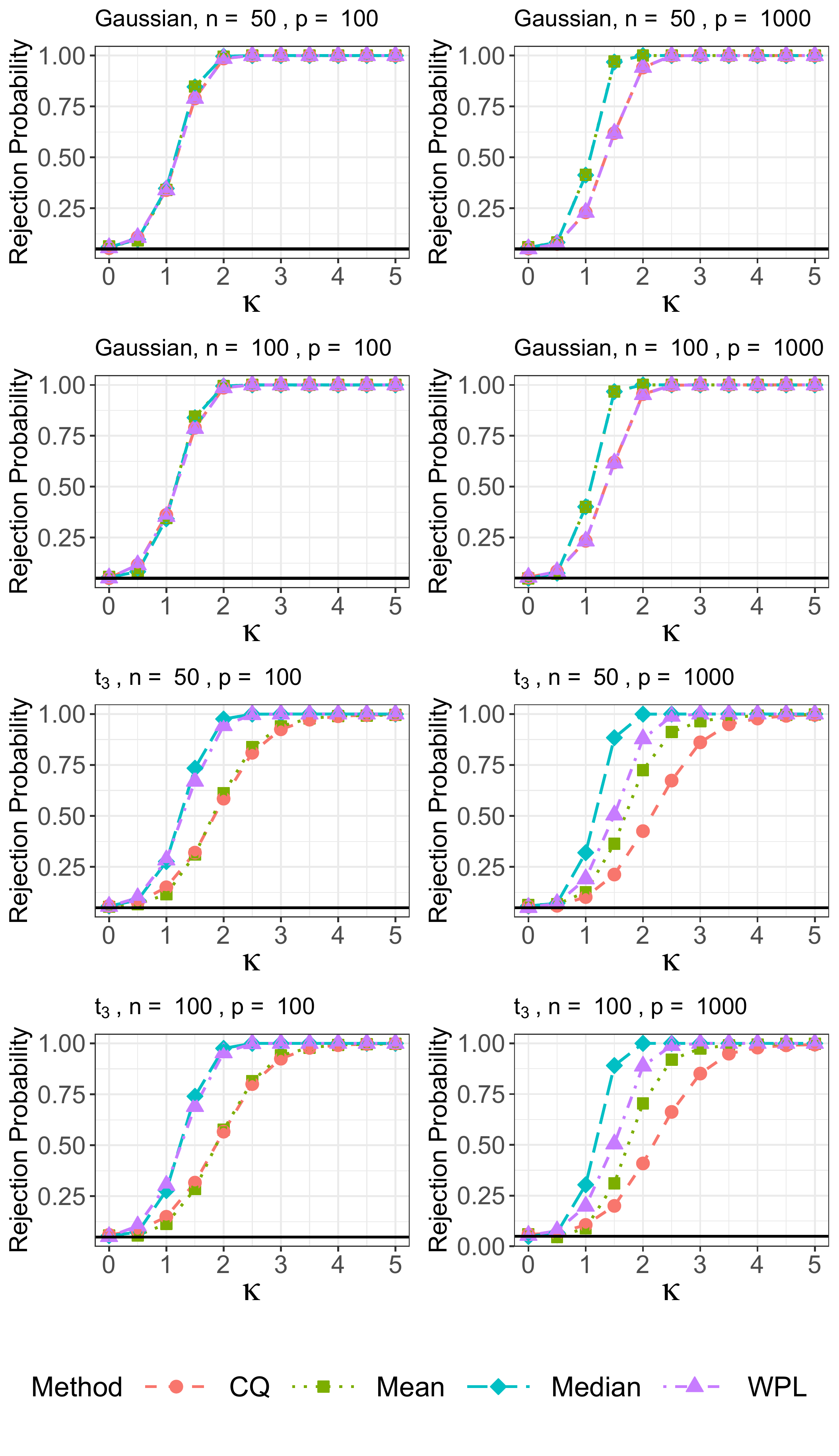}
	}
	\caption{Empirical size and power of the four tests (CQ, Mean, Median, WPL) for Models I and II with $c_0=1$ and $\rho=0.2$. The horizontal black solid line refers to the nominal $5\%$ significance level. ``Gaussian'' denotes the multivariate normal distribution, and $t_3$ denotes the multivariate $t$-distribution with $3$ degrees of freedom.}
	\label{FigS2}
\end{figure}

\begin{figure}[htbp]
	\centerline{
		\includegraphics[width=11cm]{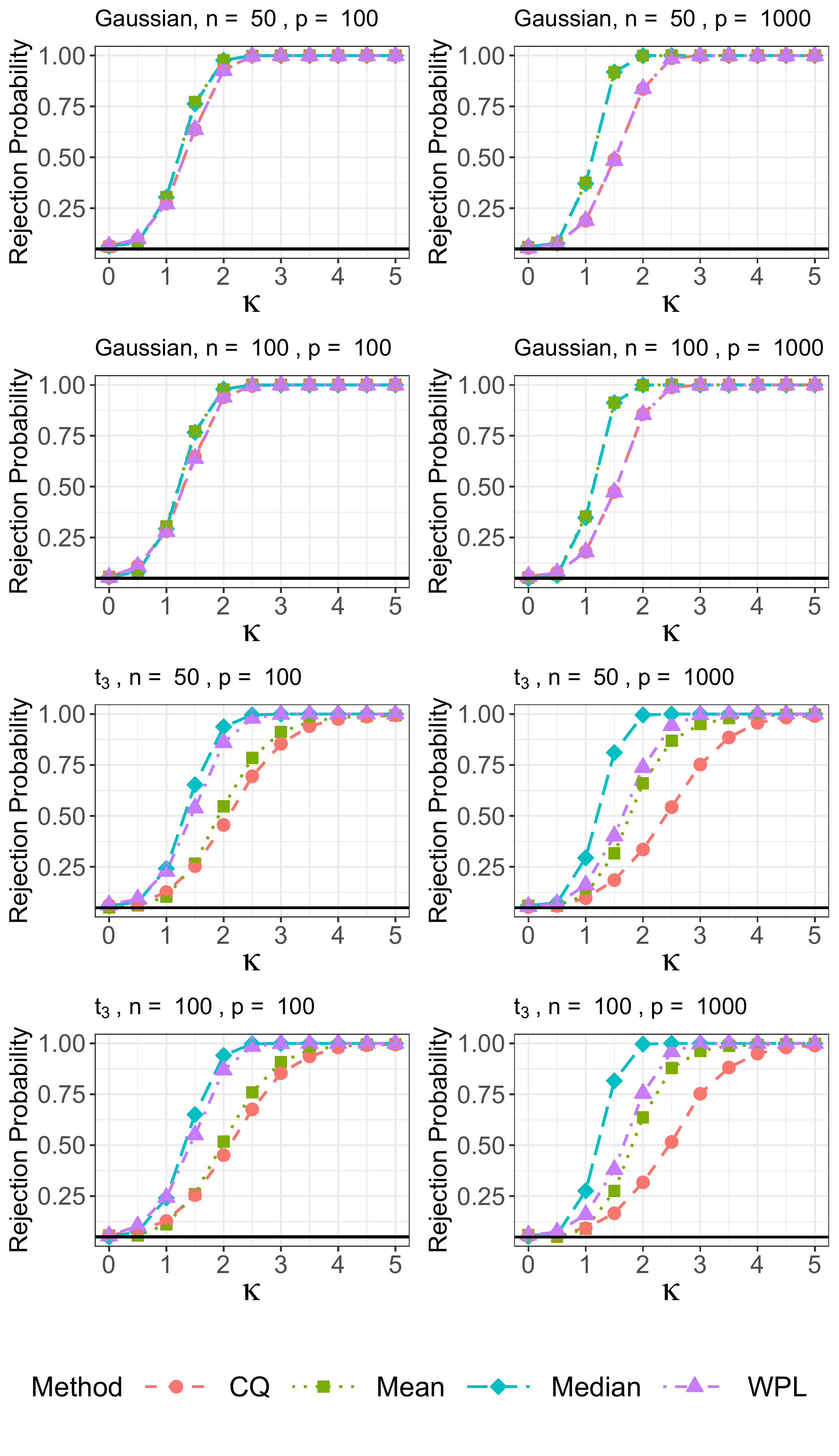}
	}
	\caption{Empirical size and power of the four tests (CQ, Mean, Median, WPL) for Models I and II with $c_0=1$ and $\rho=0.5$. The horizontal black solid line refers to the nominal $5\%$ significance level. ``Gaussian'' denotes the multivariate normal distribution, and $t_3$ denotes the multivariate $t$-distribution with $3$ degrees of freedom.}
	\label{FigS4}
\end{figure}

\begin{figure}[htbp]
	\centerline{
		\includegraphics[width=11cm]{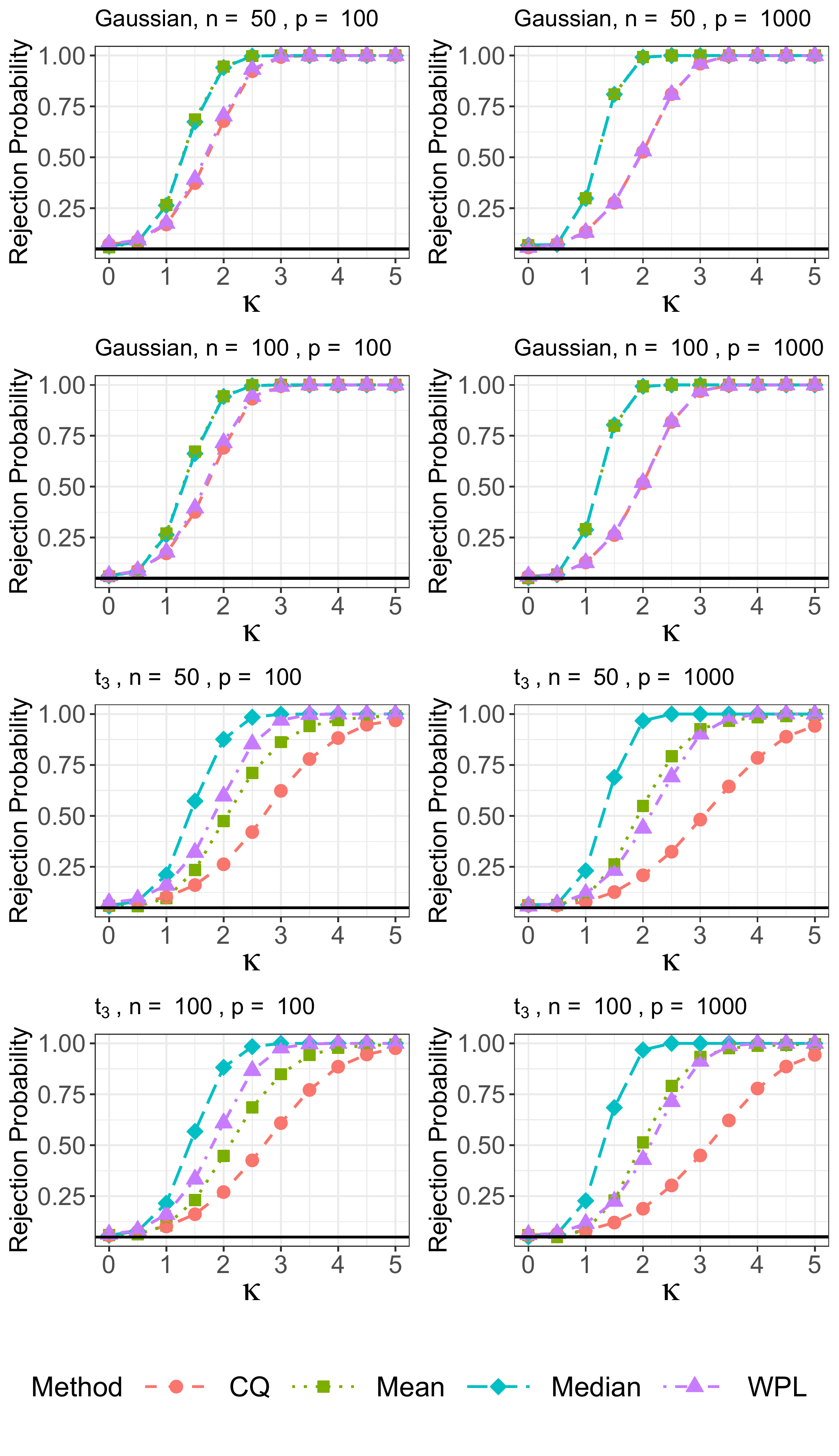}
	}
	\caption{Empirical size and power of the four tests (CQ, Mean, Median, WPL) for Models I and II with $c_0=1$ and $\rho=0.8$. The horizontal black line refers to the nominal $5\%$ significance level. ``Gaussian'' denotes the multivariate normal distribution, and $t_3$ denotes the multivariate $t$-distribution with $3$ degrees of freedom.}
	\label{FigS6}
\end{figure}

Figures \ref{Fig2}--\ref{FigS6} plot the empirical size ($\kappa=0$) and power ($\kappa\neq0$) of four (CQ, Mean, Median, and WPL) tests at the $5\%$ significance level for Models I and II. % with $c_0=0.5$ and  $\rho=0$. 
%The results for Model III and other scenarios of Models I and II are similar and are reported in the supplementary material due to the space limitation.
The results of $\kappa=0$ indicates that the empirical sizes of all these four tests are close to the nominal significance level under different case scenarios.
When $\kappa\neq 0$, the power of these tests increases as $\kappa$ increases, that is, as the signal getting stronger.
For Gaussian data, the Mean test based on $T_{\mathrm{Mean}}$ and the Median test based on $T_{n}$ have similar power performances, and they advance both the CQ test and the WPL test, which are $L_2$-norm type tests.
In addition, when the data are from multivariate $t$-distribution, the Median test outperforms the Mean test, which shows the superiority of the procedure based on the sample spatial median over that based on the sample mean under heavy-tailedness.
In summary, the Median test based on $T_n$ is preferred among the four tests when the alternative is sparse and the underlying distribution is heavy-tailed.

\begin{figure}[htbp]
	\centerline{
		\includegraphics[width=11cm]{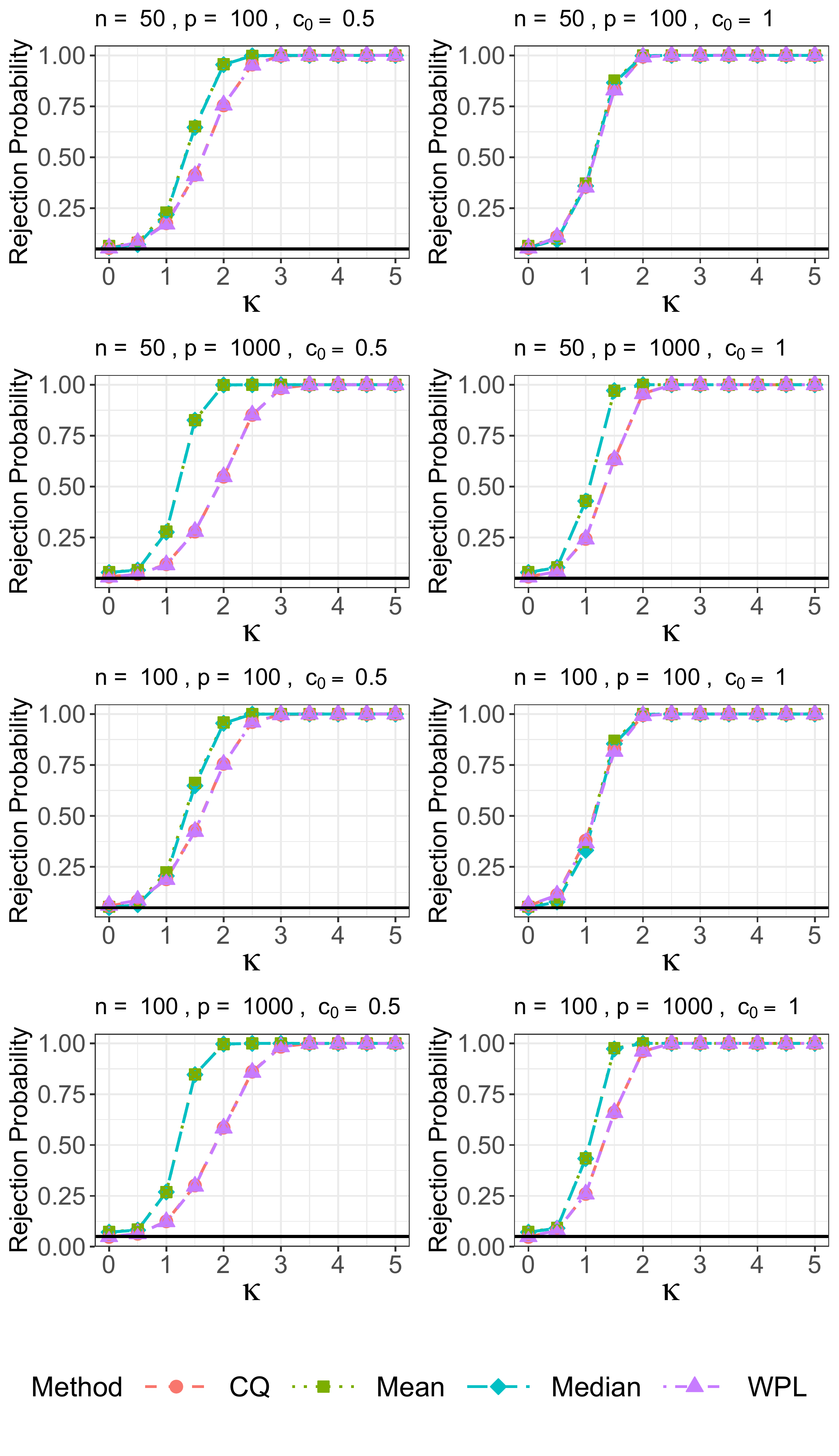}
	}
	\caption{Empirical size and power of the four tests (CQ, Mean, Median, WPL) for Model III with $\rho=0$. The horizontal black solid line refers to the nominal $5\%$ significance level. }
	\label{FigS01}
\end{figure}

Second, Figure \ref{FigS01} depicts empirical size and power of the four tests (CQ, Mean, Median, WPL) for Model III with $\rho=0$. 
It can be seen that, even Model III is not a member of the elliptical distribution family, the size of the Median test can still control the size at the nominal level $\alpha=0.05$, and this is also the case for the WPL test. 
%This shows the robustness of the Median test under model misspecification. 
We can also see that the Median test and the Mean test have better power performance than the CQ test and the WPL test, especially for $c_0=0.5$ when the number of non-zero element in $\btheta$ is relatively small.

\end{document}